\documentclass[
    aps,
    prx,
    superscriptaddress,
    twocolumn,
    a4paper,
    floatfix,
    longbibliography
]{revtex4-2}

\usepackage[american]{babel}
\usepackage[utf8x]{inputenc}

\usepackage{grffile} 

\usepackage{newtxtext,newtxmath}
\usepackage{microtype}

\usepackage{amsmath}
\usepackage{makecell}
\usepackage{bbm}
\usepackage{bm}
\usepackage{mathtools}
\usepackage{dsfont}
\usepackage{braket}
\usepackage{cancel}
\usepackage{slashed}
\usepackage{bbold}
\usepackage{dsfont}

\usepackage{physics}
\usepackage{multirow}

\usepackage{graphicx}
\usepackage[svgnames]{xcolor}

\usepackage{siunitx} 

\usepackage{ragged2e}
\usepackage{array}
\usepackage{tabularx}
\usepackage{booktabs}

\definecolor{cset-aps-blueberry}{RGB}{28,128,158}
\definecolor{cset-aps-blue}{RGB}{46,44,184}
\definecolor{cset-aps-turquoise}{RGB}{0,67,88}
\definecolor{cset-aps-limegreen}{RGB}{190,219,67}
\definecolor{cset-aps-green}{RGB}{31,138,112}
\definecolor{cset-aps-yellow}{RGB}{255,225,25}
\definecolor{cset-aps-orange}{RGB}{253,116,0}
\definecolor{cset-aps-red}{RGB}{219,0,43}

\usepackage{tikz}
\usepackage{tikz-feynman}

\usepackage{pgfplots}
\pgfplotsset{%
    every axis legend/.append style={%
        cells={anchor=west},
        at={(0.96,0.04)},
        anchor=south east,
        font=\scriptsize,
        },
    every axis/.append style={%
        yticklabel style={%
            /pgf/number format/fixed zerofill,
            /pgf/number format/precision=2},
        },
    width= \textwidth,
    height=8cm,
    xmajorgrids=true,
    xminorgrids=false,
    minor x tick num=1,
}

\usetikzlibrary{decorations}
\usetikzlibrary{arrows.meta,arrows}

\usepackage{pict2e,picture}

\makeatletter
\DeclareRobustCommand{\Arrow}[1][]{%
\check@mathfonts
\if\relax\detokenize{#1}\relax
\settowidth{\dimen@}{$\m@th\rightarrow$}%
\else
\setlength{\dimen@}{#1}%
\fi
\sbox\z@{\usefont{U}{lasy}{m}{n}\symbol{41}}%
\begin{picture}(\dimen@,\ht\z@)
\roundcap
\put(\dimexpr\dimen@-.7\wd\z@,0){\usebox\z@}
\put(0,\fontdimen22\textfont2){\line(1,0){\dimen@}}
\end{picture}%
}
\makeatother

\usepackage{hyperref}
\hypersetup{%
    colorlinks=true,
    linkcolor={cset-aps-red},
    linkbordercolor={cset-aps-red},
    filecolor={cset-aps-orange},
    filebordercolor={cset-aps-orange},
    citecolor={cset-aps-blue},
    citebordercolor={cset-aps-blue},
    urlcolor={cset-aps-green},
    urlbordercolor={cset-aps-green},
    menucolor={cset-aps-limegreen},
    menubordercolor={cset-aps-limegreen},
    breaklinks=true,
    pdfborderstyle={/S/U/W 2},
    pdfpagemode=UseOutlines,
    pdfstartpage={1},
}

\newcommand{\ee}{\operatorname{e}}
\newcommand{\ii}{\operatorname{i}}
\newcommand{\re}{\mathrm{Re}}
\newcommand{\im}{\mathrm{Im}}

\usepackage{xspace}

\newcommand{\ie}{i.\,e.,}
\newcommand{\eg}{e.\,g.,}
\newcommand{\vect}[1]{\boldsymbol{#1}}

\newcommand*{\wideOverset}[2]{\mathrel{\overset{#1}{\resizebox{\widthof{\kern1.25pt\ensuremath{\mathrel{\overset{#1}{#2}}}}}{\heightof{\ensuremath{#2}}}{\ensuremath{#2}}}}}

\usepackage{wasysym}

\usepackage{diagbox}

\usepackage{simplewick}

\usepackage[clock]{ifsym}
\usepackage{lipsum}

\newcommand{\orcid}[1]{\href{https://orcid.org/#1}{\includegraphics[width=7pt]{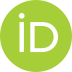}}}

\begin{document}
\title{Phase shifts and fluctuations from motional, relativistic, and dark-matter effects in ion-trap clocks}

\newcommand{\affHAN}{\address{Institut f{\"u}r Quantenoptik, Leibniz Universit{\"a}t Hannover, Welfengarten 1, D-30167 Hannover, Germany}}
\newcommand{\affULM}{\address{Institut f{\"u}r Quantenphysik and Center for Integrated Quantum Science and Technology (IQST), Universit{\"a}t Ulm, Albert-Einstein-Allee 11, D-89081 Ulm, Germany}}
\newcommand{\affTUDa}{\address{Technische Universit{\"a}t Darmstadt, Fachbereich Physik, Institut f{\"u}r Angewandte Physik, Schlossgartenstr. 7, D-64289 Darmstadt, Germany}}

\author{Alexander Bott\,\orcid{0000-0002-7986-4834}\textsuperscript{\affULM}}
\email{alexander.bott@uni-ulm.de}
\email{alexander.bott@bott.org}
\affULM

\author{Fabio Di Pumpo\,\orcid{0000-0002-6304-6183}\textsuperscript{\affULM}}
\affULM

\author{Enno Giese\,\orcid{0000-0002-1126-6352}\textsuperscript{\affTUDa}}
\affTUDa

\begin{abstract}
\noindent
Ions possess both center-of-mass motion and internal states, rendering trapped-ion optical clocks susceptible to perturbations of both degrees of freedom, which results in systematic shifts and fluctuations with motional and relativistic origins.
We develop a quantum theory that connects these effects to filter functions determined by the employed pulse sequences and show how the interrogation can be tailored to suppress or enhance specific perturbations.
Our treatment includes laser frequency noise and motional effects such as recoil and Doppler shifts, micromotion and excess micromotion, as well as state-dependent trap imperfections and accelerations, \eg{} caused by Stark shifts.
We further account for relativistic effects including time dilation and the mass defect, and show how the same framework can be applied to hypothetical dark-matter signals.
By considering ensemble-averaged phases and their variances, we characterize both systematic shifts and fluctuations, while asynchronous operation relative to the Paul-trap drive allows motional fluctuations to average down between runs.
Our results provide a unified framework for characterizing and controlling the systematic and statistical contributions that determine the trueness and precision of trapped-ion optical clocks.
\end{abstract}

\maketitle

\section{Introduction}
Atomic and ion clocks employ sequences of laser pulses to drive transitions between internal states and, in some implementations, to determine time through the resulting interference patterns~\cite{Poli2013, Ludlow2015, Safronova2018, Fortier2026}.
While these rely on a superposition of internal states, ions are quantum objects, and their center-of-mass motion is therefore also quantum mechanical, encoding their wave-packet character~\cite{Leibfried2003, Martinez2022}.
Since optical transitions of free particles lead to substantial recoil~\cite{Giese2015, Rudolph2020, Chiarotti2022, Niehof2025}, which forms the basis of light-pulse atom interferometry~\cite{Bongs2019, Abend2024}, optical clocks confine the particles to suppress this effect~\cite{Ludlow2015, Safronova2018}.
However, the motion of the particles in the trap, such as ions in a Paul trap~\cite{Paul1990, Glauber1992, Schleich2001, Leibfried2003}, must still be described quantum mechanically.

The performance of clocks is characterized by both systematic frequency shifts and stochastic frequency fluctuations. 
If not characterized, systematic shifts introduce a bias and deteriorate the trueness of the measurement, requiring theoretical estimates and experimental characterization for their correction.
The fluctuations act as statistical noise and determine the precision of the experiment, which can improve as the number of measurements increases.
To improve clock performance, hyper-Ramsey~\cite{Yudin2010, Huntemann2012, Zanon-Willette2022, Zanon-Willette2016, Hobson2016} and dynamical-decoupling~\cite{Cywinski2008, Kotler2011, Biercuk2011, Shaniv2017, Aharony2021, Zaheer2025} pulse sequences have been developed to suppress probe-induced frequency shifts and dephasing, respectively. 
Additionally, tailored interrogation protocols have been developed to reduce aliasing and the Dick effect~\cite{Dick1990, Santarelli1998, Westergaard2010, Hume2016, Schioppo2017, Schulte2020}.
These pulse-sequences primarily address errors in the control of internal states of two-level systems and do not typically account for the quantized motion or relativistic effects of ions.
In contrast, motional shifts in ion clocks have been analyzed, including excess micromotion~\cite{Berkeland1998, Keller2015, Higgins2021, Lindvall2025b, Arnold2026}, recoil effects beyond the Lamb--Dicke limit~\cite{Lizuain2007}, Stark shifts~\cite{Itano2000, Dube2013, Tofful2024, Lindvall2025b, Arnold2026}, and relativistic effects such as time dilation induced by the mass defect~\cite{Yudin2018, Sonnleitner2018, Asano2024}, which gives rise to the second-order Doppler shift~\cite{Ludlow2015, Martinez2022, Miao2022}. 
Previous work has examined relativistic motional shifts in trapped-ion clocks using quantum-mechanical treatments~\cite{Martinez2022} and has investigated generalized interrogation sequences~\cite{Zanon-Willette2022}. 
However, how the quantum state of the ion's motion shapes the phase response to general interrogation sequences, and how this response determines the associated variances and covariances, has not been addressed within a unified framework.

In this paper, we theoretically connect the quantized, micromotion-dressed wave-packet motion to dynamical decoupling and investigate the resulting fluctuations of motional shifts.
We incorporate laser frequency noise and consider signatures of a fluctuating dark-matter field in the clock signal. 
We neglect light shifts, which could be incorporated into the framework, and assume instantaneous pulses~\cite{Ufrecht20202, Ufrecht2019}, thermal motional states~\cite{Martinez2022, Leibfried2003}, and an interrogation protocol in which the pulse sequence is not synchronized with the micromotion. 
In this regime, the motional phase becomes a shot-to-shot random variable. 
Its mean gives rise to a systematic shift, while its variance contributes to the statistical uncertainty.
We develop a Fourier description of the ion's motion in the trap and combine it with a perturbative branch description of the interferometer~\cite{Ufrecht20202}.
This framework allows us to treat arbitrary pulse sequences using filter functions. 
We recover established results for Ramsey interrogation with two pulses and further extend them to periodic dynamical-decoupling sequences~\cite{Viola1998, Cywinski2008, Biercuk2011, Aharony2021}.
We find that the filter function depends on whether the number of pulses in such sequences is even or odd. 
In particular, odd, echo-type sequences~\cite{Hahn1950, Cywinski2008, Biercuk2011} suppress systematic phase shifts while coherently enhancing resonant fluctuations, which can be exploited for the detection of dark matter~\cite{Zaheer2025, Aharony2021, DiPumpo2023b}. 
We further show that dark-matter searches can be compromised when the target signal frequency overlaps resonances of the trapped-ion motion, including the secular frequency, the rf drive, and its relevant sidebands or harmonics.
Our quantum treatment of the internal and external degrees of freedom provides access to both systematic shifts, which determine the trueness of the measurement, and statistical fluctuations, which determine its precision and average down with repeated measurements. 

In Sec.~\ref{sec: phase shifts}, we introduce the Hamiltonian including the relevant perturbations, calculate the motional effects in the Paul trap, and obtain the systematic shifts from ensemble averages. The variances and covariances of the different phase contributions determine the corresponding fluctuations and are discussed in Sec.~\ref{sec: fluctuations}, before we conclude in Sec.~\ref{sec: discussion}.
Appendix~\ref{app: COMMotion} provides details of the quantum description of the center-of-mass motion, Appendix~\ref{app: PertPhase} presents the perturbative treatment of the interferometer phase, and Appendix~\ref{app: Ensemble averages} gives details of the ensemble averages, including analytic expressions for the filter function.

\section{Phase shifts in ion clocks}
\label{sec: phase shifts}
\subsection{Phase accumulation and perturbations}
\begin{figure}[htb]
    \centering
    \def\svgwidth{\columnwidth}
	\scalebox{1}{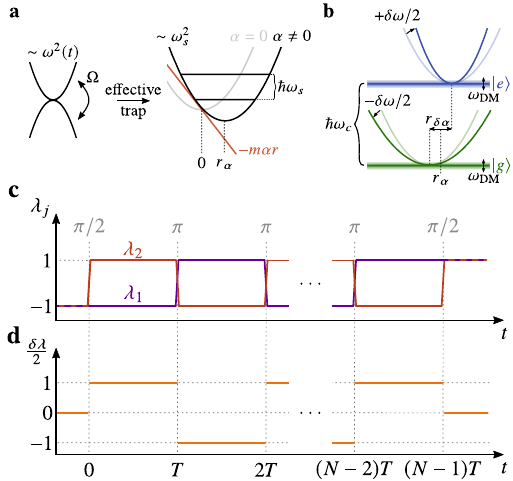}
	\caption{
    (a) The unperturbed Paul trap, with a time-dependent potential $\sim \omega^2(t)$ oscillating at radio frequency $\Omega$, gives rise to an effective trapping potential with secular frequency $\omega_s$.
    A linear acceleration $\alpha$ displaces the effective potential by the sag $\vect{r}_{\vect{\alpha}}$.
    (b) State-dependent perturbations for the two clock states $\ket{e}$ (blue) and $\ket{g}$ (green), separated by the clock energy $\hbar\omega_c$.
    A non-magic, differential trap frequency is described by $\delta\omega$, while a differential acceleration gives rise to a sag difference $\vect{r}_{\delta\vect{\alpha}}$ between the equilibrium positions of the two states.
    A dark-matter field modulates the clock energy at frequency $\omega_\text{DM}$.
    (c) Time-dependent state-response functions $\lambda_{j}(t)$ of branch $j$ describing a pulse sequence of $N$ pulses, in which two $\pi/2$ pulses sandwich equidistant $\pi$ pulses. 
    (d) Their difference gives the sensitivity function $\delta\lambda/2$, which characterizes the response to the pulse sequence.
    }
	\label{fig: Fig1}
\end{figure}

We describe the dynamics of a trapped ion of mass $m$ and internal energy splitting $\hbar \omega_{\text{c}}$ for clocks relying on interferometric phase signals through the Hamiltonian $\hat{H}_j=\sum_\kappa \hat{H}_j^{(\kappa)}$, depending on the branch $j$ of the clock.
The unperturbed part, corresponding to $\kappa=0$, is given by
\begin{align}
\begin{split}
\label{eq:H0}
    \hat{H}_{j}^{(0)}  = \lambda_j \frac{\hbar \omega_{\text{c}}}{2}+\frac{\hat{\vect{p}}^2}{2m}+\frac{m}{2}\hat{\vect{r}}^\text{T}\omega^2\hat{\vect{r}} -m\vect{\alpha}^\text{T}\hat{\vect{r}}+\dot\lambda_j\frac{\hbar\phi_{\text{L}}}{2}.
\end{split}
\end{align}
Here, the time-dependent state-response function~\cite{Niehof2025} $\lambda_j$ encodes the internal state of the ion, with $\lambda_j= + 1$ for the ion being in the excited state and $\lambda_j= - 1$ for the ground state, as shown in Fig.~\ref{fig: Fig1}.
It defines the two branches of time evolution for the two interfering components, which are solely determined by the internal state the ion is in at a given point in the interrogation sequence.
The energy splitting $\hbar\omega_c$ already incorporates constant shifts of the clock frequency, caused for example by Stark or quadrupole shifts~\cite{Ludlow2015, Arnold2026, Lindvall2025b} or relativistic effects~\cite{Martinez2022}.
The center-of-mass position and momentum operators $\hat{\vect{r}}$ and $\hat{\vect{p}}$ fulfill the canonical commutation relation $[\hat{r}_i, \hat{p}_l] = \ii \hbar \delta_{il}$ component-wise. 
During the interrogation, the ion is confined in a three-dimensional Paul trap~\cite{Leibfried2003, Schleich2001, Paul1990} centered at the origin $\vect{r}=0$. 
This trap is governed by the tensor-valued, time-dependent trap frequency $\omega^2 = \omega_\text{DC}^2+\omega_{\text{AC}}^2\cos(\Omega t + \theta_{\text{rf}})$, where the constant DC contribution $\omega_\text{DC}$ is complemented by the AC one $\omega_\text{AC}$, leading to oscillations at radio frequency (rf) $\Omega$, with a total offset $\theta_\text{rf}$ of the trap phase.
We assume the Paul trap to be ideal, such that $\omega_{\text{AC}}$ and $\omega_{\text{DC}}$ share the same principal axes.
Additionally, the ion experiences an acceleration $\vect{\alpha}$ from a linear potential, caused for example by leading-order gravity~\cite{Martinez2022} or non-vanishing static electric fields~\cite{Higgins2021, Berkeland1998} at the trap center. 
This effect induces an acceleration sag of the trap, resulting in excess micromotion~\cite{Berkeland1998,Martinez2022}.
The laser phase $\phi_\text{L}$, evaluated at the trap center, is imprinted~\cite{Niehof2025} through a branch-dependent delta distribution $\dot \lambda_j$ at each pulse that induces a transition between the two internal states. 
It includes the instantaneous laser frequency through $\dot \phi_{\text{L}}$.

The Hamiltonian includes a series of branch-dependent perturbations shown in Tab.~\ref{tab: perturbations}.
The first perturbation, denoted by $\kappa= \text{rec}$, represents the photon recoil $\pm\hbar\vect{k}$ that is instantaneously imparted to the ion upon absorption or emission of a photon from a laser pulse. 
This contribution is suppressed in the Lamb--Dicke regime~\cite{Ludlow2015, Lizuain2007}, where the wavelength of the laser is much larger than the characteristic length scales of the ion's motion in the trap, and gives rise to the first-order Doppler shift~\cite{Ludlow2015}, as shown later.

The differential acceleration $\delta\vect{\alpha}$ between both internal states constitutes another perturbation, which can be caused for example by spatial gradients of the Stark shift~\cite{Itano2000, Guevara-Bertsch2022}, black-body radiation~\cite{Haslinger2018} or magnetic fields~\cite{Mintert2001,Ivanov2017}, as well as the mass defect in combination with contributions from gravitational redshifts~\cite{Martinez2022}, or possible violations of the Einstein equivalence principle~\cite{Damour2010,DiPumpo2021,DiPumpo2022}. 
All of these effects result in a potential scaling linearly with the ion's position operator $\hat{\vect{r}}$ and will be denoted by the index $\kappa=\delta \vect{\alpha}$.

Perturbations resulting in different potentials between internal states, scaling quadratically with $\hat{\vect{r}}$, lead effectively to a shift of the trap frequency depending on the internal state. 
The main source for this effect results from the trap potential itself, which, due to the different electric polarizabilities of the internal states, gives rise to a differential quadratic Stark shift~\cite{Itano2000, Dube2013, Higgins2017, Higgins2019} and, therefore, to a differential frequency shift $\delta \omega^2$.
Here, we neglected time-dependent oscillations of the differential trap frequency, since they are additionally suppressed for large radio frequencies. 
For $\delta\omega^2=0$, the trapping potentials are state independent and can therefore be considered to be magic~\cite{Lemonde2009, Derevianko2011, Ahlheit2025}.
We use the index $\kappa = \delta \omega$ to refer to perturbations of this kind.

The index $\kappa = \text{2ds}$ denotes relativistic second-order Doppler shifts as a consequence of the mass defect of the ion~\cite{Martinez2022}, where the mass difference $\delta m = \hbar \omega_{\text{c}}/c^2$ between the ions in the excited and ground state, with $c$ being the speed of light, causes a difference in the center-of-mass energies associated with each state~\cite{Yudin2018, Asano2024, Sonnleitner2018}. 
The second-order Doppler shift in particular is caused by the resulting difference of kinetic energies between the internal states.

Finally, as an example of a beyond-Standard-Model effect one might wish to detect, we also include a state-dependent coupling of ultralight dark matter, \eg{} through a scalar Dilaton field~\cite{Damour2010, Damour1994, Alves2000, Damour2012, Hees2018, Derr2023} or an Axion field~\cite{Zhao2024, Kim2024, Flambaum2023, Bauer2025}, to the ion. 
Here, the constant $\beta$ is the mean coupling strength of the dark-matter field to the two internal states~\cite{Derr2023, DiPumpo2022}, while the field $\varrho(t)$ itself is given in units of frequency and evaluated at the origin, neglecting suppressed~\cite{Derr2023, DiPumpo2023b} spatial effects. 
This dark-matter field results in a differential modulation $\hbar\varrho$ of the energy levels of the ion, and will be represented by the index $\kappa = \text{DM}$ in the following.
We do not take into account Post-Newtonian terms~\cite{Martinez2022} and other corrections that do not depend on the internal states of the ion, since such effects are common for both branches and thus cancel to leading-order in the interferometer phase.

{\renewcommand{\arraystretch}{2.75}
\begin{table*}
\caption{\label{tab: perturbations}%
Hamiltonian contributions and the systematic phase shifts they introduce.
For each perturbation $\kappa$, we list the respective branch-dependent Hamiltonian $\hat{H}_j^{(\kappa)}$, the dimensionless factor $g_i^{(\kappa)}$ that determines its order of magnitude, the characteristic frequency $\omega_i^{(\kappa)}$, and the generator $\hat{h}_i^{(\kappa)}$.
The index $i$ denotes the different principal axes of the trap, with summation over $i$ implicitly assumed.
The ensemble-averaged systematic phase shifts $\overline{\varphi_{\kappa}}$ depend on the filter function $\mathcal{F}(\omega T) = \int\dd t\, \delta\lambda/2\exp(\ii \omega t)/T$ and are given to lowest order in $q_i$.
For an even number of pulses $N$, $\mathcal{F}(0)=1$, whereas $\mathcal{F}(0)=0$ for an odd number.
The last column gives the corresponding orders of magnitude for the shifts, based on the $^2\text{S}_{1/2}\rightarrow\,\!^2\text{D}_{5/2}$ transition~\cite{Lindvall2025b} in $^{88}\text{Sr}^+$ with $\Omega\cong2\pi\times14.4\,\text{MHz}$, $\omega_{s,i}\cong 2\pi\,\text{MHz}$ and $a_i\sim q_i^2\sim \mu_i^2$, as well as a residual electric field of $20\,\text{V}/\text{m}$ at the trap center and a characteristic time $T\cong 0.15$\,s for a Ramsey sequence~\cite{Letchumanan2004}.
To express the contributions in dimensionless form, we introduce the characteristic length scale $\ell_{s,i}=\sqrt{\hbar/(2m\omega_{s,i})}$ and the associated velocity $v_{s,i}=\omega_{s,i}\ell_{s,i}$.
The recoil-induced first-order Doppler shift (rec) scales with the Lamb--Dicke parameter $\eta_i=k_i\ell_{s,i}$, while  its second-order contribution, the Lamb--Dicke shift, scales with $\eta_i^2$.
A differential acceleration $\delta\alpha$ causes a differential sag $r_{\delta\alpha,i}=\delta\alpha_i/\omega_{s,i}^2$, which couples to the acceleration sag $r_{\alpha,i}=\alpha_i/\omega_{s,i}^2$.
A differential trapping potential $\delta\omega$ directly leads to wave-packet effects, as apparent from its dependence on the variance $\overline{\langle\hat{r}_{\text{I},i}^2\rangle}=r_{\alpha,i}^2+\ell_{s,i}^2(2n_i+1)$ to lowest order in $q_i$ and $\mu_i$, for a thermal state with excitation number $n_i$.
The mass defect, in combination with the kinetic contribution, acts similarly and gives rise to the second-order Doppler shift (2ds).
A hypothetical dark-matter (DM) field leads to fluctuations of the energy spacing, but introduces no systematic phase shift, as its mean contribution vanishes.
}
\begin{ruledtabular}
\begin{tabular}{ccccccc}
\label{tab: Phases Main}
$\kappa$ & $\hat{H}_j^{(\kappa)}$ & $g_i^{(\kappa)}$ & $\omega_i^{(\kappa)}$
    & $\hat{h}_i^{(\kappa)}$ & $\overline{\varphi_\kappa}$
    & $\big|\overline{\varphi_\kappa}\big|/(\omega_{\text{c}} T)$ \\
\colrule
$0$
    & $\displaystyle \lambda_j\, \frac{\hbar\omega_{\text{c}}}{2}
      + \dot{\lambda}_j\, \frac{\hbar\phi_\text{L}}{2}$
    & $1$
    & $\omega_{\text{c}}$
    & $\displaystyle \frac{\dot{\phi}_\text{L} - \omega_{\text{c}}}{\omega_{\text{c}}}$
    & $\displaystyle \left( \omega_\text{L} - \omega_{\text{c}} \right) T\,
      \mathcal{F}(0)$
    & $\displaystyle \frac{|\omega_\text{L} - \omega_{\text{c}}|}{\omega_{\text{c}}}\,
      $ \\
rec
    & $\displaystyle - \dot{\lambda}_j\, \frac{\hbar}{2}\,
      \vect{k}^\mathsf{T} \hat{\vect{r}}$
    & $\eta_i$
    & $\omega_{s,i}$
    & $\displaystyle - \frac{\dot{\hat{r}}_{\text{I},i}}{v_{s,i}}$
    & $0$
    & $0$ \\
$\delta\vect{\alpha}$
    & $\displaystyle - \lambda_j\, \frac{m}{2}\, \delta\vect{\alpha}^\mathsf{T}
      \hat{\vect{r}}$
    & $\displaystyle \frac{r_{\delta\alpha,i}}{2\ell_{s,i}}$
    & $\omega_{s,i}$
    & $\displaystyle \frac{\hat{r}_{\text{I},i}}{\ell_{s,i}}$
    & $\displaystyle \frac{r_{\delta\alpha,i}\, r_{\alpha,i}}{2\ell_{s,i}^2}\,
      \frac{\mu_i}{2}\, \Omega T\, \mathcal{F}(0)$
    & $10^{-19}$ \\
$\delta\omega$
    & $\displaystyle \lambda_j\, \frac{m}{4}\, \hat{\vect{r}}^\mathsf{T}
      \delta\omega^2\, \hat{\vect{r}}$
    & $\displaystyle -\frac{\delta\omega_i^2}{4\omega_{s,i}^2}
      $
    & $\omega_{s,i}$
    & $\displaystyle \frac{\hat{r}_{\text{I},i}^2}{\ell_{s,i}^2}$
    & $\displaystyle - \frac{\delta\omega_i^2}{4\omega_{s,i}^2}\,
      \frac{\langle\overline{\hat{r}_{\text{I},i}^2}\rangle}{\ell_{s,i}^2}\,
      \frac{\mu_i}{2}\, \Omega T\, \mathcal{F}(0)$
    & $10^{-19}$ \\
2ds
    & $\displaystyle - \lambda_j\, \frac{\delta m}{2m}\,
      \frac{\hat{\vect{p}}^2}{2m}$
    & $\displaystyle \frac{v_{s,i}^2}{2c^2} $
    & $\omega_{\text{c}}$
    & $\displaystyle \frac{\dot{\hat{r}}_{\text{I},i}^2}{v_{s,i}^2}$
    & $\displaystyle \frac{v_{s,i}^2}{2c^2}
      \left(2 n_i + 1 + \frac{q_i^2}{2\mu_i^2} \frac{\langle\overline{\hat{r}_{\text{I},i}^2}\rangle}{\ell_{s,i}^2}\right)
      \omega_{\text{c}} T\, \mathcal{F}(0)$
    & $10^{-17}$ \\
DM
    & $\displaystyle \lambda_j\, \beta\, \frac{\hbar\varrho(t)}{2}$& $\beta$
    & $\varrho_0$
    & $\displaystyle - \frac{\varrho(t)}{\varrho_0}$
    & $0$
    & $0$ \\
LD
    &  ---
    & $\eta_i^2 $
    & $\omega_{s,i}$
    & $\displaystyle
      \frac{\ii\, [\hat{r}_{\text{I},i}(0), \dot{\hat{r}}_{\text{I},i}]}{2\ell_{s,i} v_{s,i}}$
    & $\displaystyle -\eta_i^2\omega_{s,i} T\re\mathcal{F}(\omega_{s,i}T)$
    & $10^{-17}$ \\
\end{tabular}
\end{ruledtabular}
\end{table*}}

We associate with each branch of the total Hamiltonian $\hat{H}_j$ an effective~\cite{Ufrecht20202, Ufrecht2019} time evolution through $\hat{U}_j$.
The interferometer signal is then determined by the expectation value of the overlap between both branches at the detection time $t_d$.
This overlap $ \langle\hat{U}_{1}^{\dagger}\hat{U}_{2}\rangle = \mathcal{V}\exp(\ii\varphi)$ results in a visibility $\mathcal{V}$ and a phase $\varphi$, where the quantum-mechanical expectation value $\langle \bullet \rangle$ is taken with respect to the initial center-of-mass state prepared in the effective trap generated by the rf potential (where $\vect{\alpha}=0$).
To evaluate the overlap, we first change into the interaction picture with respect to $\hat{H}_j^{(0)}$, described in Appendix~\ref{app: COMMotion}, and then apply the perturbative treatment~\cite{Ufrecht2019, Ufrecht20202} outlined in Appendix~\ref{app: PertPhase}.
By moving into the interaction picture we factorize the time-evolution operator through $\hat{U}_j=\hat{U}_{j}^{(0)}\hat{U}_{Ij}$, with an unperturbed part $\hat{U}_{j}^{(0)}$ generated by $\hat{H}_j^{(0)}$, such that the unperturbed contribution $\hat{U}_{1}^{(0)\dagger}\hat{U}_{2}^{(0)}$ arises.
It includes internal-state-dependent contributions which are independent of the center-of-mass position and momentum of the ion, leading to a cancellation of every effect except the clock contribution and the laser phase in the unperturbed signal.
These remaining terms lead to the complex number
\begin{equation}
    \hat{U}_1^{(0)\dagger}\hat{U}_2^{(0)} = \ee^{\ii \varphi_0}= \exp\left\lbrace- \ii \int \dd t \frac{\delta \lambda(t)}{2} (\omega_{\text{c}} - \dot \phi_L)\right\rbrace,
\end{equation}
encoding the unperturbed phase $\varphi_0$, where integration by parts was used for the laser phase term.
We show in Appendix~\ref{app: PertPhase} that the phases from all effects considered so far can be cast into the form 
\begin{equation}
\varphi_{\kappa} = g_i^{(\kappa)} \omega_i^{(\kappa)} T \int\!\! \frac{\dd t}{T} \frac{\delta \lambda}{2}\langle \hat{h}_i^{(\kappa)}\rangle,
\end{equation}
where a summation over spatial principal axes $i$ of the trap is implicitly assumed.
The dimensionless factor $g_i^{(\kappa)}$ indicates the order of magnitude of each phase, while the frequency scale of the phase is given by $\omega_i^{(\kappa)}$, and $\hat{h}_i^{(\kappa)}$ is its generator. 
The latter contains any time-dependency of each phase contribution, thus encoding any effects of the center-of-mass motion of the ion in the trap.
Each phase contribution includes the sensitivity function $\delta \lambda /2= (\lambda_2-\lambda_1)/2$, which is the difference of the state-response functions of each branch, and encodes the pulse sequence and thus alternates between $\pm1$ due to internal-state transitions.
The sensitivity function vanishes before and after the last pulse in the total sequence, since both branches then populate the same internal state.
Finally, we introduced the characteristic timescale of the interrogation sequence $T$. 
For equidistant laser pulses, $T$ is given by the time that elapses between two pulses.

In our treatment, the unperturbed phase $\varphi_0$ arises from the unperturbed part of the overlap, while all perturbative phases are determined by an expansion of the overlap in the interaction picture to first order in each perturbation, see Appendix~\ref{app: PertPhase}.
The only exception is the perturbation caused by the photon recoil, where the second-order effect is of similar order as the first-order phases generated by all other perturbations.
To obtain the explicit time dependence for the generators of the perturbative phases, we consider the center-of-mass motion in the interaction picture. 
A detailed calculation can be found in Appendix~\ref{app: COMMotion} and is summarized in the following.

\subsection{Atomic motion in Paul trap}
The unperturbed Hamiltonian includes a harmonic trapping potential with a time-dependent trap frequency, so that $\omega^2(t)$ can be decomposed into static (DC) and oscillating (AC) components.
In the interaction picture, the resulting motion can be separated into a rapidly oscillating micromotion at $\Omega$ and a slowly varying secular motion, whose dynamics are governed by an effective trapping frequency that is shifted relative to that of the corresponding static trap.
More precisely, assuming that the principal axes $i$ of the DC and AC components of the trap-frequency tensor are aligned, the secular motion is determined by the secular frequency $\omega_{s,i}=\mu_i\Omega/2$, where $\mu_i$ is the characteristic exponent of the corresponding Mathieu equation~\cite{Leibfried2003, Abramowitz1964, Meixner1954}.
In the regime of rapid rf drive, the exponent is approximately $\mu_i^2\cong a_i+q_i^2/2$, with $a_i = 4\omega_{\text{DC},i}^2/\Omega^2$ and $q_i = -2\omega_{\text{AC},i}^2/\Omega^2$.
For a rapidly oscillating rf field we have $a_i,q_i^2\ll1$.

The motion in this interaction picture can be split into a classical trajectory $\bm{r}$, which contains the excess micromotion from the linear acceleration $\bm{\alpha}$ inducing a sag, and excitations of the quantized motion $\hat{\bm{x}}$ in the effective trap, such that $\hat{\bm{r}}_\text{I} =  \bm{r}+\hat{\bm{x}}$, see Appendix~\ref{app: COMMotion} for details. 
We assume quasi-stationary micromotion for the classical trajectory, \ie{} $\bm{r}(t+2\pi/\Omega)=\bm{r}(t)$, so that the secular oscillations of the classical motion have been cooled or damped away and only the micromotion at $\Omega$, driven by the rf field, remains.
In this regime, the classical trajectory is, to first order in $q_i$, given by
\begin{equation}
    \label{eq:r_i}
    r_{i}(t)\cong \Big[1-q_i\cos(\Omega t+\theta_{\text{rf}}) /2 \Big]r_{\vect{\alpha},i},
\end{equation}
where we defined the displacement $ r_{\vect{\alpha},i} = \alpha_i/\omega_{s,i}^2 $ as the sag caused by the acceleration $\vect{\alpha}$. 
The classical trajectory therefore describes the excess micromotion caused by residual forces on the ion in the trap center.

In contrast, the quantized motion is mainly governed by the secular motion.
In the direction of the principal axis $i$, the quantized part of the ion's position can be decomposed as
\begin{equation}
    \label{eq:x_i}
    \hat{x}_{i}(t) = \ell_{s,i} \epsilon_i (t)\hat{b}_i^{\dagger} + \text{ h.\,c.},
\end{equation}
where $\hat{b}_i$ and $\hat{b}_i^{\dagger}$ are the annihilation and creation operators of excitations in the effective trapping potential, with commutation relation $[\hat{b}_i,\hat{b}_i^{\dagger}]=1$.
The frequency of the effective trap is given by the secular frequency, hence the ground state of the effective trapping potential sets the length scale $\ell_{s,i}=\sqrt{\hbar/(2m\omega_{s,i})}$.
In this picture, the dimensionless complex amplitudes $\epsilon_i$ encode the secular motion, oscillating at $\omega_{s,i}$, as well as all deviations of the quantized part of the trajectory induced by the rf field at frequency $\Omega$.
To first order in $q_i$, the amplitudes are given by $\epsilon_i\cong\exp(\ii\omega_{s,i} t)[1-q_i\cos(\Omega t+\theta_{\text{rf}} )/2]$.
Higher-order corrections can be obtained from the expansion in harmonics of the trap phase $\Omega t+\theta_{\text{rf}}$, with the corresponding coefficients up to second order provided in Appendix~\ref{app: COMMotion}.

\subsection{Systematic phase shifts}
As the clock is typically operated over multiple shots to average down statistical fluctuations~\cite{Arnold2026, Lindvall2025b, Steinel2023}, we introduce the ensemble average $\overline{\bullet}$ to account for shot-to-shot random elements.
These include laser frequency noise, the rf phase $\theta_\text{rf}$ when interrogating asynchronously with respect to the rf drive, and possible dark-matter candidates, which are all assumed to fluctuate independently.
The ensemble average removes the random contributions while retaining their systematic effect on the measured signal.
Hence, the averaged phases $\overline{\varphi_\kappa}$ contribute to the observed signal and account for systematic shifts.
\begin{figure}[!h]
    \centering
    \begin{tikzpicture}

\begin{axis}[
    name=colorbar,
    xtick pos=top,
    xticklabel pos=upper,
    width=0.95\columnwidth,
    height=0.225\columnwidth,
    ymin=0, ymax=1,
    xmin=-6.5, xmax=6.5,
    axis lines=box,
    ytick=\empty,
    xtick={-6,0,6},
    xticklabel style={
        /pgf/number format/fixed,
    },
    title= $\mathrm{Re}\mathcal{F}(\omega T)$,
    title style={yshift=0.15cm},
]
    \addplot graphics[
        xmin=-6.5, xmax=6.5,
        ymin=0, ymax=1,
    ] {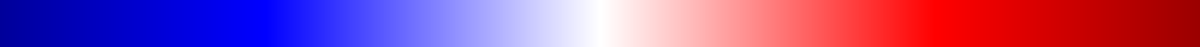};
\end{axis}

\pgfmathsetmacro{\piVal}{pi}
\pgfmathsetmacro{\twoPiVal}{2*pi}
\pgfmathsetmacro{\threePiVal}{3*pi}
\pgfmathsetmacro{\fourPiVal}{4*pi}
\begin{axis}[
   at={(colorbar.south west)},
    anchor=north west,
    yshift=-2mm,
    name=ReF,
    width=0.95\columnwidth,
    height=0.59\columnwidth,
    xmin=2, xmax=15,
    xtick = {2,3,4,5,6,7,8,9,10,11,12,13,14,15},
	xticklabels={,,,,,,,,,,,,},
    ymin=-0.28, ymax=\fourPiVal,
    ylabel={$\omega T$},
    ytick = {0,\piVal,\twoPiVal,\threePiVal,\fourPiVal},
	yticklabels={$0$, $\pi$, $2\pi$, $3\pi$, $4\pi$},
    yticklabel style={align=right,xshift=-0.1cm,inner sep=0pt,outer sep=0pt,inner ysep=0.05ex},
	ylabel style={yshift=-0.4cm,inner sep=0pt,outer sep=0pt,inner ysep=0.05ex},
    axis lines=box,
    enlargelimits=false,
]
    \addplot graphics[
        xmin=2, xmax=15,
        ymin=-0.28, ymax=\fourPiVal,
    ] {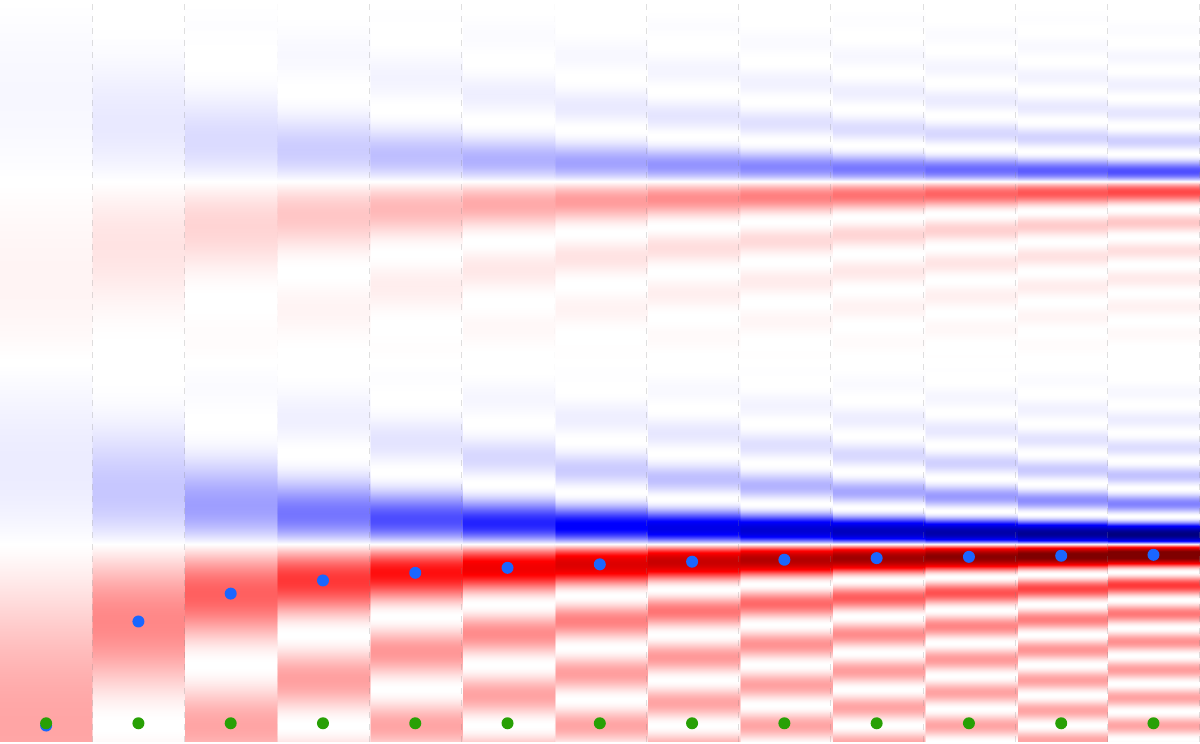};
\end{axis}

\begin{axis}[
    at={(ReF.south west)},
    anchor=north west,
    name=F0,
    width=0.95\columnwidth,
    height=0.35\columnwidth,
    xmin=2, xmax=15,
    xtick = {2,3,4,5,6,7,8,9,10,11,12,13,14,15},
	xticklabels={,,,,,,,,,,,,,},
    ymin=-0.2, ymax=1.2,
    ytick = {0,1},
	yticklabels={0,1},
    ylabel={$\mathcal{F}(0)$},
    ylabel style={yshift=-0.4cm},
    ]
    \addplot graphics[
        xmin=2, xmax=15,
        ymin=-0.2, ymax=1.2,
    ] {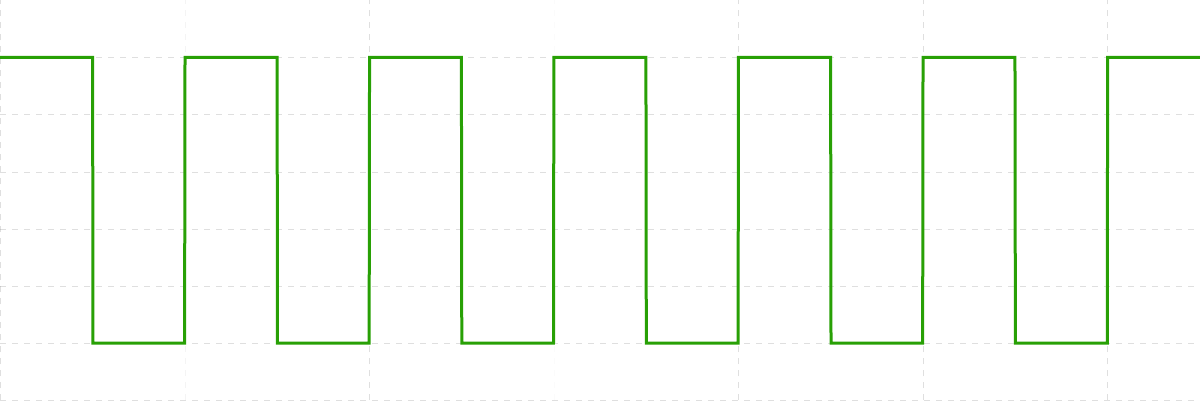};
\end{axis}

\begin{axis}[
    at={(F0.south west)},
    name=minReF,
    anchor=north west,
    width=0.95\columnwidth,
    height=0.35\columnwidth,
    xmin=2, xmax=15,
    xtick = {2,3,4,5,6,7,8,9,10,11,12,13,14,15},
	xticklabels={2,,4,,6,,8,,10,,12,,14,},
    xticklabel style={align=center,yshift=-0.1cm,inner sep=0pt,outer sep=0pt},
    x tick label as interval,
    xlabel={$N$},
    xlabel style={yshift=0.15cm},
    ymin=0, ymax=7,
    ytick = {1,6},
	yticklabels={1,6},
    ylabel={$\max\mathrm{Re}\mathcal{F}$},
    ylabel style={yshift=-0.4cm},
    ]
    \addplot graphics[
        xmin=2, xmax=15,
        ymin=0, ymax=7,
    ] {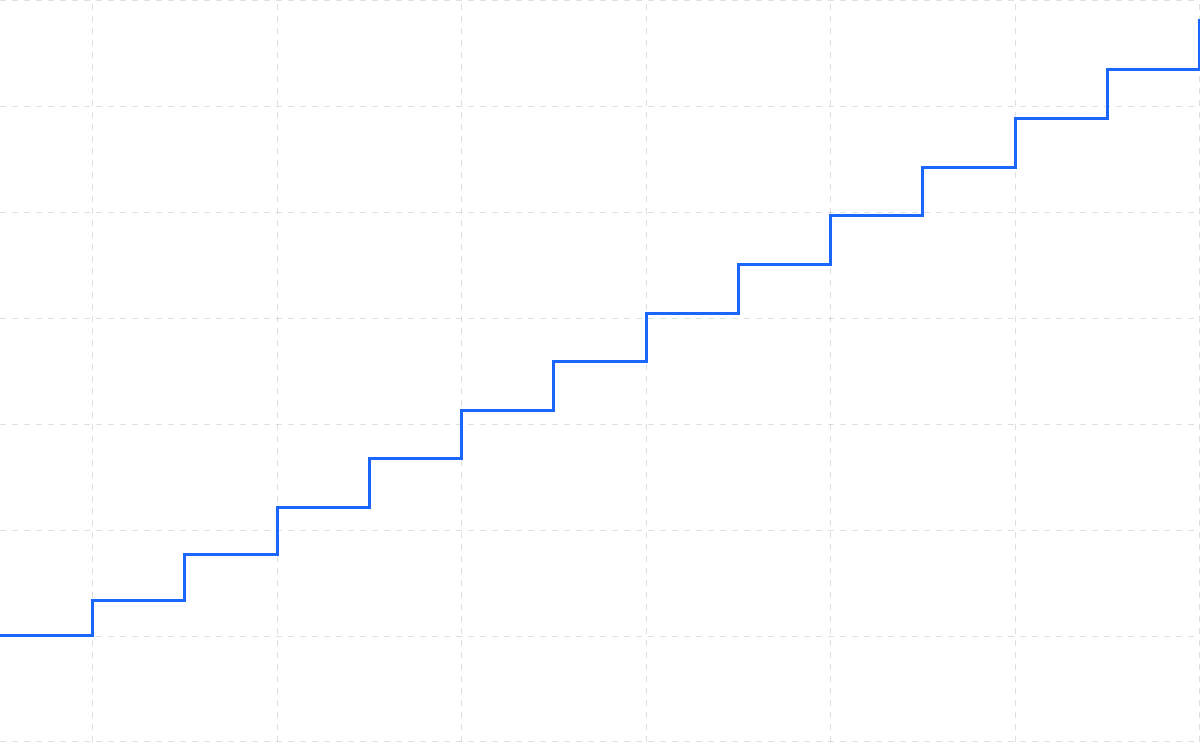};
\end{axis}

\node[black, font=\bfseries, xshift = 2.5pt, yshift = -0.5pt] at (ReF.north west) [below right] {a};
\node[black, font=\bfseries, xshift = 2.5pt, yshift = -5pt] at (F0.north west) [below right] {b};
\node[black, font=\bfseries, xshift = 2.5pt, yshift = -0.5pt] at (minReF.north west) [below right] {c};

\end{tikzpicture}
    \caption{
    Density plot (a) of the filter function $\re\mathcal{F}$, showing resonances approaching odd multiples of $\omega T=\pi$.
    While $\mathcal{F}(0)$ (green dots) alternates between $1$ and $0$ for even and odd pulse numbers $N$, respectively (b), the amplitude of the filter function at resonance (blue dots) increases approximately linearly with the number of pulses (c).
    }
    \label{fig:ReF}
\end{figure}

The dominant contribution, and the phase of primary interest, is the clock phase itself.
We consider a laser with frequency noise around the carrier frequency $\omega_\text{L}=\overline{\dot{\phi}_\text{L}} $.
As the carrier frequency is time-independent, its contribution is directly captured by the ensemble-averaged phase $ \overline{\varphi}_0= - (\omega_{\text{c}}-\omega_L) T \mathcal{F}(0)$.
Here, we introduce the dimensionless filter function $\mathcal{F}(\omega T) = \int\dd t\, \delta\lambda/2\exp(\ii \omega t)/T $ defined as the Fourier transform of the sensitivity function $\delta \lambda /2$, determined by the applied pulse sequence.
In the following, we consider a Ramsey sequence, possibly extended to a periodic  sequence used for dynamical decoupling~\cite{Biercuk2011, Aharony2021} by inserting evenly spaced $\pi$ pulses between the two $\pi/2$ pulses, with all pulses separated by a duration $T$.
We assume that the ion is in the ground state at the beginning of the sequence and that the read-out of the interferometer phase is described by a projective measurement on the excited state.
In this case, the sensitivity function vanishes before and after the last pulse in the total sequence, since both branches then populate the same internal state.
Without loss of generality, we further assume that the first $\pi/2$ pulse is initialized at $t=0$.
The corresponding sensitivity function is shown in Fig.~\ref{fig: Fig1}\,(d).
In Fig.~\ref{fig:ReF}\,(a), we plot the real part of the filter function for this pulse sequence, which exhibits distinct resonances approaching $\omega T = \pi$ with increasing number of pulses.
The amplitude of these resonances increases with the number of pulses.
Panel (b) of the same figure shows that $\mathcal{F}(0)=1$  for an even number of pulses $N$, while $\mathcal{F}(0)=0$ for an odd number.
Consequently, a Ramsey sequence with an even number of pulses measures the clock phase, whereas it is suppressed for an echo sequence with an odd number of pulses.
Additional phase shifts can arise from center-of-mass perturbations, which may fluctuate from shot to shot. 
As the micromotion occurs at the rf, operating the clock synchronously with this frequency causes the resulting motional phase shifts to add coherently and therefore not average out. 
To suppress such effects, we suggest an asynchronous interrogation that is not synchronized with the micromotion.
This amounts to averaging over the initial rf phase $\theta_\text{rf}$ within the ensemble average.

With the help of Eqs.~\eqref{eq:r_i} and \eqref{eq:x_i}, we find for an initially thermal quantum state of motion in the effective trap that $\langle\overline{\hat{\vect{r}}_\text{I}}\rangle =  \vect{r}_{\vect{\alpha}}$, since  $\langle\hat{\vect{x}}\rangle =0$ and $\overline{\vect{r}}=\vect{r}_{\alpha}$.
Consequently, the systematic shift caused by state-dependent accelerations $\delta\vect{\alpha}$ depends on both the acceleration sag $r_{ \vect{\alpha},i}$ and on the differential sag $r_{\delta \vect{\alpha},i}= \delta \alpha_i/\omega_{s,i}^2$, which set the characteristic scale of the corresponding phase.
Table~\ref{tab: Phases Main} shows the respective phase contribution and demonstrates that, as for most systematic shifts, it depends on the same filter function as the clock phase, namely $\mathcal{F}(0)$. 
This shift caused by a differential sag, for example caused by gravity in combination with the mass defect~\cite{Martinez2022} or differential Stark potentials arising from trap-field gradients or from spatially varying black-body radiation~\cite{Haslinger2018, Wineland1998}, is compared to the secular oscillator length scale $\ell_{s,i}$.
Such a shift can also encode hypothetical violations of the universality of free fall and the Einstein equivalence principle, if different internal states experience different gravitational accelerations~\cite{Zhang2020,DiPumpo2022,  Derr2023}. 
Such state-dependent gravitational accelerations are closely connected to tests of the universality of gravitational redshift~\cite{DiPumpo2021, DiPumpo2023}.
For a thermal state, we also find that $\langle\overline{\dot{\hat{\vect{r}}}}\rangle = 0$, since again the quantum-mechanical expectation value of the quantized motion vanishes, while the time-independent accelerational sag drops out of the classical velocity and the remaining oscillating terms average out.
Consequently, the first-order Doppler shift from the recoil perturbation vanishes, because its generator after partial integration is proportional to $\overline{\langle\dot{\hat{\vect{r}}}\rangle}$, see Tab.~\ref{tab: Phases Main}.

For static state-dependent perturbations to the harmonic potential, \ie{} a non-magic trap, we find to lowest order in $q_i$ the expression $\langle\overline{\hat{r}_{\text{I},i}^2}\rangle = r_{\vect{\alpha},i}^2+ \ell_{s,i}^2 (2n_i+1) $, where we used  $\overline{ |\epsilon_i|^2} = 1 $ and $\overline{r_i^2}=r_{\alpha,i}^2$.
Here, we again assume a thermal state for the secular motion, with the thermal phonon number $n_i=\langle\hat{b}_i^{\dagger}\hat{b}_i\rangle$ in direction $i$.
Hence, the phase contribution from a differential trap frequency $\delta\omega$ is not only caused by excess micromotion, but also contains a wave-packet contribution from the quantized motion, as summarized in Tab.~\ref{tab: Phases Main}.
Similarly, we find from Appendix~\ref{app: COMMotion} for the second moment of the velocity $\langle\overline{\dot{\hat{r}}_{\text{I},i}^2}\rangle / v_{s,i}^2 = 2 n_i + 1 + (q_i^2/2\mu_i^2) \langle\overline{\hat{r}_{\text{I},i}^2}\rangle/\ell_{s,i}^2 $.
The first term captures the contribution from the wave-packet width associated with the secular motion, while the second term describes an additional modification caused by micromotion, scaling as $\sim q_i^2/(2\mu_i^2)$.
Here, $v_{s,i}= \ell_{s,i} \omega_{s,i}$ is the characteristic secular velocity associated with the ground state of the effective trap.
As a consequence, the mass defect gives rise to the second-order Doppler shift due to time dilation, see Tab.~\ref{tab: Phases Main}.
In addition, since  $\langle\overline{\hat{r}_{\text{I},i}^2}\rangle = r_{\vect{\alpha},i}^2+ \ell_{s,i}^2 (2n_i+1) $ this shift also includes a contribution from gravitational sag included in $r_{\alpha,i}$, as noted previously~\cite{Martinez2022}.

While the first-order Doppler shift can be dominant for interrogation synchronous with the rf, it vanishes upon ensemble averaging in the case of asynchronous interrogation considered here. 
However, the second-order contribution from this recoil effect does not vanish upon averaging, but gives a contribution to the overall phase that can be of the same order of magnitude as first-order contributions from the other perturbations considered above.
We refer to this second-order recoil contribution as the Lamb--Dicke term, whose magnitude scales quadratically with the Lamb--Dicke parameter $\eta_i = k_i \ell_{s,i}$.
For this higher-order shift, the motion of the ion enters through the commutator of position and velocity evaluated at different times, as derived in Appendix~\ref{app: PertPhase}.
For a sequence of equally spaced pulses, the position contributes only from the first pulse, such that the Lamb--Dicke phase is determined by the generator $[\hat{r}_{\text{I},l}(0), \dot{\hat{r}}_{\text{I},i}(t)] = -2\ii\delta_{il} \ell_{s,i}^2\im[ \epsilon_i(0) \dot\epsilon_i^*(t)]$.
This expression is independent of the initial quantum state due to the canonical commutation relations reducing the commutator to a complex number.
Because the amplitudes $\epsilon_i$ are evaluated at different times, the commutator contains oscillations at the secular frequency and at rf sidebands.
To lowest order in $q_i$, only the secular motion survives the ensemble averaging, see Tab.~\ref{tab: Phases Main}.
Including higher-order contributions, however, the filter function is also evaluated at the rf sidebands $\omega_i^{(k)}=\omega_{s,i}+k\Omega$, see Appendix~\ref{app: COMMotion}.
In contrast to the other systematic phase shifts, which are caused by the static part of the perturbations and scale with $\mathcal{F}(0)$, taking the values $0$ or $1$ depending on the pulse sequence, the Lamb--Dicke shift scales dominantly with  $|\re\mathcal{F}(\omega_{s,i}T)|$.
On resonance, this filter function grows with increasing number $N$ of pulses, making the Lamb--Dicke phase increasingly sensitive to the secular motion, see Fig.~\ref{fig:ReF}\,(c).
However, for an off-resonant pulse sequence satisfying $\omega_{s,i}=2l\pi$ for integer $l$, the filter function $\mathcal{F}(\omega_{s,i}T)=0$, suppressing the dominant contribution to the Lamb--Dicke shift. 
The behavior of the filter function as a function of $\omega T$ is shown in Fig.~\ref{fig:ReF}\,(a).
Hence, the phase shift can be suppressed by choosing a suitable interrogation time such that the secular frequency is off-resonant with the pulse sequence.

Apart from laser frequency noise and the rf phase, the phase induced by a stochastic dark-matter field also fluctuates from shot to shot and is assumed to be independent of the other processes.
We assume that the dark-matter field has no coherent static component and consequently a vanishing mean $\overline{\varrho}=0$, so that it induces no systematic phase shift, see Tab.~\ref{tab: Phases Main}.
However, dark-matter field fluctuations can still be observable if the field's coherence time exceeds the duration of the interrogation sequence.
The corresponding fluctuations are discussed together with the fluctuations from the other noise sources in Sec.~\ref{sec: fluctuations}.

Table~\ref{tab: Phases Main} also provides orders of magnitude for the different systematic shifts to give some intuition for their relevance. 
As an example, we consider the $^2\text{S}_{1/2}\rightarrow\,\!^2\text{D}_{5/2}$ transition~\cite{Lindvall2025b} in $^{88}\text{Sr}^+$.
We further assume the ion to be subject to an acceleration caused by a residual electric field with its largest component $E_i\cong20\,\text{V}/{\text{m}}$ at the trap center, driving excess micromotion. 
The contributions $\overline{\varphi}_{\delta\omega}$ and $\overline{\varphi}_{\delta\alpha}$ are dominated by differential Stark shifts caused by the trap potential and its gradient, respectively.
We assume $\Omega\cong 2\pi\times14.4\,\text{MHz}$ and $\omega_{s,i}\cong 2\pi\,\text{MHz}$ for the rf and secular frequency~\cite{Lindvall2025b}, as well as $a_i\sim q_i^2\sim \mu_i^2$.
Furthermore, the ion is assumed to be cooled to the ground state in all directions so that $n_i=0$, and we consider a Ramsey sequence assuming an interrogation time~\cite{Letchumanan2004} of $T\cong 0.15$\,s.

\section{Fluctuations}
\label{sec: fluctuations}
So far, we have studied systematic shifts that survive averaging over different shots.
However, the phases fluctuate from shot to shot about the means provided in Tab.~\ref{tab: Phases Main}.
{\renewcommand{\arraystretch}{2.8}
\begin{table}[htb]
\caption{\label{tab: main variances}%
Leading-order terms of dominating contributions to the phase fluctuations.
The variance $\Delta\varphi_0^2$ of the unperturbed phase contribution is determined by the laser-frequency noise spectrum $S_{ \dot \phi_\text{L}}(\omega)$, weighted by $|\mathcal{F}|^2$, and is statistically independent of the other fluctuations.
The leading motional contribution is the recoil term $\Delta\varphi_\text{rec}^2$, whose covariance with all other contributions vanishes except for covariance with the Lamb--Dicke phase $\operatorname{Cov}_{\text{rec,LD}}$.
This covariance probes the filter function at the rf, secular, and higher-harmonic frequencies through
$\mathcal{T}_i^{(1)}=-q_i\re\mathcal{F}(\omega_{s,i}T)/2-q_i/(2\mu_i)[\mathcal{F}((\omega_{s,i}+\Omega)T)-\mathcal{F}^*((\omega_{s,i}-\Omega)T)]$.
The indices $i$  and $l$ denote the principal axes, with summation over them implicitly assumed.
The next relevant motional contribution is the fluctuation of the Lamb--Dicke phase.
In contrast to these noise contributions, dark matter described by a stationary, monochromatic field with fluctuating phase gives rise to a characteristic signal, with the filter function evaluated at the dark-matter frequency $\omega_\text{DM}$.
}
\begin{ruledtabular}
\begin{tabular}{ccc}
\label{tab: Cov Main}
$\Delta\varphi_0^2$
    & $\displaystyle \int \frac{\dd\omega}{2\pi}\,
      S_{\dot{\phi}_\text{L}}(\omega)
      \left| T\, \mathcal{F}(\omega T) \right|^2$
   \\
$\Delta\varphi_\text{rec}^2$
    & $\displaystyle \frac{(\Omega T)^2}{8}  
      \Big|q_i k_i r_{\alpha,i}\mathcal{F}(\Omega T) \Big|^2$
   \\
$\operatorname{Cov}_\text{rec,LD}$
    & $\displaystyle \frac{q_i}{4}k_ir_{\alpha,i}\eta_l^2\Omega\omega_{s,l}T^2\im[\mathcal{F}(\Omega T)\mathcal{T}_l^{(1)*}]$ 
      \\
$\Delta\varphi_\text{LD}^2$
    & $\displaystyle \frac{1}{2}\left|\eta_i^2\omega_{s,i}T\mathcal{T}_i^{(1)}\right|^2$
    \\
$\Delta\varphi_\text{DM}^2$
    & $\displaystyle \frac{(\beta\, \varrho_0 T )^2}{2}
      \left| \mathcal{F}(\omega_\text{DM} T) \right|^2$
\end{tabular}
\end{ruledtabular}
\end{table}}
These fluctuations are characterized by the covariances  $\operatorname{Cov}_{\kappa,\kappa'}
= \overline{\varphi_\kappa \varphi_{\kappa'}} - \overline{\varphi_\kappa}\,
\overline{\varphi_{\kappa'}}$ and variances $\Delta\varphi_\kappa^2 =
\operatorname{Cov}_{\kappa,\kappa}$ obtained from the ensemble average.

\begin{figure}[!h]
    \centering
    \begin{tikzpicture}

\pgfmathsetmacro{\piVal}{pi}
\pgfmathsetmacro{\twoPiVal}{2*pi}
\pgfmathsetmacro{\threePiVal}{3*pi}
\pgfmathsetmacro{\fourPiVal}{4*pi}

\begin{axis}[
    name=colorbar,
    xtick pos=top,
    xticklabel pos=upper,
    width=0.95\columnwidth,
    height=0.225\columnwidth,
    ymin=0, ymax=1,
    xmin=0, xmax=1500,
    axis lines=box,
    ytick=\empty,
    xtick={0,750,1500},
    xticklabels={0,750,1500},
    title= $\left|\omega T\mathcal{F}(\omega T)\right|^2$,
    title style={yshift=0.15cm},
]
    \addplot graphics[
        xmin=0, xmax=1500,
        ymin=0, ymax=1,
    ] {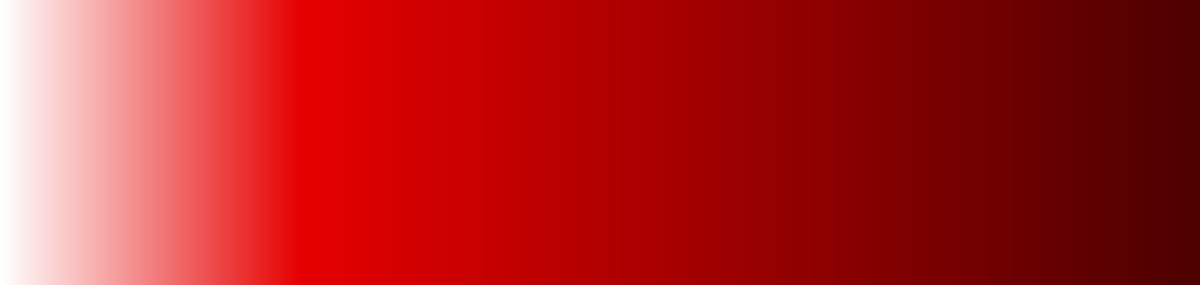};
\end{axis}

\begin{axis}[
   at={(colorbar.south west)},
    anchor=north west,
    yshift=-2mm,
    name=AbsF2,
    width=0.95\columnwidth,
    height=0.59\columnwidth,
    xmin=2, xmax=21,
    xtick = {2,3,4,5,6,7,8,9,10,11,12,13,14,15,16,17,18,19,20,21},
	xticklabels={,,,,,,,,,,,,,,,,,,,},
    xticklabel style={align=center,yshift=-0.1cm,inner sep=0pt,outer sep=0pt},
    x tick label as interval,
    ymin=-0.28, ymax=\fourPiVal,
    ylabel={$\omega T$},
    ytick = {0,\piVal,\twoPiVal,\threePiVal,\fourPiVal},
	yticklabels={$0$, $\pi$, $2\pi$, $3\pi$, $4\pi$},
    yticklabel style={align=right,xshift=-0.1cm,inner sep=0pt,outer sep=0pt,inner ysep=0.05ex},
	ylabel style={yshift=-0.4cm,inner sep=0pt,outer sep=0pt,inner ysep=0.05ex},
    axis lines=box,
    enlargelimits=false,
]
    \addplot graphics[
        xmin=2, xmax=21,
        ymin=-0.28, ymax=\fourPiVal,
    ] {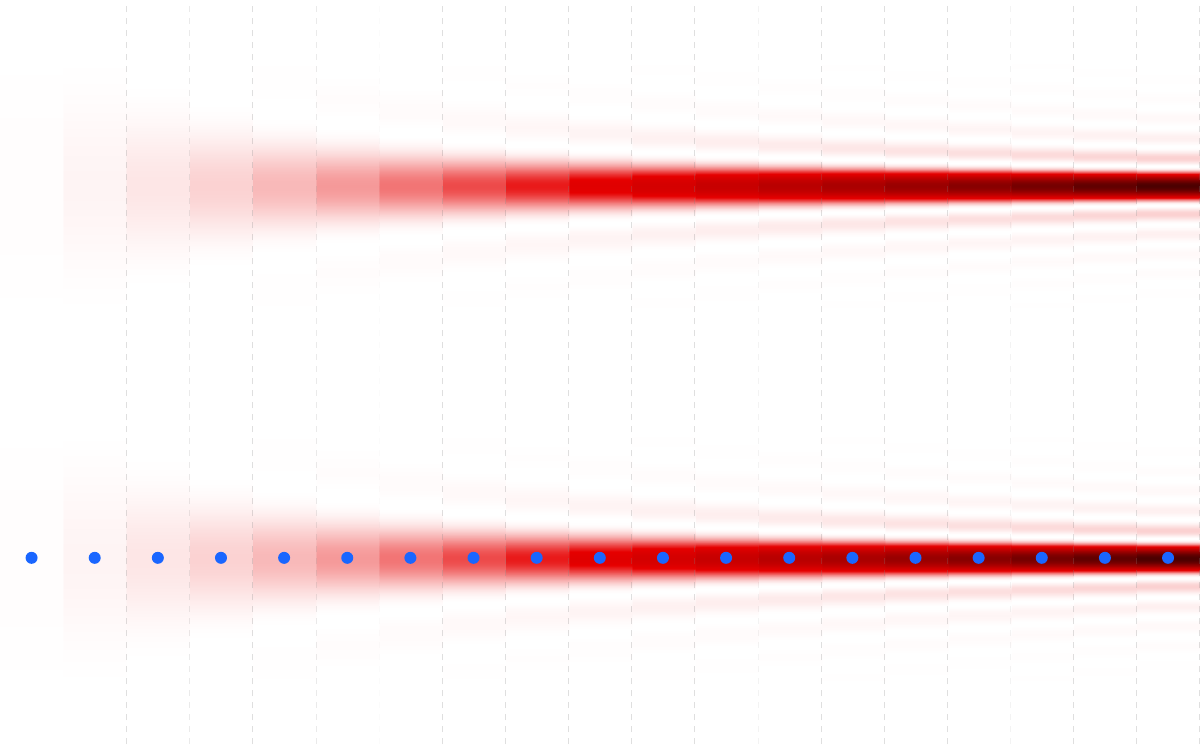};
\end{axis}

\begin{axis}[
    at={(AbsF2.south west)},
    name=AbsF2Inset,
    anchor=north west,
    width=0.95\columnwidth,
    height=0.35\columnwidth,
    xmin=2, xmax=21,
    xtick = {2,3,4,5,6,7,8,9,10,11,12,13,14,15,16,17,18,19,20,21},
	xticklabels={2,,4,,6,,8,,10,,12,,14,,16,,18,,20,},
    xticklabel style={align=center,yshift=-0.1cm,inner sep=0pt,outer sep=0pt},
    x tick label as interval,
    xlabel={$N$},
    xlabel style={yshift=0.15cm},
    ymin=-300, ymax=1800,
    ytick = {0,750,1500},
	yticklabels={0,750,1500},
    ylabel={$\left|\pi\mathcal{F}(\pi)\right|^2$},
    ylabel style={yshift=-0.2cm},
    ]
    \addplot graphics[
        xmin=2, xmax=21,
        ymin=-300, ymax=1800,
    ] {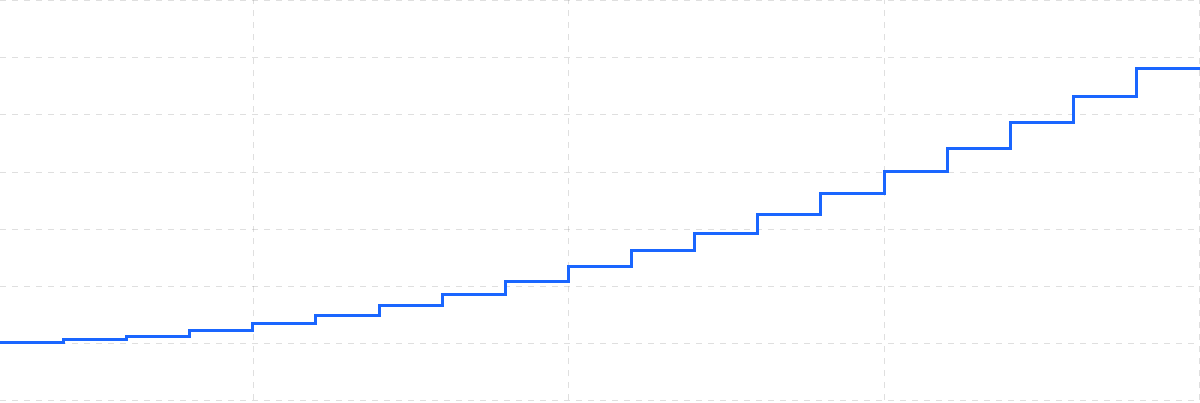};
\end{axis}

\node[black, font=\bfseries, xshift = 2.5pt, yshift = -0.5pt] at (AbsF2.north west) [below right] {a};
\node[black, font=\bfseries, xshift = 2.5pt, yshift = -0.5pt] at (AbsF2Inset.north west) [below right] {b};

\end{tikzpicture}
    \caption{
    Filter function $|\omega T\mathcal{F}(\omega T)|^2$ for fluctuations (a).
    Distinct resonances at $\omega T=\pi$ and its odd multiples arise (blue dots).
    The filter function (b) at resonance scales quadratically with the pulse number $N$.
    }
    \label{fig:AbsF2}
\end{figure}
Most variances depend on the modulus square of the filter function $|\mathcal{F}(\omega T)|^2$, evaluated at the characteristic frequency $\omega$ of the respective noise perturbation.
The overall behavior of $|\mathcal{F}(\omega T)|^2$ is shown in Fig.~\ref{fig:AbsF2}\,(a).
Here, we observe resonances at $\omega T = \pi$ and its odd multiples, where $|\omega T \mathcal{F}(\omega T)|^2\sim (N-1)^2$ scales quadratically with the number of pulses, for both even and odd numbers.
Panel (b) of the figure demonstrates this scaling behavior.
At the same time, the width of these resonances scales as $\propto 1/(N-1)$, such that they become increasingly narrow with an increasing number of pulses.

In Sec.~\ref{sec: phase shifts}, we identified three independent random quantities, fluctuating from shot to shot:
Laser frequency noise, the rf phase $\theta_\text{rf}$, as the interrogation sequence does not start synchronously with the driving field, and the dark-matter field.
In this section, we show that for white laser frequency noise, this scaling behavior reduces to a linear one, as expected.
We further study the dominant fluctuations arising from motional shifts, but refer to Appendix~\ref{app: Ensemble averages} for a complete discussion.
Resonances in the filter function should be avoided when they amplify unwanted perturbations, while they can be exploited to enhance periodic signals from possible dark-matter fields, even when successive experimental runs are mutually incoherent. 
 We therefore turn to their use for amplifying such signals in Sec.~\ref{subsec: Dark matter fluct}.

\subsection{Laser-frequency fluctuations}
The fluctuating contribution of the generator of the unperturbed clock phase listed in Tab.~\ref{tab: perturbations}, namely $\dot{\phi}_\text{L} - \omega_\text{L}$, is the laser-frequency noise.
Assuming a stationary process, its correlation function enters through the power spectral density $S_{\dot{\phi}_\text{L}}(\omega)$, such that the variance of the clock phase is given by the overlap of this noise spectrum with the filter function $|\mathcal{F}(\omega T)|^2$, see Tab.~\ref{tab: Cov Main}.
While  $|\mathcal{F}|^2$ may scale quadratically with the number of pulses on resonance, the overlap with a power spectral density does not necessarily show the same scaling, as the width of the resonances decreases as well.

In the following, we assume Ornstein-Uhlenbeck frequency noise~\cite{Uhlenbeck1930, Gardiner2004, Sanchez2025}, which is characterized by $\overline{\dot{\phi}_\text{L}(t) \dot{\phi}_\text{L}(t')} - \omega_\text{L}^2 =
\sigma_\text{L}^2\, \ee^{-\gamma_\text{L}|t - t'|}$, where $\sigma_\text{L}$ is the rms frequency fluctuation and $\gamma_\text{L}$ the relaxation rate.
For the fluctuations to be uncorrelated between shots, the frequency-noise correlation time $1/\gamma_\text{L}$  must be shorter than the cycle time between successive shots.
For this type of noise, the overlap integral reduces to
\begin{align}
\begin{split}
\Delta\varphi_0^2 =&\, \frac{2\sigma_\text{L}^2}{\gamma_\text{L}^2} \bigg\{
\left( N - 1 \right) \Big[ \gamma_\text{L} T - 2 \tanh\frac{\gamma_\text{L}T}{2}
\Big] \\%
&\, + \tanh^2 \frac{\gamma_\text{L}T}{2}
\Big[ 1 - \left( -1 \right)^{N-1} \ee^{-\gamma_\text{L}\left( N - 1 \right)T}
\Big] \bigg\}.
\label{eq: OU variance}
\end{split}
\end{align}
In the limit of $\gamma_\text{L} T \gg 1$, the noise is effectively white and $\Delta\varphi_0^2 \to 2 (N-1)T\sigma_\text{L}^2/\gamma_\text{L}$ scales linearly with the duration of the pulse sequence.
Hence, pulse sequences offer no protection against white frequency noise, but we also observe no resonant amplification.
 
\subsection{Asynchronous interrogation}
For the calculation of motional effects, we consider asynchronous interrogation, where only the static components of the induced phases yield systematic phase shifts.
The oscillating components of the motion do however influence the associated phase fluctuations and can be even larger than the systematic shifts.
As several motional perturbations are correlated, their covariances also contribute to the total phase variance.
The dominant fluctuation is the variance $\Delta\varphi_{\text{rec}}^2$ of the first-order Doppler shift, which causes no systematic shift.
It arises from excess micromotion induced by the acceleration sag displacing the ion from the unperturbed motion in the trap.
As the corresponding variance is proportional to the modulus square of the filter function $|\mathcal{F}|^2$, it scales quadratically with $N-1$ on resonance at $\Omega T =\pi$ and odd multiples thereof, see Fig.~\ref{fig:AbsF2}\,(b), and thus grows more rapidly than the contribution from white laser frequency noise.
The covariances of the recoil fluctuations with all other phase perturbations, except for the Lamb--Dicke phase, vanish, as we demonstrate in Appendix~\ref{app: Ensemble averages}.
Since the remaining covariances not involving the recoil term are much smaller, we collect only the dominant terms in Tab.~\ref{tab: Cov Main}, while the complete set is given in Appendix~\ref{app: Ensemble averages}.
These dominant terms are the variance of the Lamb--Dicke phase and the covariance of the recoil and Lamb--Dicke phases.
They contain linear combinations of filter functions, namely $\mathcal{T}_i^{(1)}=-(q_i/2)\re\mathcal{F}(\omega_{s,i}T)-q_i/(2\mu_i)[\mathcal{F}((\omega_{s,i}+\Omega)T)-\mathcal{F}^*((\omega_{s,i}-\Omega)T)]$.
Therefore interrogation should avoid not only resonance with the rf, but also with the secular frequency and  the rf sidebands to suppress these fluctuations.

\subsection{Dark matter}
\label{subsec: Dark matter fluct}
Since we assume that the mean of the stochastic dark-matter field vanishes, no systematic phase shift can be observed in such experiments.
However, its characteristic signatures in the induced phase fluctuations can be used for detection~\cite{DiPumpo2023b}.
We describe ultralight dark matter as a stationary, monochromatic field with frequency $\omega_\text{DM}$ and a fluctuating phase~\cite{Derr2023, DiPumpo2022}.
For the special case of scalar dark matter, we find a coupling amplitude~\cite{Derr2023}
\begin{equation}
    \varrho_0 =  4\omega_c\frac{m_\text{P} c^2}{\hbar \omega_\text{DM}} \sqrt{ \pi \frac{\varrho_\text{DM} L_\text{P}^3}{m_\text{P} c^2}}
\end{equation}
in units of frequency for a scalar field, where $\varrho_\text{DM}$ is the local dark-matter energy density and $L_\text{P}$ and $m_\text{P}$ are the Planck length and mass, respectively.
For other dark-matter models, \eg{} axions~\cite{Zhao2024, Kim2024, Flambaum2023, Bauer2025}, this amplitude differs.
Here, we assume that both internal states couple with the same dimensionless coupling constant $\beta$. 
A differential coupling can also be included, which changes the amplitude and scaling of the induced clock-frequency modulation~\cite{Derr2023}.
For such a dark-matter field, when successive experimental runs are mutually incoherent, the power spectral density reduces to a single peak.
The resulting fluctuations then take the form shown in Tab.~\ref{tab: Cov Main}, with the filter function $|\mathcal{F}(\omega_\text{DM}T)|^2$ evaluated at the frequency of the field.
While the fluctuations discussed before contribute to the statistical uncertainty, those induced by dark matter can be used for its detection.
As a consequence, one ideally operates on resonance.
Hence, for $\omega_\text{DM} T = (2\ell+1)\pi$ we find
\begin{equation}
\Delta\varphi_\text{DM}^2 = 2 \left[ \frac{\beta\, \varrho_0
\left( N - 1 \right) T}{\left( 2\ell + 1 \right) \pi} \right]^2 ,
\label{eq: dm resonance}
\end{equation}
growing quadratically with the applied number of pulses, see Fig.~\ref{fig:AbsF2}\,(b).
The first resonance $\ell=0$ provides the largest response.
These fluctuations compete with other noise sources, although white laser frequency noise scales only linearly with the number of pulses.
However, center-of-mass motion in the trap gives rise to noise terms that depend on the same filter function.
A dark-matter search is therefore unfavorable at frequencies close to the secular frequency, the rf, or their harmonics, where the dark-matter signal and motional contributions can become resonant simultaneously. 
To avoid this case, the trap parameters could be changed along with the pulse separation to prevent resonant enhancement of the motional fluctuations.
In a realistic dark-matter search, one relies on a differential geometry~\cite{DiPumpo2023b, Safronova2018, Filzinger2023} to suppress laser frequency noise and other noise sources, and the impact of such a scheme on the motional contributions needs to be considered as well.

\section{Discussion}
\label{sec: discussion}
Our coherent quantum-mechanical treatment yields both systematic shifts and fluctuations of trapped-ion optical clocks from a single filter function.
The shifts arise from the static response after ensemble averaging. 
To improve the trueness of the measurement, they must be quantified and can then be corrected in post-processing once the trap parameters, such as $q_i$, $\mu_i$, and $n_i$, are sufficiently characterized.
In contrast, the fluctuations, determined by the variances and covariances, limit the precision of the measurement.
The residual bias from the systematic shifts is set by the uncertainties of the involved parameters and can therefore be reduced by improving their characterization.
Depending on the application, different pulse sequences can be chosen, which in turn determine a specific filter function evaluated at different frequencies for the respective perturbations.
For evenly spaced pulse sequences, an even number of pulses retains the full static response and therefore operates as a clock.
An odd number of pulses decouples from static phase shifts and is therefore insensitive to the clock phase and most systematic shifts considered here, while still introducing statistical noise.
Hence, such sequences can be used to either suppress or enhance fluctuations and can, in principle, be tailored further by adjusting the pulse spacing.
To remove the dependence on the initial conditions of the ion's center-of-mass motion and prevent motional phase shifts from adding up coherently, we suggest interrogating subsequent runs asynchronously with respect to the rf.
Alternatively, in a multi-pulse sequence, the pulse spacing can be chosen to be off-resonant with the trap frequency to suppress such shifts.
Fluctuations of the motion are minimized by choosing the pulse spacing such that neither the rf drive, its sidebands, nor the secular motion fall on a resonance, thereby suppressing trap-induced noise.
However, this choice also determines the sensitivity to laser frequency noise and dark-matter spectra.
As expected, white laser frequency noise scales linearly with the duration of the sequence.
In contrast, for dark-matter detection, one aims to operate on resonance, where the induced fluctuations scale quadratically with the sequence duration.
While this scaling can serve as a lever to enhance the dark-matter signature relative to the noise floor, the signal still competes with other noise sources.

Differential measurements using correlating clocks provide a route to suppress common noise.
Comparing two transitions or two different ions or isotopes interrogated by the same light field removes laser phase noise, while a differential signal induced by dark matter can survive~\cite{Arvanitaki2018, Derevianko2014, Filzinger2025}.  
For suppression of motional phases, the two systems additionally need to share the same trap drive and experience matched trap parameters, such that correlated motional phases cancel.
In principle, differential measurements at different locations could also be employed, in analogy to large-baseline atom interferometers~\cite{Schlippert2020, Canuel2018, Abend2024, Zhan2020, Abe2021}.
Such a configuration would retain differential motional noise, but could still suppress common noise sources, as proposed for gravitational-wave detection~\cite{Graham2013, Yu2011, Kolkowitz2016} and dark-matter searches~\cite{DiPumpo2023b, Abend2024, Arvanitaki2018, Derevianko2014, Graham2016}.

The variances have been computed assuming independent fluctuations between runs, encoded in the ensemble average, such that the noise correlation time is shorter than the cycle time.
The rf phase is assumed to be independent from shot to shot through asynchronous interrogation.
The trap-driven variances then average down inversely with the number of measurements without aliasing, but would add up coherently for synchronous interrogation.
Noise with a correlation time longer than the cycle time instead leads to aliasing, known as the Dick effect~\cite{Dick1990, Santarelli1998, Westergaard2010, Hume2016, Schioppo2017, Schulte2020}.
Our treatment does not include a correction for this contribution.
The treatment could, however, be generalized by including all runs in an effective filter function.
In addition, interleaved~\cite{Schioppo2017, Biedermann2013} operation or reducing the dead time~\cite{Dick1990, Westergaard2010, Schioppo2017} can mitigate such aliasing.
Moreover, if the dark-matter field remains coherent over multiple runs~\cite{Filzinger2025}, its phase is correlated between successive interrogations and can therefore add coherently, potentially allowing sensitivities beyond the limit discussed here.

Effects such as light shifts induced by the laser pulses, as well as their fluctuations and time dependence, could be included in our treatment as perturbations to the clock frequency without changing the formalism.
As a possible extension, one could include the finite duration of the pulses~\cite{Bertoldi2019} and derive a modified filter function that accounts for such effects.
Moreover, this would allow for the inclusion of hyper-Ramsey~\cite{Yudin2010} and more elaborate sequences~\cite{Zanon-Willette2022, Zanon-Willette2016, Hobson2016} beyond encoding them in the state-response function, and enable the study of fluctuations on the scale of the pulse duration.
Our treatment could also be combined with auto-balanced Ramsey spectroscopy~\cite{Sanner2018} to additionally cancel interrogation-induced errors.
Including motional decoherence~\cite{Wineland1998} and heating~\cite{Brownnutt2015}, as well as finite lifetimes~\cite{Clements2020}, requires extending our treatment to open quantum systems and the associated framework~\cite{Schlosshauer2019}.

Our formalism also transfers to other platforms whose motion is quantized and driven or confined, \eg{} multi-ion crystals, optical lattice clocks~\cite{Ludlow2015, Fortier2026}, as well as trapped~\cite{Haensel2001, Roati2004, Petrucciani2026}, guided~\cite{Arlt2005, Wu2007}, or tractor~\cite{Duspayev2021} atom interferometers and quantum-clock interferometry~\cite{Meltzer2024} in traps used for fundamental-physics applications~\cite{Cronin2009, Safronova2018, Bongs2019}.

\begin{acknowledgments}
We are grateful to W. P. Schleich for his continuing support.
We also thank D. Derr, A. Friedrich, A. Wolf, as well as the QUANTUS team for fruitful and interesting discussions.
The authors also acknowledge contributions in the form of discussions from the Terrestrial Very-Long-Baseline Atom Interferometry (TVLBAI) proto-collaboration, as well as networking support from COST Actions CA23130 - Bridging high and low energies in search of quantum gravity (BridgeQG) and CA23115 - Relativistic Quantum Information (RQI).
The QUANTUS project is supported by the German Space Agency at the German Aerospace Center (Deutsche Raumfahrtagentur im Deutschen Zentrum f\"ur Luft- und Raumfahrt, DLR) with funds provided by the Federal Ministry for Economic Affairs and Climate Action (Bundesministerium f\"ur Wirtschaft und Klimaschutz, BMWK) due to an enactment of the German Bundestag under Grant No. 50WM2450D and No. 50WM2450E (QUANTUS-VI).
\end{acknowledgments}

\section*{Author declarations}
\subsection*{Conflict of interest}

\noindent The authors have no conflicts to disclose.

\subsection*{Author contributions}
 \noindent{\sffamily\small\textbf{Alexander Bott}} Software (lead); Validation (lead); Formal Analysis (lead); Investigation (lead); Data Curation (lead); Writing -- Original Draft (equal); Writing -- Review \& Editing (equal); Visualization (lead).
 {\sffamily\small\textbf{Fabio Di Pumpo}} Conceptualization (equal); Methodology (lead); Validation (supporting); Formal Analysis (supporting); Writing –- Original Draft (equal); Writing –- Review \& Editing (equal); Supervision (equal).
 {\sffamily\small\textbf{Enno Giese}} Conceptualization (equal); Methodology (supporting); Validation (supporting); Formal Analysis (supporting); Investigation (supporting); Data Curation (supporting); Writing -- Original Draft (equal); Writing -- Review \& Editing (equal); Visualization (supporting); Supervision (equal).

\section*{Data availability}
\noindent The symbolic computational scripts that support the findings of this study are available on GitHub~\cite{BottScriptMathematica2026} and will be archived on Zenodo upon publication.

\section*{Tools and Resources}
The authors used generative artificial-intelligence tools, namely OpenAI ChatGPT (5.6 and 6 Luna) and Anthropic Claude (Fable 5.0), to independently reproduce and cross-check the analytic results of this article, including symbolic checks, and to copyedit the language.
All output was reviewed and verified by the authors, who take full responsibility for the entire content of this article.

\appendix

\section{Center-of-mass motion of ions in Paul traps}
\label{app: COMMotion}
In this Appendix, we derive the interaction-picture operators $\hat{\vect{r}}_\text{I} = \hat{U}_j^{(0)\dagger}\hat{\vect{r}}\hat{U}_j^{(0)}$ and $\hat{\vect{p}}_\text{I} = \hat{U}_j^{(0)\dagger}\hat{\vect{p}}\hat{U}_j^{(0)}$, where the time-evolution operator $\hat{U}_j^{(0)}$ is generated by $\hat{H}_{j}^{(0)}$ from Eq.~\eqref{eq:H0} in the main text and initialized at $t=0$.
These transformed operators describe the motion of the ions in the effective trap and will allow us to determine systematic shifts and fluctuations from motional effects.
Since the branch dependence of $\hat{H}_j^{(0)}$ is only encoded in a c-number, the transformed phase-space operators are state independent and common for both interferometer branches.
For these operators, we find the Heisenberg equations
\begin{align}
    \hat{\vect{p}}_\text{I} = m\dot{\hat{\vect{r}}}_\text{I} \quad\text{ and }\quad
    \ddot{\hat{\vect{r}}}_\text{I} = -\omega^2\hat{\vect{r}}_\text{I}+\vect{\alpha}.
\end{align}
This form is similar to a harmonic oscillator with a constant force, but with time-dependent trapping frequency $\omega^2 = \omega_{\text{DC}}^2+\omega_{\text{AC}}^2\cos(\Omega t+\theta_{\text{rf}})$.
In the following, we split the transformed position operator into $\hat{\vect{r}}_{\text{I}}=\vect{r}+\hat{\vect{x}}$, with a classical trajectory $\vect{r}\in\mathds{R}^3$ defined by $\ddot{\vect{r}}+\omega^2\vect{r}=\vect{\alpha}$ and a quantized part $\hat{\vect{x}}$, and similarly for $\hat{\vect{p}}_{\text{I}}=\vect{p}+m\dot{\hat{\vect{x}}}$.

\subsection{Quantized motion}
The Heisenberg equations lead to the equation of motion
\begin{align}
    \ddot{\hat{\vect{x}}} &= -\omega^2\hat{\vect{x}}
\end{align}
for the quantized part of the transformed position operator.
We assume the ion to behave as a trapped particle for the whole duration of the experiment, leading to the ansatz~\cite{Glauber1992,Leibfried2003,Schleich2001}
\begin{align}\label{eq: x_i(t)}
    \hat{x}_i(t) = \ell_{s,i} \epsilon_i (t) \hat{b}_i^{\dagger}+ \text{h.c.},
\end{align}
with excitations of an effective trapping harmonic oscillator generated or eliminated by creation or annihilation operators $\hat{b}_i^{\dagger}/\hat{b}_i$, where $i$ denotes the direction of each spatial principal trap axis.
Here,  $\ell_{s,i}=\sqrt{\hbar/(2m\omega_{s,i})}$ is the length scale of the ground state of the effective harmonic oscillator with effective trap frequency $\omega_{s,i}$, \ie{} the secular frequency.
Since $\hat{U}_j^{(0)}(0)=\mathds{1}$, the Schrödinger and the interaction picture coincide at $t=0$.
Consequently, the initial conditions of the quantized part of the operators $\hat{\vect{x}}(0) = \hat{\vect{r}}-\vect{r}(0)$ and $m\dot{\hat{\vect{x}}}(0) = \hat{\vect{p}}-m\dot{\vect{r}}(0)$ fix the relation between the annihilation and creation operators and the center-of-mass position and momentum operators of the ion up to a constant shift induced by the initial conditions of the classical trajectory.
If the ion is initially prepared in a diagonal state in the Fock basis, \eg{} a thermal state~\cite{Martinez2022, Leibfried2003}, with $\langle \hat{b}_i^{\dagger} \rangle = 0 = \langle \hat{b}_i \rangle$, $\langle \hat{b}_i^{\dagger 2} \rangle = 0 = \langle \hat{b}_i^2 \rangle$, and phonon number $\langle \hat{b}^\dagger \hat{b}\rangle = n_i$ in principal direction $i$, the expectation values $\langle \hat{\vect{x}}(t)\rangle = 0 = \langle \dot{\hat{\vect{x}}}(t)\rangle$ for the quantized part vanish for all times.
Similarly, we observe
\begin{subequations}
 \begin{align}
    \langle \hat{x}_i(t)\hat{x}_l(t) \rangle &= \delta_{il} \,\ell_{s,i}^2 |\epsilon_i|^2(2n_i + 1) \label{eq: <x_q^2> und <dot{x}_q^2>} \\
    \langle \dot{\hat{x}}_i(t)\dot{\hat{x}}_l(t)\rangle &= \delta_{il} \,\ell_{s,i}^2 |\dot\epsilon_i|^2(2n_i + 1) \label{eq: <x_q^2> und <dot{x}_q^2>2}
\end{align}   
\end{subequations}
for cross terms.

We find from the Heisenberg equations that the time-dependent amplitude $\epsilon_i$ has to fulfill the differential equation $\ddot \epsilon_i + \omega_i^2(t) \epsilon_i =0$, which can be cast into the form of a Mathieu equation~\cite{McLachlan1947,Meixner1954,Abramowitz1964} 
\begin{equation}
    \epsilon_i''(\phi) +  (a_i -2 q_i \cos 2\phi)\epsilon_i(\phi)=0,
\end{equation}
with $2\phi= \Omega t + \theta_\text{rf}$, $a_i= 4 \omega_{\text{DC},i}^2/\Omega^2$, and $q_i = - 2 \omega_{\text{AC},i}^2 / \Omega^2$.
The prime denotes a differentiation with respect to $\phi$, while $\epsilon_i$ and $\epsilon_i^*$ are linearly independent solutions~\cite{Meixner1954}.
If we assume the motion in the effective trap to be cooled, dampened, or decayed sufficiently strongly, only the oscillations due to oscillating rf-field remain observable, such that $\langle \hat{x}_i^2(t+2\pi/\Omega) \rangle=\langle \hat{x}_i^2(t) \rangle$, which implies $|\epsilon_i(t+ 2\pi/\Omega)|^2 =|\epsilon_i(t)|^2$.
Hence, the $\pi$-periodicity of the Mathieu equation allows the application of Floquet's theorem~\cite{Abramowitz1964,Meixner1954,Leibfried2003,Schleich2001}, which yields
\begin{equation}\label{eq: Floquet}
    \epsilon_i(\phi) = \ee^{\ii \theta_{s,i}}\ee^{\ii \mu_i \phi} \sum_k \xi_i^{(k)}\,\ee^{\ii 2k\phi},
\end{equation}
with Fourier components $\xi_i^{(k)}\in\mathds{R}$.
By writing $\epsilon_i$ as a single Floquet solution instead of a sum of positive and negative frequency parts, we implement the periodicity condition on the absolute value.
An additional constant phase $\theta_{s,i}$ allows for rotations in phase space and the characteristic exponent $\mu_i$ accounts for deviations from perfect periodicity through a global phase.
Moreover, $\epsilon_i(t)$ includes oscillations at rf-frequency $\Omega$, resulting in micromotion~\cite{Leibfried2003, Martinez2022}, but also oscillations at secular frequency $\mu_i\Omega/2= \omega_{s,i}$, leading to secular motion~\cite{Martinez2022, Leibfried2003}.
With this definition of the secular frequency we obtain $\ee^{\ii\mu_i\phi} = \ee^{\ii\omega_{s,i}t}\ee^{\ii\mu_i\theta_\text{rf}/2}$, while micromotion corrections to the quantized part $\hat{x}_i$ are confined to $\epsilon_i$.
Finally, we can absorb the constant phase $\mu_i\theta_{\text{rf}}/2$ into $\theta_{s,i}$.
Next, we determine the Fourier coefficients $\xi_i^{(k)}$ and the characteristic exponent $\mu_i$.
Inserting Eq.~\eqref{eq: Floquet} into the Mathieu equation yields the recursion relation
\begin{equation}\label{eq: recursion}
    q_i[\xi_i^{(k+1)}+\xi_i^{(k-1)}] = [a_i - (2k+\mu_i)^2]\xi_i^{(k)},
\end{equation}
where $k=0$ determines the characteristic exponent $\mu_i$, with $\mu_i^2 \cong a_i + q_i^2/2$~\cite{Leibfried2003, Abramowitz1964}.
Here, we assumed the rf to be large compared to the trapping frequencies, such that $a_i,q_i^2\ll1$.
The recursion relation determines the ratios $\xi_i^{(k)}/\xi_i^{(0)}$, which can be obtained to desired order in $q_i$ by a truncation due to $\xi_i^{(k)}/\xi_i^{(0)}=\mathcal{O}(q_i^{|k|})$~\cite{Abramowitz1964,Meixner1954}.
We use the rf-dominated power counting $q_i/\mu_i=\mathcal{O}(1)$ in the following, such that terms like $q_i^3/\mu_i$ and $q_i^4/\mu_i^2$ contribute at order $\mathcal{O}(q_i^2)$.
These truncation orders apply away from the limit $\mu_i\to0$, where this expansion is not uniform.
The conjugate momentum $m \dot{\hat{x}}_i(t)$ must satisfy the canonical commutator relation $[\hat{x}_i, m\dot{\hat{x}}_l] = \ii \hbar \delta_{il}$, which, together with $[\hat{b}_i, \hat{b}_l^\dagger] =\delta_{il}$, implies a Wronskian condition $\epsilon_i^* \dot \epsilon_i - \epsilon_i \dot \epsilon_i ^* = 2 \ii \omega_{s,i}$.
Inserting the Fourier series for $\epsilon_i$ and the definition for the secular frequency into the Wronskian, before averaging over one rf period, gives
\begin{align}
    \xi_i^{(0)} &= \Big[\sum_k \big(\xi_i^{(k)}/\xi_i^{(0)}\big)^2 (1+2k/\mu_i)\Big]^{-1/2}.
\end{align}
Moreover, it is convenient to define sideband frequencies $ \omega_i^{(k)} = \omega_{s,i}+k\Omega$ and the dimensionless comb of the velocity $w_i^{(k)} = \omega_i^{(k)} \xi_i^{(k)}/\omega_{s,i} = (1 +2k/\mu_i)\xi_i^{(k)}$, so that 
\begin{equation}\label{eq: deps/dt}
    \dot\epsilon_i = \ii\,\omega_{s,i}\,\ee^{\ii(\omega_{s,i}t+\theta_{s,i})}\sum_k w_i^{(k)}\ee^{2\ii k\phi}
\end{equation}
follows for the temporal derivative of $\epsilon_i$.
Finally, both $\xi_i^{(k)}$ and $w_i^{(k)}$ are listed in Tab.~\ref{tab: Floquet coefficients} up to $\mathcal{O}(q^2)$.
{\renewcommand{\arraystretch}{2.75}
\begin{table}[htb]
\caption{
    Floquet coefficients of the quantized and classical positions and velocities, $\hat{x}_i$, $\dot{\hat{x}}_i$, $r_i$, and $\dot{r}_i$ in terms of the Mathieu parameters and characteristic exponent.
    The index $i$ denotes the different principal axes of the trap.
    The coefficients are given up to $\mathcal{O}(q_i^2)$; coefficients with $|k|>3$ are at least of order $\mathcal{O}(q_i^3)$.
}
\begin{ruledtabular}\label{tab: Floquet coefficients}
\begin{tabular}{cccccc}
$k$ & $\xi_i^{(k)}$ & $w_i^{(k)}$ & $\rho_i^{(k)}$ & $v_i^{(k)}$ \\
\colrule
$0$
    & $\displaystyle 1+\frac{3q_i^2}{16}$
    & $\displaystyle 1+\frac{3q_i^2}{16}$
    & $\displaystyle 1+\frac{3q_i^2}{8}$
    & $\displaystyle 0$ \\
$\pm1$
    & $\displaystyle -\frac{q_i}{4}\pm\frac{q_i\mu_i}{4}$
    & $\displaystyle \mp\frac{q_i}{2\mu_i}+\frac{q_i}{4}$
    & $\displaystyle -\frac{q_i}{4}$
    & $\displaystyle \mp\frac{q_i}{2\mu_i}$
\\[-2pt]
    & & $\displaystyle \mp\frac{5q_i^3}{128\mu_i}\mp\frac{q_i\mu_i}{4}$
    & & $\displaystyle \mp\frac{17q_i^3}{128\mu_i}\mp\frac{q_i\mu_i}{8}$ \\
$\pm2$
    & $\displaystyle \frac{q_i^2}{64}$
    & $\displaystyle \pm\frac{q_i^2}{16\mu_i}-\frac{5q_i^2}{64}$
    & $\displaystyle \frac{q_i^2}{64}$
    & $\displaystyle \pm\frac{q_i^2}{16\mu_i}$ \\
$\pm3$
    & $\displaystyle \mathcal{O}(q_i^3)$
    & $\displaystyle \mp\frac{q_i^3}{384\mu_i}$
    & $\displaystyle \mathcal{O}(q_i^3)$
    & $\displaystyle \mp\frac{q_i^3}{384\mu_i}$ \\
\end{tabular}
\end{ruledtabular}
\end{table}}

\subsection{Classical motion}
The classical center-of-mass trajectory obeys the driven equation of motion
\begin{equation}
\ddot{r}_i(t) + \omega_i^2(t)\, r_i(t) = \alpha_i,
\label{eq: cl eom}
\end{equation}
which yields with $2\phi = \Omega t + \theta_{\text{rf}}$ the inhomogeneous Mathieu equation
\begin{equation}
r_i''+ \left(a_i - 2 q_i \cos 2\phi\right) r_i = 4\alpha_i/\Omega^2 = \mu_i^2\, r_{\alpha,i}.
\label{eq: cl mathieu}
\end{equation}
Here, primes denote differentiation with respect to $\phi$ and we introduced the accelerational sag $r_{\alpha,i} =\alpha_i/\omega_{s,i}^2$,
i.\,e.\ the static displacement the ion would acquire in a harmonic trap of the secular frequency $\omega_{s,i} = \mu_i \Omega /2$ alone.
The general solution of \eqref{eq: cl mathieu} is given by a particular solution combined with the two homogeneous Floquet solutions $\epsilon_i$ and $\epsilon_i^*$ derived above.
We require the classical trajectory to be rf-periodic $r_i(t + 2\pi/\Omega) = r_i(t)$, such that secular motion is assumed to be cooled away and only the micromotion drive remains.
Because of $\epsilon_i(\phi + \pi) = \ee^{\ii \pi \mu_i} \epsilon_i(\phi)$, the periodicity implies that the solution consists only of a particular solution if $\exp(\ii\pi \mu_i) \neq 1$, which holds throughout the stability region of the trap considered here, where $\mu\ll 1$.
From the Wronskian we obtain directly the corresponding Green's function, such that the steady-state response to the constant acceleration is
\begin{equation}
r_i(t) = r_{\alpha,i}\omega_{s,i} \lim_{\gamma \to 0^+} \int_{-\infty}\limits^{t} \dd t' \ee^{-\gamma (t - t')}\im\left[ \epsilon_i(t)\epsilon_i^*(t') \right].
\label{eq: cl steady}
\end{equation}
Using the solution of the homogeneous equation $\epsilon_i(t) = \sum_k \xi_i^{(k)} \ee^{\ii ( \omega_i^{(k)} t + k \theta_{\text{rf}}+\theta_{s,i} )}$ from Eq.~\eqref{eq: Floquet} with $\omega_i^{(k)} = \omega_{s,i} + k \Omega$, we find
\begin{align}
r_i = \frac{r_{\alpha,i}}{2 } \sum_{k,k'}
    \xi_i^{(k)} \xi_i^{(k')}
    \left( \frac{\omega_{s,i}}{\omega_i^{(k)}} + \frac{\omega_{s,i}}{\omega_i^{(k')}} \right)
    \ee^{2\ii (k-k') \phi}.
\label{eq: cl double sum sym}
\end{align}
Collecting the terms with fixed harmonic index $l = k - k'$ yields 
\begin{subequations}
\begin{align}
r_i &=r_{\alpha,i} \sum_l \rho_i^{(l)}\, \ee^{2\ii l \phi}\,,
\label{eq: cl comb}
\\
\rho_i^{(l)} &= \mu_i \sum_k \frac{\xi_i^{(k)} \xi_i^{(k+l)}\left(\mu_i + 2k + l\right)}{\left( \mu_i + 2k \right) \left( \mu_i + 2k + 2l \right)}\,,
\label{eq: rho from xi}
\end{align}
\end{subequations}
with $\rho_i^{(-l)} = \rho_i^{(l)} \in \mathds{R}$, such that the classical trajectory reduces to a pure cosine series in $2\phi$, describing excess micromotion~\cite{Berkeland1998, Keller2015, Martinez2022} caused by the acceleration $\vect{\alpha}$.
The values of the coefficients $\rho_i^{(l)}$ for the relevant contributions up to order $\mathcal{O}(q^2)$ can be found in Tab.~\ref{tab: Floquet coefficients}.
The classical velocity  
\begin{equation} \label{eq: cl velocity}
\dot{r}_i = \ii \omega_{s,i} r_{\alpha,i}\sum_l v_i^{(l)} \ee^{2\ii l \phi},
\end{equation}
follows through differentiation, with dimensionless velocity comb $v_i^{(l)} =2 l\rho_i^{(l)}/\mu_i$.
Note that $v_i^{(0)} = 0$ and, similar to the Fourier coefficients of $\epsilon_i$, we find $\rho_i^{(l)}=\mathcal{O}(q^{|l|})$, but at the same time $v_i^{(l)} = \mathcal{O}(q^{|l|-1})$.

\subsection{Quantum expectation values in the interaction picture}
At this point, both parts of $\hat{\vect{r}}_\text{I} = \vect{r} + \hat{\vect{x}}$ have been cast into the form of a Fourier series with respect to the trap phase $2\phi = \Omega t + \theta_\text{rf}$, which means that every statistical moment of $\hat{\vect{r}}_\text{I}$ can also be brought into this form.
In the following, we focus our attention on a state which is diagonal in a Fock basis with respect to $\hat{b}_i^\dagger\hat{b}_i$, \eg{} a thermal state~\cite{Martinez2022}.
For such a state, the quantized part of the motion exhibits vanishing expectation values $\langle\hat{\vect{x}}\rangle = 0 =
\langle\dot{\hat{\vect{x}}}\rangle$, resulting in the first moment being purely classical and given by Eqs.~\eqref{eq: cl comb} and \eqref{eq: cl velocity}.
For the second moments, the correlations of the motion $\sim\langle\hat{\vect{r}}_{I}^T\dot{\hat{\vect{r}}}_{I}\rangle$ are not required since the Hamiltonian does not contain terms $\sim \hat{\vect{r}}^T\hat{\vect{p}}$ that would generate motional squeezing~\cite{Sutherland2021}.
Hence, we focus on the mean squares $\langle\hat{\vect{r}}_{I}^2\rangle = \vect{r}^2+\langle\hat{\vect{x}}^2\rangle$ and $\langle\hat{\dot{\vect{r}}}_{I}^2\rangle = \dot{\vect{r}}^2+\langle\dot{\hat{\vect{x}}}^2\rangle$, including no cross term since they vanish for a state diagonal in the Fock basis.
The second moments of the quantized part of the phase-space motion do not vanish, and are instead determined by $|\epsilon_i|^2$, as evident from Eqs.~\eqref{eq: <x_q^2> und <dot{x}_q^2>} and \eqref{eq: <x_q^2> und <dot{x}_q^2>2}.
Using the Fourier series for $\epsilon_i$ from Eq.~\eqref{eq: Floquet}, $|\epsilon_i|^2$ can be written as a Fourier series
\begin{subequations}
\begin{align}
\left| \epsilon_i \right|^2 &= \sum_k \mathcal{X}_i^{(k)2} \ee^{2\ii k \phi},
\qquad
\mathcal{X}_i^{(k)2} = \sum_l \xi_i^{(l)} \xi_i^{(l-k)},
\label{eq: X2}
\\
\frac{\left| \dot\epsilon_i \right|^2}{\omega_{s,i}^2} &=
    \sum_k \mathcal{W}_i^{(k)2} \ee^{2\ii k \phi},
\qquad
\mathcal{W}_i^{(k)2} = \sum_l w_i^{(l)} w_i^{(l-k)},
\label{eq: W2}
\end{align}
\end{subequations}
in the same spirit as the classical trajectory in Eq.~\eqref{eq: cl comb}.
Similarly, the classical trajectory is given by the Fourier series in Eqs.~\eqref{eq: cl comb} and~\eqref{eq: cl velocity}. 
Hence, the squares can be written as
\begin{subequations}
\begin{align}
r_i^2 &= r_{\alpha,i}^2\sum_k \mathcal{R}_i^{(k)2}\ee^{2\ii k \phi},
\qquad
\mathcal{R}_i^{(k)2} \equiv \sum_l \rho_i^{(l)} \rho_i^{(l-k)},
\label{eq: R2}
\\%
\frac{\dot{r}_i^2}{\omega_{s,i}^2} &= \, r_{\alpha,i}^2\sum_k \mathcal{V}_i^{(k)2}\ee^{2\ii k \phi},
\qquad
\mathcal{V}_i^{(k)2} \equiv \sum_l v_i^{(l)} v_i^{(l-k)},
\label{eq: V2}
\end{align}
\end{subequations}
with dimensionless velocity coefficients $v_i^{(l)} =2 l\rho_i^{(l)}/\mu_i$.
Since $w_i^{(k)} = \mathcal{O}(q^{|k|-1})=v_i^{(k)}$ and $\mathcal{W}_i^{(k)2}=\mathcal{O}(q^{|k|-2}) = \mathcal{V}_i^{(k)2}$, we need to consider all contributions from $\mathcal{R}_i^{(k)2}$ and $\mathcal{V}_i^{(k)2}$ up to order $|k|=4$, which are listed in Tab.~\ref{tab: Floquet coefficients2}.
{\renewcommand{\arraystretch}{2.5}
\begin{table}[htb]
\caption{
    Fourier coefficients $R_i^{(k)2}$ and $V_i^{(k)2}$ of $\langle\hat{r}_{I,i}^2\rangle/\ell_{s,i}^2$ and $\langle\dot{\hat{r}}_{I,i}^2\rangle/(\omega_{s,i}^2\ell_{s,i}^2)$, respectively, expanded in the rf phase $\phi$ and given up to $\mathcal{O}(q_i^2)$.
    The index $i$ denotes the different principal axes of the trap.
    Coefficients with $|k|>4$ are at least of order $\mathcal{O}(q_i^3)$.
    The dimensionless position variance for the static component, independent of the phase $2\phi=\Omega t + \theta_\text{rf}$, is $R_i^{(0)2}=(2n_i+1)(1+q_i^2/2)+(r_{\alpha,i}^2/\ell_{s,i}^2)(1+7q_i^2/8)$.
    The contribution $\sim 2n_i+1$ arises from the quantized motion and contains the thermal phonon number $n_i$, while the contribution $\sim r_{\alpha,i}^2/\ell_{s,i}^2$ arises from the classical motion due to the acceleration sag.
   With this definition, higher-order contributions should be truncated consistently with the order in $q_i$ considered.
}
\begin{ruledtabular}\label{tab: Floquet coefficients2}
\begin{tabular}{ccc}
$k$ & $R_i^{(k)2}$ & $V_i^{(k)2}$ \\
\colrule
$0$
    & $\displaystyle R_i^{(0)2}$
    & $\displaystyle \left(\frac{q_i^2}{2\mu_i^2}+\frac{q_i^2}{4}-\frac{21q_i^4}{128\mu_i^2}\right)R_i^{(0)2}+(2n_i+1)\left(1+\frac{3q_i^2}{4}\right)$
    \\
$\pm1$
    & $\displaystyle -\frac{q_i}{2}R_i^{(0)2}$
    & $\displaystyle \frac{q_i}{2}(2n_i+1)-\frac{q_i^3}{16\mu_i^2}R_i^{(0)2}$
   \\
$\pm2$
    & $\displaystyle \frac{3q_i^2}{32}R_i^{(0)2}$
    & $\displaystyle -\frac{7q_i^2}{32}(2n_i+1)-\left(\frac{q_i^2}{4\mu_i^2}+\frac{q_i^2}{8}-\frac{17q_i^4}{192\mu_i^2}\right)R_i^{(0)2}$
    \\
$\pm3$
    & $\displaystyle \mathcal{O}(q_i^3)$
    & $\displaystyle \frac{q_i^3}{16\mu_i^2}R_i^{(0)2}$
    \\
$\pm4$
    & $\displaystyle \mathcal{O}(q_i^3)$
    & $\displaystyle -\frac{q_i^4}{256\mu_i^2}R_i^{(0)2}$
     \\
\end{tabular}
\end{ruledtabular}
\end{table}}

{\renewcommand{\arraystretch}{2.75}
\begin{table}
\caption{
    Fourier coefficients $\Theta_i^{(k,l)}$ of $\im\left[\epsilon_i(t')\dot{\epsilon}_i^*(t)\right]$, associated with the commutator contribution to the Lamb--Dicke phase, expanded in the rf phase $2(l-k)\phi$ and given up to $\mathcal{O}(q_i^2)$.
    The index $i$ denotes the different principal axes of the trap.
    Coefficients with $|k|>2$ or $|l|>3$ are at least of order $\mathcal{O}(q_i^3)$.
    For entries beyond the expansion order, the order of the leading omitted term is given to distinguish them from coefficients that vanish.
}
\begin{ruledtabular}
\begin{tabular}{c|cccc}
$k\, \backslash\, l$ & $0$ & $\pm1$ & $\pm2$ & $\pm3$ \\
\colrule
$-2$
    & $\displaystyle \frac{q_i^2}{64}$
    & $\displaystyle \mp\frac{q_i^3}{128\mu_i}$
    & $\displaystyle \mathcal{O}(q_i^3)$
    & $\displaystyle \mathcal{O}(q_i^3)$ \\
$-1$
    & $\displaystyle -\frac{q_i}{4}-\frac{q_i\mu_i}{4}$
    & $\displaystyle \pm\frac{q_i^2}{8\mu_i}-\frac{q_i^2}{16}(1\mp2)
       $
    & $\displaystyle \mp\frac{q_i^3}{64\mu_i}$
    & $\displaystyle \mathcal{O}(q_i^3)$ \\
$0$
    & $\displaystyle 1+\frac{3q_i^2}{8}$
    & $\displaystyle
       \mp\frac{q_i}{2\mu_i}+\frac{q_i}{4}$
    & $\displaystyle
       \pm\frac{q_i^2}{16\mu_i}-\frac{5q_i^2}{64}$
    & $\displaystyle \mp\frac{q_i^3}{384\mu_i}$ \\
    &&$\displaystyle \mp\frac{17q_i^3}{128\mu_i}
       \mp\frac{q_i\mu_i}{4}$&&\\
$1$
    & $\displaystyle -\frac{q_i}{4}+\frac{q_i\mu_i}{4}$
    & $\displaystyle \pm\frac{q_i^2}{8\mu_i}-\frac{q_i^2}{16}(1\pm2)$
    & $\displaystyle \mp\frac{q_i^3}{64\mu_i}$
    & $\displaystyle \mathcal{O}(q_i^3)$ \\
$2$
    & $\displaystyle \frac{q_i^2}{64}$
    & $\displaystyle \mp\frac{q_i^3}{128\mu_i}$
    & $\displaystyle \mathcal{O}(q_i^3)$
    & $\displaystyle \mathcal{O}(q_i^3)$ \\
\end{tabular}
\end{ruledtabular}
\end{table}

By combining the quantized and classical motion, we obtain
\begin{subequations}
\begin{align}\label{eq: Rk Vk}
\langle \hat{r}_{\text{I},i}^2 \rangle &=  \ell_{s,i}^2 \sum_k R_i^{(k)2} \ee^{2\ii k \phi}\\
\langle \dot{\hat{r}}_{\text{I},i}^2 \rangle &= \omega_{s,i}^2 \ell_{s,i}^2 \sum_k V_i^{(k)2} \ee^{2\ii k \phi},
\end{align}
\end{subequations}
with combined Fourier coefficients
\begin{subequations}
\begin{align}
    R_i^{(k)2} &= \left( 2n_i+1 \right) \mathcal{X}_i^{(k)2} + \frac{r_{\alpha,i}^2}{\ell_{s,i}^2} \mathcal{R}_i^{(k)2}\\
    V_i^{(k)2} &= \left( 2n_i+1 \right) \mathcal{W}_i^{(k)2} + \frac{r_{\alpha,i}^2}{\ell_{s,i}^2} \mathcal{V}_i^{(k)2}
\end{align}
\end{subequations}
The second-order recoil term, giving rise to the Lamb--Dicke phase, couples the position at one instant of time to the velocity at another one through the commutator $\left[\hat{r}_{\text{I},l}(t') , \dot{\hat{r}}_{\text{I},i}(t) \right]$.
Since the classical trajectory is a c-number and has no effect on the commutator, the latter is determined solely by the quantized trajectory from Eq.~\eqref{eq: x_i(t)} as
\begin{equation}
\left[\hat{r}_{\text{I},l}(t'), \dot{\hat{r}}_{\text{I},i}(t)\right] = -2\ii\delta_{il} \ell_{s,i}^2\im\left[ \epsilon_i(t') \dot\epsilon_i^*(t)\right],
\label{eq: comm}
\end{equation}
rendering the commutator a complex number due to $[\hat{b}_i,\hat{b}_l^{\dagger}]=\delta_{il}$.
Therefore, it is independent of the initial quantum state of motion.
By inserting the Floquet solution for $\epsilon_i$ and $\dot{\epsilon}_i$, we find
\begin{equation}
\frac{[\hat{r}_{\text{I},l}(t'), \dot{\hat{r}}_{\text{I},i}(t)]}{2\ii\omega_{s,i}\ell_{s,i}^2} = \delta_{il}\re\left[\sum_{k,k'}\Theta_i^{(k,k')}\ee^{\ii\omega_i^{\left(\frac{k+k'}{2}\right)}\delta t+2\ii(k'-k)\bar{\phi}}\right],
\label{eq: comm double}
\end{equation}
with $\Theta_i^{(k,k')} =  \xi_i^{(k)} w_i^{(k')}$, $\delta t = t-t'$, and $\bar{\phi}=\theta_{\text{rf}}/2+\Omega(t+t')/4$.
Note that the additional constant Floquet phase $\theta_{s,i}$ has dropped out everywhere, since all observable moments for a diagonal initial state are determined by products of the form $\epsilon_i(t)\epsilon_l^*(t')$ with $i=l$.
Such phases are only relevant for non-diagonal motional states where $\langle\hat{b}_i^2\rangle \neq 0$, \eg{} squeezed or coherent states, becoming an additional random phase alongside $\theta_\text{rf}$.

\section{Perturbative expansion of the interferometer phase}
\label{app: PertPhase}
In this Appendix, we transfer a branch-dependent operator formalism for atom interferometers~\cite{Ufrecht2019, Ufrecht20202} to interferometric atomic clocks.
For clocks, the two branches $j=1,2$ are characterized by the Hamiltonian $\hat{H}_j$ that effectively acts only on the center-of-mass motion of each branch, but also includes atom-light interactions and pulse sequences through effective time-dependent internal energies.
The two interfering branches are distinguished by the state-response function ~\cite{Niehof2025} $\lambda_j(t)=\pm 1$, which denotes whether the atom is in the excited ($+1$) or the ground state ($-1$), encoding the pulse sequence.
Hence, each branch is governed by an effective time evolution represented by $\hat{U}_j$.
The overlap $\langle\hat{U}_{1}^{\dagger}\hat{U}_{2}\rangle = \mathcal{V}\exp(\ii\varphi)$ of both time-evolved branches yields the  visibility (contrast) $\mathcal{V}$ of the interference pattern, together with a phase $\varphi$.
Here, the quantum-mechanical expectation value $\langle \bullet \rangle$ is computed with respect to the initial state prepared in the ideal rf-trap without excess micromotion, as defined by the quantized motion calculated in Appendix~\ref{app: COMMotion}.
Moreover, the Hamiltonian $\hat{H}_j = \hat{H}_j^{(0)}+\sum_\kappa\hat{H}_j^{(\kappa)}$, leading to the effective time evolution $\hat{U}_j$ of each branch, is given by a sum of one unperturbed contribution and several perturbations described through the index $\kappa$.
Then, changing into an interaction picture with respect to $\hat{H}_j^{(0)}$ splits the time-evolution operator into $\hat{U}_j=\hat{U}_{j}^{(0)}\hat{U}_{Ij}$, with the unperturbed part $\hat{U}_{j}^{(0)}$ generated by $\hat{H}_j^{(0)}$. 
The unperturbed overlap operator $\hat{U}_{1}^{(0)\dagger}\hat{U}_{2}^{(0)} = \exp(\ii \varphi_0)$ gives rise to the phase $\varphi_0 = - \int \dd t (\hat{H}_2^{(0)}-\hat{H}_1^{(0)})/\hbar$, which is only the c-number
\begin{equation}
 \hat{U}_{1}^{(0)\dagger}\hat{U}_{2}^{(0)} = \ee^{\ii \varphi_0}= \exp\left\lbrace- \ii \int \dd t \frac{\delta \lambda(t)}{2} (\omega_{\text{c}} - \dot \phi_L)\right\rbrace.
\end{equation}
Thus, the overlap including perturbations reduces to $\hat{U}_{1}^{\dagger}\hat{U}_{2}= \exp(\ii \varphi_0)\hat{U}_{I1}^{\dagger}\hat{U}_{I2}$.
Here, the time evolution $\hat{U}_{Ij}$ in the interaction picture is generated by the perturbations $\hat{H}_{Ij}=\sum_{\kappa}\hat{H}_j^{(\kappa)}(\hat{\vect{r}}_I,\hat{\vect{p}}_I)$, being evaluated at the unperturbed trajectories $\hat{\vect{r}}_I=\hat{U}_{j0}^{\dagger}\hat{\vect{r}}\hat{U}_{j0}$ and $\hat{\vect{p}}_I=\hat{U}_{j0}^{\dagger}\hat{\vect{p}}\hat{U}_{j0}$ in the interaction picture.
In contrast to the analogous treatment for atom interferometers~\cite{Ufrecht2019, Ufrecht20202}, both branches of the clock follow the same unperturbed center-of-mass motion, such that $\hat{\vect{r}}_I,\hat{\vect{p}}_I$ are independent of $j$.
Next, we remove the time ordering from the time-evolution operators with the help of a Magnus expansion~\cite{Blanes2010, Magnus1954, Ufrecht20202}, and then explicitly compute the expectation value through a Cumulant expansion~\cite{Cramer1999, Ufrecht20202}.
To this end, we introduce mean and differential perturbative Hamiltonians $\hat{H}^{(\kappa)}_{\pm}$ through $\hat{H}^{(\kappa)}_{1/2}(\hat{\vect{r}}_I,\hat{\vect{p}}_I)= \hat{H}^{(\kappa)}_{+}\mp \hat{H}^{(\kappa)}_{-}/2$.

With this definition, we find up to second order in the perturbations the phase $\varphi\cong \varphi_0+\sum_{\kappa}\varphi_{\kappa}+\sum_{\kappa,\kappa'}\varphi_{\kappa,\kappa'}$, with first-order contribution
\begin{subequations}
\begin{align}
\varphi_{\kappa} =&\, - \frac{1}{\hbar} \int\!\!\dd t\,
    \langle \hat{H}_-^{(\kappa)}(t) \rangle  =g_i^{(\kappa)} \omega_i^{(\kappa)} T \int\!\!  \frac{\dd t}{T} \frac{\delta \lambda}{2}\langle \hat{h}_i^{(\kappa)}\rangle,
\label{eq: first order}
\end{align}
where we sum over indices $i$ associated with the spatial principal axes, and defined the sensitivity function $\delta \lambda /2 = (\lambda_2-\lambda_1)/2= \pm 1$, which is zero before the first and after the last pulse.
Here, $T$ is the time scale associated with the pulse sequence, \eg{} for evenly spaced pulses it represents the elapsed time between them.
We also defined the frequency scale $\omega_i^{(\kappa)}$ associated with each perturbation, as well as the dimensionless prefactor $g_i^{(\kappa)}$ quantifying the order of magnitude of the perturbations and the generator $\langle \hat{h}_i^{(\kappa)}\rangle$ of the differential perturbations that may depend on time and on initial conditions, as listed in Tab.~\ref{tab: phases}.
The second-order contribution to the phase is
\begin{align}
\varphi_{\kappa,\kappa'} =&\, - \frac{\ii}{\hbar^2}
    \int\limits_0^{t_d}\!\!\dd t \int\limits_0^{t}\!\!\dd t'\,
    \langle [ \hat{H}_{+}^{(\kappa)}(t'),
    \hat{H}_{-}^{(\kappa')}(t) ] \rangle ,
\label{eq: second order}
\end{align}
\end{subequations}
where $t_d>t_N$ is the detection time, observing that higher orders contain nested commutators composed of $\hat{H}_{\pm}^{(\kappa)}$.
Even for higher orders in the expansions, the innermost commutator always contains $\hat{H}_{-}^{(\kappa)}$, which justifies the omission of path-independent terms, \eg{} post-Newtonian terms~\cite{Martinez2022}, from the perturbative Hamiltonian, since the differential Hamiltonian $\hat{H}_-^{(\kappa)}$ associated with them vanishes.

Most perturbations in the Hamiltonian depend on $\lambda_j$.
However, $\hat{H}_j^{(\text{rec})}= - \hbar \delta \dot \lambda \vect{k}^\text{T} \hat{\vect{r}}_\text{I}/2$ and therefore depends on  $ \dot \lambda_j$ instead, representing an instantaneous momentum recoil at the time of each pulse instead of encoding the internal state for the time in between.
However, we can use the product rule to obtain $\hat{H}_-^{(\text{rec})}=- \hbar \partial_t [\delta \lambda \vect{k}^\text{T} \hat{\vect{r}}_\text{I}]/2 + \hbar  \delta  \lambda \vect{k}^\text{T} \dot{\hat{\vect{r}}}_\text{I}/2$.
Both branches occupy the same internal state at the beginning and end of the sequence, so $\delta\lambda(0) =0= \delta\lambda(t_d)$. 
Consequently, the first term in $\hat{H}_-^{(\text{rec})}$ cancels when integrated over a full interferometer sequence.
The second term gives the conventional form of the recoil phase with generator $\hat{h}_i^{\text{(rec)}}= - \dot{\hat{\vect{r}}}_{\text{I},i} / v_{s,i}$, order of magnitude $g_i^{\text{(rec)}}=\eta_i = k_i \ell_{s,i}$, and frequency scale $\omega_i^\text{(rec)}=\omega_{s,i}$.
We introduced the length scale $\ell_{s,i}=\sqrt{\hbar/(2m\omega_{s,i})}$ and the velocity scale $v_{s,i}=\omega_{s,i}\ell_{s,i}$ along the principal axis $i$ in the effective trap with trap frequency $\omega_{s,i}$, as well as the Lamb--Dicke parameter $\eta_i$.
The resulting recoil phase corresponds to the first-order Doppler shift~\cite{Ludlow2015}, determined by the wave vector times the velocity.
Since the Lamb--Dicke parameter $\eta_i$ can be orders of magnitude larger than the corresponding parameters of other perturbations for optical transitions, we additionally compute the second-order Magnus contribution from recoil.
We refer to this contribution as the Lamb–Dicke phase
\begin{equation}
\varphi_\text{LD} = \varphi_\text{rec,rec} = \eta_i \eta_l\omega_{s,i}T\int\limits_0^{t_d}\!\! \frac{\dd t}{T}  \frac{\delta \lambda(t)}{2} \frac{\langle \ii [\hat{r}^+_l(t),\dot{\hat{r}}_{\text{I},i}(t)] \rangle}{\ell_{s,l}v_{s,i}}, 
\label{eq: LD step1b}
\end{equation}
with $\hat{r}^+_i(t)=\int_0^{t}\!\!\dd t'\dot\lambda_+(t')\hat{r}_{\text{I},i}(t')/2$  and $\dot{\lambda}_+ = (\dot{\lambda}_2+\dot{\lambda}_1)/2$.
Here, $\dot{\lambda}_+$ is only non-zero at $\pi/2$-pulses. 
For a sequence of $N$ instantaneous pulses, with $\pi/2$-pulses at the beginning and end, it can therefore be written as $\dot{\lambda}_+=\delta(t)+\delta(t-t_N)$.
The sensitivity function $\delta\lambda$ in the outer integral of $\varphi_{\text{LD}}$ confines the outer time to $t < t_N$, so the delta function at $t_N$ lies outside the domain of the inner integration and does not contribute. 
The delta function at $t=0$, however, does contribute, since the domain of the inner integration includes the initial pulse, yielding $\hat{r}^+_i(t)= \hat{r}_{\text{I},i}(0)/2$.
Accordingly, the Lamb--Dicke phase assumes the same form as the first-order phases given by Eq.~\eqref{eq: first order}, and we obtain $g_{i}^\text{(LD)} = \eta_i^2$, $\omega_i^\text{(LD)}= \omega_{s,i}$, and the generator $\hat{h}_{i}^\text{(LD)} = \ii [ \hat{r}_{\text{I},i}(0), \dot{\hat{r}}_{\text{I},i}(t) ]/(2 \ell_{s,i}v_{s,i})$ since the commutator vanishes for $i \neq l $, as shown in Appendix~\ref{app: COMMotion}.
Notably, all contributions of the perturbations considered here are therefore integrals over the sensitivity function $\delta \lambda/2$.

\section{Ensemble averages}
\label{app: Ensemble averages}
Three independent random elements enter the phases, fluctuating from shot to shot: the laser-frequency noise, the initial conditions of the center-of-mass motion of the ion, encoded in the trap phase $\theta_\text{rf}$, and the dark-matter field.
Hence, we consider ensemble averages, obtained by averaging over many shots and denoted by $\overline{\bullet}$.
Because the three sources are independent, their contributions can be treated separately.
In addition, we assume coherence times of the respective noises to be smaller than the cycle time of the experiment, rendering these noise sources independent from shot to shot.
Hence, we observe systematic phase shifts $\overline{\varphi_\kappa}$ for each contribution $\kappa$.
The overall phase noise is characterized by covariances $ \operatorname{Cov}_{\kappa,\kappa'}=\overline{\varphi_\kappa\varphi_{\kappa'}} - \overline{\varphi_\kappa} \,\overline{\varphi_{\kappa'}}$ and variances $\Delta \varphi_\kappa^2 = \operatorname{Cov}_{\kappa,\kappa}$.

{\renewcommand{\arraystretch}{2.8}
\begin{table*}[htb]
\caption{
    List of the individual motional perturbations $\kappa$, the dimensionless factor $g_i^{(\kappa)}$ that determines their order of magnitude, the quantum-mechanical expectation value of the generator $\langle\hat{h}_i^{(\kappa)}\rangle$, and its Fourier decomposition in the Mathieu phase $2\phi=\Omega t+\theta_\text{rf}$.
    We further give the Fourier coefficients $\Phi_\kappa^{(n)}$ of the interferometric phase before the ensemble average defined by $\varphi_\kappa=\sum_n\Phi_\kappa^{(n)}\ee^{\ii n\theta_\text{rf}}$ and their static contribution $\Phi_\kappa^{(0)}$, the latter to lowest order in $q_i$.
    The index $i$ denotes the different principal axes of the trap, with summation over $i$ implicitly assumed in the third and fourth column.
    The coefficients  $\Phi_\kappa^{(n)}$ depend on the filter function $\mathcal{F}(\omega T)=\int\dd t\,\delta\lambda/2\,\exp(\ii\omega t)/T$, evaluated at harmonics of the rf $\Omega$ or the secular frequency $\omega_{s,i}$, as well as their combinations $\mathcal{T}_i^{(n)}$ given by Eq.~\eqref{eq: T def}.
}
\begin{ruledtabular}\label{tab: phases}
\begin{tabular}{ccccc}
$\kappa$ & $g_i^{(\kappa)}$
    & $\langle\hat{h}_i^{(\kappa)}\rangle$
    & $\Phi_\kappa^{(n)}$  & $\Phi_\kappa^{(0)}$ \\
\colrule
rec
    & $\displaystyle \eta_i$
    & $\displaystyle - \frac{\langle \dot{\hat{r}}_{\text{I},i}\rangle}{v_{s,i}}
      = - \ii\, \frac{r_{\alpha,i}}{\ell_{s,i}} \sum_n v_i^{(n)}\,
      \ee^{2\ii n \phi}$
    & $\displaystyle - \ii\, \eta_i\, \frac{r_{\alpha,i}}{\ell_{s,i}}\, v_i^{(n)}\;
      \frac{\mu_i}{2}\, \Omega T\, \mathcal{F}(n \Omega T)$
    & $0$ \\
$\delta\vect{\alpha}$
    & $\displaystyle \frac{r_{\delta\alpha,i}}{2\ell_{s,i}}$
    & $\displaystyle \frac{\langle\hat{r}_{\text{I},i}\rangle}{\ell_{s,i}}
      = \frac{r_{\alpha,i}}{\ell_{s,i}} \sum_n \rho_i^{(n)}\,
      \ee^{2\ii n \phi}$
    & $\displaystyle \frac{r_{\delta\alpha,i}\, r_{\alpha,i}}{2\ell_{s,i}^2}\,
      \rho_i^{(n)}\; \frac{\mu_i}{2}\, \Omega T\, \mathcal{F}(n \Omega T)$
    & $\displaystyle \frac{r_{\delta\alpha,i}\, r_{\alpha,i}}{2\ell_{s,i}^2}\;
      \frac{\mu_i}{2}\, \Omega T\, \mathcal{F}(0)$ \\
$\delta\omega$
    & $\displaystyle - \frac{\delta\omega_i^2}{4\omega_{s,i}^2}$
    & $\displaystyle \frac{\langle\hat{r}_{\text{I},i}^2\rangle}{\ell_{s,i}^2}
      = \sum_n R_i^{(n)2}\,
      \ee^{2\ii n \phi}$
    & $\displaystyle - \frac{\delta\omega_i^2}{4\omega_{s,i}^2}\, R_i^{(n)2}\;
      \frac{\mu_i}{2}\, \Omega T\, \mathcal{F}(n \Omega T)$
    & $\displaystyle - \frac{\delta\omega_i^2}{4\omega_{s,i}^2}
      \bigg( 2n_i + 1+\frac{r_{\alpha,i}^2}{\ell_{s,i}^2} \bigg)
      \frac{\mu_i}{2}\, \Omega T\, \mathcal{F}(0)$ \\
2ds
    & $\displaystyle \frac{v_{s,i}^2}{2c^2}$
    & $\displaystyle \frac{\langle\dot{\hat{r}}_{\text{I},i}^2\rangle}{v_{s,i}^2}
      = \sum_n V_i^{(n)2}\,
      \ee^{2\ii n \phi}$
    & $\displaystyle \frac{\omega_{\text{c}}}{\Omega}\, \frac{v_{s,i}^2}{2c^2}\, V_i^{(n)2}\;
      \Omega T\mathcal{F}(n \Omega T)$
    & $\displaystyle \omega_{\text{c}} T\, \frac{v_{s,i}^2}{2c^2}
      \bigg( 2n_i + 1 + \frac{q_i^2}{2\mu_i^2}\, R_i^{(0)2} \bigg)
      \mathcal{F}(0)$ \\
LD
    & $\displaystyle \eta_i^2$
    &$\displaystyle 
      \frac{\ii\langle[\hat{r}_i^+,\dot{\hat{r}}_{\text{I},i}]\rangle}
      {\ell_{s,i} v_{s,i}} = -\re \sum_{k,k'} \Theta_i^{(k,k')}
      \ee^{\ii [\left( k' - k \right) \theta_\text{rf}
      +  \omega_i^{(k')} t] }$
    & $\displaystyle - \frac{\eta_i^2 \mu_i}{4}\, \Omega T\; \mathcal{T}_i^{(n)}$
    & $\displaystyle - \frac{\eta_i^2 \mu_i}{2} \Omega T
       \re\mathcal{F}\big( \omega_{s,i} T \big)$
\end{tabular}
\end{ruledtabular}
\end{table*}}

\subsection{Laser-frequency fluctuations}
We first consider the unperturbed clock phase with frequency scale $\omega^{(0)} = \omega_{\text{c}}$, order of magnitude $g^{(0)} = 1$, and generator $h^{(0)} = (\dot{\phi}_\text{L} - \omega_{\text{c}})/\omega_{\text{c}}$ as introduced in Appendix~\ref{app: PertPhase}, but now explicitly assuming a fluctuating laser phase.
This contribution exerts an averaged phase shift
\begin{align}
\overline{\varphi_0} =& \int \dd t \frac{\delta\lambda}{2} (\overline{\dot\phi_\text{L}}-\omega_{\text{c}} ) = (\omega_\text{L} -\omega_{\text{c}}) T \mathcal{F}(0),
\label{eq: phi0}
\end{align}
assuming that $\dot\phi_\text{L}$ fluctuates around the carrier laser frequency $\omega_{\text{L}}$.
We defined $\mathcal{F}(0)= \int \dd t \delta\lambda(t)/(2T)$, where $T$ is a characteristic time scale of the interrogation sequence.
For periodic dynamical-decoupling sequences, with $N-2$ equidistant $\pi$-pulses sandwiched between two $\pi/2$-pulses for initialization and readout, the characteristic time $T$ is the time between two of these pulses and $\delta \lambda$ flips between $\pm 1$ at each pulse, leading to
\begin{equation}
\mathcal{F}(0) =
    \begin{cases}
    1 , & N \text{ even} , \\
    \hphantom{-} 0 , & N \text{ odd}.
    \end{cases}
\label{eq: filter dc}
\end{equation}
Fluctuations of $\dot\phi_\text{L}$ are statistically independent of fluctuations of the other two variables, resulting in $\operatorname{Cov}_{0,\kappa'} = 0 $ for $\kappa'\neq 0$ and only the variance
\begin{equation}
\Delta \varphi_0^2  =\int\!\! \dd t \int\!\! \dd t' \frac{\delta\lambda(t)}{2}\frac{\delta\lambda(t')}{2} \left( \overline{\dot\phi_\text{L}(t)\dot\phi_\text{L}(t')} - \omega_\text{L}^2 \right)
\end{equation}
remains.
For stationary processes, the correlation function only depends on $\delta t = t-t'$ and can be written through the power spectral density $\overline{\dot\phi_\text{L}(t)\dot\phi_\text{L}(t')} - \omega_\text{L}^2= \int \dd\omega S_{\dot \phi_\text{L}}(\omega) \ee^{\ii\omega(t-t')}/(2\pi)$, which implies that the unperturbed fluctuations are given by the overlap
\begin{equation}
    \Delta \varphi_0^2 = \int \frac{\dd\omega}{2\pi} S_{\dot \phi_\text{L}}(\omega)\big| T \mathcal{F}(\omega T) \big|^2
\end{equation}
between this noise spectrum and the filter function 
\begin{align}
\mathcal{F}(x)  &= \int\limits\!\!\dd t\frac{\delta\lambda(t)}{2T} \ee^{\ii x t/T}= \frac{1}{x} \tan\frac{x}{2} \left[ 1 + (-1)^N \ee^{\ii (N-1) x} \right]. 
\label{eq: filter}
\end{align}
To obtain the closed form, we again assumed a periodic dynamical-decoupling sequence starting at $t=0$. 
The modulus square of the filter function is then given as
\begin{align}
\left| \mathcal{F}(x) \right|^2 = \frac{4}{x^2} \tan^2\frac{x}{2}
    \begin{cases}
    \cos^2\frac{ \left( N - 1 \right) x}{2} , & N \text{ even} , \\
    \sin^2\frac{ \left( N - 1 \right) x}{2} , & N \text{ odd}.
    \end{cases}
\label{eq: filter abs}
\end{align}
Additionally, we find $\mathcal{F}(2m\pi)=0$ for any integer $m\geq 1$ and $\mathcal{F}((2m+1)\pi)=2\ii(N-1)/[(2m+1)\pi]$ for $m\geq 0$ indicating a suppression at even multiples of $\pi$. 

\subsection{Asynchronous interrogation }
In this subsection, we  compute the average phase contributions apart from the dominant laser phase and the effect of dark matter, focusing on the latter in the next subsection.
As shown in Appendix~\ref{app: PertPhase}, each of these phases is given by an overlap of the sensitivity function $\delta\lambda$ with a generator $\langle\hat{h}_{i}^{(\kappa)}\rangle$, with a characteristic frequency $\omega_i^{(\kappa)}$ and all couplings collected in a dimensionless constant prefactor $g_{i}^{(\kappa)}$.
The corresponding generators are time dependent due to the center-of-mass motion of the ion in the interaction picture, calculated in Appendix~\ref{app: COMMotion} as Fourier series in harmonics of the trap phase $\Omega t +\theta_{\text{rf}}$.
Hence, we also obtain a Fourier series for the generators, which in turn leads to an expansion of the phases in harmonics of $\theta_{\text{rf}}$, yielding
\begin{align}
\varphi_\kappa =&\sum_n \Phi_\kappa^{(n)} \ee^{\ii n \theta_\text{rf}} ,
\qquad
\Phi_\kappa^{(-n)} = \Phi_\kappa^{(n)*} ,
\label{eq: rf series}
\end{align}
where the individual coefficients $\Phi_{\kappa}^{(n)}$ are in Tab.~\ref{tab: phases}.
These coefficients introduce an overlap of the oscillating motion of the generators with $\delta\lambda$, introducing the filter function $\mathcal{F}(n\Omega T)$ as defined in Eq.~\eqref{eq: filter} into the corresponding phases.
As a special case, the Lamb--Dicke contribution samples at the sidebands $\omega_i^{(k)}=\omega_{s,i}+k\Omega$, associated with a linear combination of filter functions
\begin{align}
\mathcal{T}_i^{(n)} =&\, \sum_k \Big[ \Theta_i^{(k-n,k)}\mathcal{F}\big( \omega_i^{(k)}T \big)+ \Theta_i^{(k+n,k)}\mathcal{F}^*\big( \omega_i^{(k)}T \big) \Big].
\label{eq: T def}
\end{align}

In the following, we assume an asynchronous implementation of the experiment, where the interrogation sequences start randomly compared to the trap oscillation, such that $\theta_{\text{rf}}$ is a random variable for each shot.
The ensemble average of $\ee^{\ii(n-n')\theta_\text{rf}}$ leads then to $\delta_{nn'}$, which results in
\begin{align}
\overline{\varphi_\kappa} =&\, \Phi_\kappa^{(0)}.
\label{eq: rf mean}
\end{align}
Thus, the average recoil phase vanishes because the velocity of the ion contains no constant contribution, \ie{} $v_i^{(0)}=0$. 
However, the contributions arising from $\delta \vect{a}, \delta \omega$, and $\text{2ds}$ depend on $\mathcal{F}(0)$ and need not vanish.
The second moments are given by the overlap of two Fourier expansions
\begin{align}
\operatorname{Cov}_{\kappa,\kappa'} = 2 \sum_{n >0} \re[ \Phi_\kappa^{(n)} \Phi_{\kappa'}^{(n)*} ] ,
\label{eq: rf cov}
\end{align}
with the variance $\Delta \varphi_\kappa^2  = \operatorname{Cov}_{\kappa,\kappa}=2\sum_{n>0}|\Phi_\kappa^{(n)}|^2$ constituting the diagonal element.
Because the product $\Phi_{\text{rec}}^{(n)} \Phi_{\kappa'}^{(n)*}$ of the recoil contribution with the other perturbations is purely imaginary, the corresponding covariances vanish, except for the covariance with the Lamb–Dicke contribution.
For optical transitions, the recoil leads to the dominant perturbation, followed by Lamb--Dicke, while others are more strongly suppressed.
Due to this hierarchy, we compute the variance of the recoil phase up to fourth order in $q_i$, and the other variances to lowest order for completeness.
Moreover, covariances other than $\operatorname{Cov}_{\text{rec},\text{LD}}$ are not provided because they lie beyond our order of consideration. 
The remaining contributions for the variances and covariances are listed in Tab.~\ref{tab: cov q2}.
We also identify the relation
\begin{align}
    \mathcal{T}^{(1)}_i \simeq  -\frac{q_i}{2}\re\mathcal{F}(\omega_{s,i}T) - \frac{q_i}{2\mu_i}\Big( \mathcal{F}(\omega_i^{(1)}T) - \mathcal{F}^*(\omega_i^{(-1)}T) \Big) ,
\end{align}
for the leading term at the secular frequency and at the first rf sideband, which is useful for the calculation of the (co-)variances involving the Lamb--Dicke phase.
{\renewcommand{\arraystretch}{3}
\begin{table}
\caption{
    Variances of the individual motional perturbations.
    The indices $i,l$ denote the principal axes of the trap, with summation implicitly assumed.
    The variances depend on the filter function $|\mathcal{F}(\omega T)|^2=|\int\dd t\,\delta\lambda/2\,\exp(\ii\omega t)/T|^2$, evaluated at $\Omega$, $2\Omega$, and combinations with the secular frequency $\omega_{s,i}$ through $\mathcal{T}_i^{(1)}$ given by Eq.~\eqref{eq: T def}.
    The dominant covariance is given as well.
    The variance of the recoil phase is given to $\mathcal{O}(q^4)$, while all others are given to lowest order.
}
\begin{ruledtabular}\label{tab: cov q2}
\begin{tabular}{cc}
$\Delta\varphi_\text{rec}^2$
    & $\displaystyle \frac{(\Omega T)^2}{8} \left|q_ik_i r_{\alpha,i}\left(1 + \frac{17 q_i^2}{64}+\frac{\mu_i^2}{4} \right)\mathcal{F}(\Omega T) \right|^2 $
       \\

    & $\displaystyle +  \frac{(\Omega T)^2}{512} |q_i^2k_i r_{\alpha,i}\mathcal{F}(2\Omega T) |^2$
       \\
$\Delta\varphi_{\delta\vect{\alpha}}^2$
    & $\displaystyle \frac{(\Omega T)^2}{128}  \,
      \left|q_i \mu_i\frac{r_{\delta\alpha,i}\, r_{\alpha,i}}{\ell_{s,i}^2}
      \mathcal{F}(\Omega T)\right|^2$ \\
$\Delta\varphi_{\delta\omega}^2$
    & $\displaystyle \frac{(\Omega T)^2}{128} \left|q_i\mu_i\frac{\delta\omega_i^2}{\omega_{s,i}^2}\frac{\overline{\langle\hat{r}_{\text{I},i}^2\rangle}}{\ell_{s,i}^2}
     \mathcal{F}(\Omega T) \right|^2$ \\
$\Delta\varphi_\text{2ds}^2$
    & $\displaystyle \frac{(\omega_{\text{c}} T)^2}{32}\left|\frac{v_{s,i}^2}{c^2}
      \frac{\overline{\langle\hat{r}_{\text{I},i}^2\rangle}}{\ell_{s,i}^2}\frac{q_i^2}{\mu_i^2}  \mathcal{F}(2\Omega T) \right|^2$ \\
$\Delta\varphi_\text{LD}^2$
    & $\displaystyle \frac{1}{8}
      \Big| \eta_i^2 \mu_i \,\Omega T \mathcal{T}_i^{(1)} \Big|^2$ \\
$\operatorname{Cov}_\text{rec,LD}$
    & $\displaystyle 
      \frac{q_i}{4} k_i r_{\alpha,i} 
      \eta_l^2 \Omega\omega_{s,l}T^2\im\!\left[ \mathcal{F}(\Omega T)
       \mathcal{T}_l^{(1)*}  \right]$ \\
\end{tabular}
\end{ruledtabular}
\end{table}}

\subsection{Dark matter}
For dark matter, we use the same phase representation as for the other phases, with $g^\text{(DM)} = \beta$ representing the coupling of the field to the internal energy, and with generator $\hat{h}^\text{(DM)}= - \varrho(t)/\varrho_0$ with frequency scale $\omega^\text{(DM)}=\varrho_0$ , corresponding to a coupling amplitude given in the main text.
Hence, we find the dark-matter phase
\begin{equation}
    \varphi_\text{DM}=-\beta \int \!\! \dd t \frac{\delta \lambda(t)}{2} \varrho(t), 
\end{equation}
where the dark-matter fluctuations exhibit a vanishing mean $\bar{\varrho}=0$ for coherence times smaller than the cycle time, such that $\overline{\varphi_\text{DM}}=0$ follows.
Since dark matter is assumed to be statistically independent of the laser-phase noise and the trap, no covariances between them survive.
Similar to the laser phase, we use the power spectral density of a stationary dark-matter field $\overline{\varrho(t)\varrho(t')}=\int\dd\omega S_\text{DM}(\omega)\exp[\ii\omega (t-t')]/(2\pi)$ to find the fluctuation of the dark-matter phase
\begin{equation}
    \Delta \varphi_\text{DM}^2 =\beta^2 \int\! \frac{\dd \omega}{2\pi} S_\text{DM}(\omega)  \left|T \mathcal{F}(\omega T)\right|^2.
\end{equation}
For a monochromatic field whose coherence time exceeds the interrogation duration, but whose phase is uncorrelated between successive experimental runs,  the power spectral density reduces to
\begin{align}
S_{\text{DM}}(\omega) =&\, \pi \varrho_0^2
    [ \delta( \omega - \omega_\text{DM} )+\delta( \omega + \omega_\text{DM} )]/2,
\end{align}
leading to
\begin{align}
\Delta\varphi_\text{DM}^2
    =&\, \frac{\left( \beta\, \varrho_0\, T \right)^2}{2}\,
    \big| \mathcal{F}\big( \omega_\text{DM} T \big) \big|^2,
\end{align}
as presented in the main text.
\bibliography{Literatur}

\begin{thebibliography}{113}%
\makeatletter
\providecommand \@ifxundefined [1]{%
 \@ifx{#1\undefined}
}%
\providecommand \@ifnum [1]{%
 \ifnum #1\expandafter \@firstoftwo
 \else \expandafter \@secondoftwo
 \fi
}%
\providecommand \@ifx [1]{%
 \ifx #1\expandafter \@firstoftwo
 \else \expandafter \@secondoftwo
 \fi
}%
\providecommand \natexlab [1]{#1}%
\providecommand \enquote  [1]{``#1''}%
\providecommand \bibnamefont  [1]{#1}%
\providecommand \bibfnamefont [1]{#1}%
\providecommand \citenamefont [1]{#1}%
\providecommand \href@noop [0]{\@secondoftwo}%
\providecommand \href [0]{\begingroup \@sanitize@url \@href}%
\providecommand \@href[1]{\@@startlink{#1}\@@href}%
\providecommand \@@href[1]{\endgroup#1\@@endlink}%
\providecommand \@sanitize@url [0]{\catcode `\\12\catcode `\$12\catcode
  `\&12\catcode `\#12\catcode `\^12\catcode `\_12\catcode `\%12\relax}%
\providecommand \@@startlink[1]{}%
\providecommand \@@endlink[0]{}%
\providecommand \url  [0]{\begingroup\@sanitize@url \@url }%
\providecommand \@url [1]{\endgroup\@href {#1}{\urlprefix }}%
\providecommand \urlprefix  [0]{URL }%
\providecommand \Eprint [0]{\href }%
\providecommand \doibase [0]{https://doi.org/}%
\providecommand \selectlanguage [0]{\@gobble}%
\providecommand \bibinfo  [0]{\@secondoftwo}%
\providecommand \bibfield  [0]{\@secondoftwo}%
\providecommand \translation [1]{[#1]}%
\providecommand \BibitemOpen [0]{}%
\providecommand \bibitemStop [0]{}%
\providecommand \bibitemNoStop [0]{.\EOS\space}%
\providecommand \EOS [0]{\spacefactor3000\relax}%
\providecommand \BibitemShut  [1]{\csname bibitem#1\endcsname}%
\let\auto@bib@innerbib\@empty
\bibitem [{\citenamefont {Poli}\ \emph {et~al.}(2013)\citenamefont {Poli},
  \citenamefont {Oates}, \citenamefont {Gill},\ and\ \citenamefont
  {Tino}}]{Poli2013}%
  \BibitemOpen
  \bibfield  {author} {\bibinfo {author} {\bibfnamefont {N.}~\bibnamefont
  {Poli}}, \bibinfo {author} {\bibfnamefont {C.~W.}\ \bibnamefont {Oates}},
  \bibinfo {author} {\bibfnamefont {P.}~\bibnamefont {Gill}},\ and\ \bibinfo
  {author} {\bibfnamefont {G.~M.}\ \bibnamefont {Tino}},\ }\bibfield  {title}
  {\bibinfo {title} {Optical atomic clocks},\ }\href
  {https://doi.org/10.1393/ncr/i2013-10095-x} {\bibfield  {journal} {\bibinfo
  {journal} {Riv. Nuovo Cimento}\ }\textbf {\bibinfo {volume} {036}},\ \bibinfo
  {pages} {555} (\bibinfo {year} {2013})}\BibitemShut {NoStop}%
\bibitem [{\citenamefont {Ludlow}\ \emph {et~al.}(2015)\citenamefont {Ludlow},
  \citenamefont {Boyd}, \citenamefont {Ye}, \citenamefont {Peik},\ and\
  \citenamefont {Schmidt}}]{Ludlow2015}%
  \BibitemOpen
  \bibfield  {author} {\bibinfo {author} {\bibfnamefont {A.~D.}\ \bibnamefont
  {Ludlow}}, \bibinfo {author} {\bibfnamefont {M.~M.}\ \bibnamefont {Boyd}},
  \bibinfo {author} {\bibfnamefont {J.}~\bibnamefont {Ye}}, \bibinfo {author}
  {\bibfnamefont {E.}~\bibnamefont {Peik}},\ and\ \bibinfo {author}
  {\bibfnamefont {P.~O.}\ \bibnamefont {Schmidt}},\ }\bibfield  {title}
  {\bibinfo {title} {Optical atomic clocks},\ }\href
  {https://doi.org/10.1103/RevModPhys.87.637} {\bibfield  {journal} {\bibinfo
  {journal} {Rev. Mod. Phys.}\ }\textbf {\bibinfo {volume} {87}},\ \bibinfo
  {pages} {637} (\bibinfo {year} {2015})}\BibitemShut {NoStop}%
\bibitem [{\citenamefont {Safronova}\ \emph {et~al.}(2018)\citenamefont
  {Safronova}, \citenamefont {Budker}, \citenamefont {DeMille}, \citenamefont
  {Kimball}, \citenamefont {Derevianko},\ and\ \citenamefont
  {Clark}}]{Safronova2018}%
  \BibitemOpen
  \bibfield  {author} {\bibinfo {author} {\bibfnamefont {M.~S.}\ \bibnamefont
  {Safronova}}, \bibinfo {author} {\bibfnamefont {D.}~\bibnamefont {Budker}},
  \bibinfo {author} {\bibfnamefont {D.}~\bibnamefont {DeMille}}, \bibinfo
  {author} {\bibfnamefont {D.~F.~J.}\ \bibnamefont {Kimball}}, \bibinfo
  {author} {\bibfnamefont {A.}~\bibnamefont {Derevianko}},\ and\ \bibinfo
  {author} {\bibfnamefont {C.~W.}\ \bibnamefont {Clark}},\ }\bibfield  {title}
  {\bibinfo {title} {Search for new physics with atoms and molecules},\ }\href
  {https://doi.org/10.1103/RevModPhys.90.025008} {\bibfield  {journal}
  {\bibinfo  {journal} {Rev. Mod. Phys.}\ }\textbf {\bibinfo {volume} {90}},\
  \bibinfo {pages} {025008} (\bibinfo {year} {2018})}\BibitemShut {NoStop}%
\bibitem [{\citenamefont {Fortier}\ \emph {et~al.}(2026)\citenamefont
  {Fortier}, \citenamefont {Luiten},\ and\ \citenamefont
  {Margolis}}]{Fortier2026}%
  \BibitemOpen
  \bibfield  {author} {\bibinfo {author} {\bibfnamefont {T.~M.}\ \bibnamefont
  {Fortier}}, \bibinfo {author} {\bibfnamefont {A.~N.}\ \bibnamefont
  {Luiten}},\ and\ \bibinfo {author} {\bibfnamefont {H.~S.}\ \bibnamefont
  {Margolis}},\ }\bibfield  {title} {\bibinfo {title} {Optical atomic clocks:
  defining the future of time and frequency metrology},\ }\href
  {https://doi.org/10.1364/OPTICA.575770} {\bibfield  {journal} {\bibinfo
  {journal} {Optica}\ }\textbf {\bibinfo {volume} {13}},\ \bibinfo {pages}
  {143} (\bibinfo {year} {2026})}\BibitemShut {NoStop}%
\bibitem [{\citenamefont {Leibfried}\ \emph {et~al.}(2003)\citenamefont
  {Leibfried}, \citenamefont {Blatt}, \citenamefont {Monroe},\ and\
  \citenamefont {Wineland}}]{Leibfried2003}%
  \BibitemOpen
  \bibfield  {author} {\bibinfo {author} {\bibfnamefont {D.}~\bibnamefont
  {Leibfried}}, \bibinfo {author} {\bibfnamefont {R.}~\bibnamefont {Blatt}},
  \bibinfo {author} {\bibfnamefont {C.}~\bibnamefont {Monroe}},\ and\ \bibinfo
  {author} {\bibfnamefont {D.}~\bibnamefont {Wineland}},\ }\bibfield  {title}
  {\bibinfo {title} {Quantum dynamics of single trapped ions},\ }\href
  {https://doi.org/10.1103/RevModPhys.75.281} {\bibfield  {journal} {\bibinfo
  {journal} {Rev. Mod. Phys.}\ }\textbf {\bibinfo {volume} {75}},\ \bibinfo
  {pages} {281} (\bibinfo {year} {2003})}\BibitemShut {NoStop}%
\bibitem [{\citenamefont {Mart\'{\i}nez-Lahuerta}\ \emph
  {et~al.}(2022)\citenamefont {Mart\'{\i}nez-Lahuerta}, \citenamefont {Eilers},
  \citenamefont {Mehlst\"aubler}, \citenamefont {Schmidt},\ and\ \citenamefont
  {Hammerer}}]{Martinez2022}%
  \BibitemOpen
  \bibfield  {author} {\bibinfo {author} {\bibfnamefont {V.~J.}\ \bibnamefont
  {Mart\'{\i}nez-Lahuerta}}, \bibinfo {author} {\bibfnamefont {S.}~\bibnamefont
  {Eilers}}, \bibinfo {author} {\bibfnamefont {T.~E.}\ \bibnamefont
  {Mehlst\"aubler}}, \bibinfo {author} {\bibfnamefont {P.~O.}\ \bibnamefont
  {Schmidt}},\ and\ \bibinfo {author} {\bibfnamefont {K.}~\bibnamefont
  {Hammerer}},\ }\bibfield  {title} {\bibinfo {title} {Ab initio quantum theory
  of mass defect and time dilation in trapped-ion optical clocks},\ }\href
  {https://doi.org/10.1103/PhysRevA.106.032803} {\bibfield  {journal} {\bibinfo
   {journal} {Phys. Rev. A}\ }\textbf {\bibinfo {volume} {106}},\ \bibinfo
  {pages} {032803} (\bibinfo {year} {2022})}\BibitemShut {NoStop}%
\bibitem [{\citenamefont {Giese}(2015)}]{Giese2015}%
  \BibitemOpen
  \bibfield  {author} {\bibinfo {author} {\bibfnamefont {E.}~\bibnamefont
  {Giese}},\ }\bibfield  {title} {\bibinfo {title} {Mechanisms of matter-wave
  diffraction and their application to interferometers},\ }\href
  {https://doi.org/10.1002/prop.201500020} {\bibfield  {journal} {\bibinfo
  {journal} {Fortschr. Phys.}\ }\textbf {\bibinfo {volume} {63}},\ \bibinfo
  {pages} {337} (\bibinfo {year} {2015})}\BibitemShut {NoStop}%
\bibitem [{\citenamefont {Rudolph}\ \emph {et~al.}(2020)\citenamefont
  {Rudolph}, \citenamefont {Wilkason}, \citenamefont {Nantel}, \citenamefont
  {Swan}, \citenamefont {Holland}, \citenamefont {Jiang}, \citenamefont
  {Garber}, \citenamefont {Carman},\ and\ \citenamefont {Hogan}}]{Rudolph2020}%
  \BibitemOpen
  \bibfield  {author} {\bibinfo {author} {\bibfnamefont {J.}~\bibnamefont
  {Rudolph}}, \bibinfo {author} {\bibfnamefont {T.}~\bibnamefont {Wilkason}},
  \bibinfo {author} {\bibfnamefont {M.}~\bibnamefont {Nantel}}, \bibinfo
  {author} {\bibfnamefont {H.}~\bibnamefont {Swan}}, \bibinfo {author}
  {\bibfnamefont {C.~M.}\ \bibnamefont {Holland}}, \bibinfo {author}
  {\bibfnamefont {Y.}~\bibnamefont {Jiang}}, \bibinfo {author} {\bibfnamefont
  {B.~E.}\ \bibnamefont {Garber}}, \bibinfo {author} {\bibfnamefont {S.~P.}\
  \bibnamefont {Carman}},\ and\ \bibinfo {author} {\bibfnamefont {J.~M.}\
  \bibnamefont {Hogan}},\ }\bibfield  {title} {\bibinfo {title} {Large momentum
  transfer clock atom interferometry on the 689 nm intercombination line of
  strontium},\ }\href {https://doi.org/10.1103/PhysRevLett.124.083604}
  {\bibfield  {journal} {\bibinfo  {journal} {Phys. Rev. Lett.}\ }\textbf
  {\bibinfo {volume} {124}},\ \bibinfo {pages} {083604} (\bibinfo {year}
  {2020})}\BibitemShut {NoStop}%
\bibitem [{\citenamefont {Chiarotti}\ \emph {et~al.}(2022)\citenamefont
  {Chiarotti}, \citenamefont {Tinsley}, \citenamefont {Bandarupally},
  \citenamefont {Manzoor}, \citenamefont {Sacco}, \citenamefont {Salvi},\ and\
  \citenamefont {Poli}}]{Chiarotti2022}%
  \BibitemOpen
  \bibfield  {author} {\bibinfo {author} {\bibfnamefont {M.}~\bibnamefont
  {Chiarotti}}, \bibinfo {author} {\bibfnamefont {J.~N.}\ \bibnamefont
  {Tinsley}}, \bibinfo {author} {\bibfnamefont {S.}~\bibnamefont
  {Bandarupally}}, \bibinfo {author} {\bibfnamefont {S.}~\bibnamefont
  {Manzoor}}, \bibinfo {author} {\bibfnamefont {M.}~\bibnamefont {Sacco}},
  \bibinfo {author} {\bibfnamefont {L.}~\bibnamefont {Salvi}},\ and\ \bibinfo
  {author} {\bibfnamefont {N.}~\bibnamefont {Poli}},\ }\bibfield  {title}
  {\bibinfo {title} {Practical limits for large-momentum-transfer clock atom
  interferometers},\ }\href {https://doi.org/10.1103/PRXQuantum.3.030348}
  {\bibfield  {journal} {\bibinfo  {journal} {PRX Quantum}\ }\textbf {\bibinfo
  {volume} {3}},\ \bibinfo {pages} {030348} (\bibinfo {year}
  {2022})}\BibitemShut {NoStop}%
\bibitem [{\citenamefont {Niehof}\ \emph {et~al.}(2025)\citenamefont {Niehof},
  \citenamefont {Derr},\ and\ \citenamefont {Giese}}]{Niehof2025}%
  \BibitemOpen
  \bibfield  {author} {\bibinfo {author} {\bibfnamefont {C.}~\bibnamefont
  {Niehof}}, \bibinfo {author} {\bibfnamefont {D.}~\bibnamefont {Derr}},\ and\
  \bibinfo {author} {\bibfnamefont {E.}~\bibnamefont {Giese}},\ }\bibfield
  {title} {\bibinfo {title} {Finite-speed-of-light effects in atom
  interferometry: Diffraction mechanisms and resonance conditions},\ }\href
  {https://doi.org/10.1116/5.0284803} {\bibfield  {journal} {\bibinfo
  {journal} {AVS Quantum Sci.}\ }\textbf {\bibinfo {volume} {7}},\ \bibinfo
  {pages} {034404} (\bibinfo {year} {2025})}\BibitemShut {NoStop}%
\bibitem [{\citenamefont {Bongs}\ \emph {et~al.}(2019)\citenamefont {Bongs},
  \citenamefont {Holynski}, \citenamefont {Vovrosh}, \citenamefont {Bouyer},
  \citenamefont {Condon}, \citenamefont {Rasel}, \citenamefont {Schubert},
  \citenamefont {Schleich},\ and\ \citenamefont {Roura}}]{Bongs2019}%
  \BibitemOpen
  \bibfield  {author} {\bibinfo {author} {\bibfnamefont {K.}~\bibnamefont
  {Bongs}}, \bibinfo {author} {\bibfnamefont {M.}~\bibnamefont {Holynski}},
  \bibinfo {author} {\bibfnamefont {J.}~\bibnamefont {Vovrosh}}, \bibinfo
  {author} {\bibfnamefont {P.}~\bibnamefont {Bouyer}}, \bibinfo {author}
  {\bibfnamefont {G.}~\bibnamefont {Condon}}, \bibinfo {author} {\bibfnamefont
  {E.~M.}\ \bibnamefont {Rasel}}, \bibinfo {author} {\bibfnamefont
  {C.}~\bibnamefont {Schubert}}, \bibinfo {author} {\bibfnamefont {W.~P.}\
  \bibnamefont {Schleich}},\ and\ \bibinfo {author} {\bibfnamefont
  {A.}~\bibnamefont {Roura}},\ }\bibfield  {title} {\bibinfo {title} {Taking
  atom interferometric quantum sensors from the laboratory to real-world
  applications},\ }\href {https://doi.org/10.1038/s42254-019-0117-4} {\bibfield
   {journal} {\bibinfo  {journal} {Nat. Rev. Phys.}\ }\textbf {\bibinfo
  {volume} {1}},\ \bibinfo {pages} {731} (\bibinfo {year} {2019})}\BibitemShut
  {NoStop}%
\bibitem [{\citenamefont {Abend}\ \emph {et~al.}(2024)\citenamefont {Abend}
  \emph {et~al.}}]{Abend2024}%
  \BibitemOpen
  \bibfield  {author} {\bibinfo {author} {\bibfnamefont {S.}~\bibnamefont
  {Abend}} \emph {et~al.},\ }\bibfield  {title} {\bibinfo {title} {Terrestrial
  very-long-baseline atom interferometry: {W}orkshop summary},\ }\href
  {https://doi.org/10.1116/5.0185291} {\bibfield  {journal} {\bibinfo
  {journal} {AVS Quantum Sci.}\ }\textbf {\bibinfo {volume} {6}},\ \bibinfo
  {pages} {024701} (\bibinfo {year} {2024})}\BibitemShut {NoStop}%
\bibitem [{\citenamefont {Paul}(1990)}]{Paul1990}%
  \BibitemOpen
  \bibfield  {author} {\bibinfo {author} {\bibfnamefont {W.}~\bibnamefont
  {Paul}},\ }\bibfield  {title} {\bibinfo {title} {Electromagnetic traps for
  charged and neutral particles},\ }\href
  {https://doi.org/10.1103/RevModPhys.62.531} {\bibfield  {journal} {\bibinfo
  {journal} {Rev. Mod. Phys.}\ }\textbf {\bibinfo {volume} {62}},\ \bibinfo
  {pages} {531} (\bibinfo {year} {1990})}\BibitemShut {NoStop}%
\bibitem [{\citenamefont {Glauber}(1992)}]{Glauber1992}%
  \BibitemOpen
  \bibfield  {author} {\bibinfo {author} {\bibfnamefont {R.~J.}\ \bibnamefont
  {Glauber}},\ }\bibfield  {title} {\bibinfo {title} {The quantum mechanics of
  trapped wave packets},\ }in\ \href@noop {} {\emph {\bibinfo {booktitle}
  {Laser Manipulation of Atoms and Ions, Proceedings of the International
  School of Physics ‘‘{E}nrico {F}ermi’’ Course 118}}},\ \bibinfo
  {editor} {edited by\ \bibinfo {editor} {\bibfnamefont {E.}~\bibnamefont
  {Arimondo}}, \bibinfo {editor} {\bibfnamefont {W.~D.}\ \bibnamefont
  {Phillips}},\ and\ \bibinfo {editor} {\bibfnamefont {F.}~\bibnamefont
  {Strumia}}}\ (\bibinfo  {publisher} {North-Holland Elsevier Science
  Publishers B.V.},\ \bibinfo {address} {Amsterdam},\ \bibinfo {year} {1992})\
  p.\ \bibinfo {pages} {643}\BibitemShut {NoStop}%
\bibitem [{\citenamefont {Schleich}(2001)}]{Schleich2001}%
  \BibitemOpen
  \bibfield  {author} {\bibinfo {author} {\bibfnamefont {W.~P.}\ \bibnamefont
  {Schleich}},\ }\href {https://doi.org/10.1002/3527602976} {\emph {\bibinfo
  {title} {Quantum Optics in Phase Space}}}\ (\bibinfo  {publisher} {John Wiley
  \& Sons, Ltd},\ \bibinfo {year} {2001})\BibitemShut {NoStop}%
\bibitem [{\citenamefont {Yudin}\ \emph {et~al.}(2010)\citenamefont {Yudin},
  \citenamefont {Taichenachev}, \citenamefont {Oates}, \citenamefont {Barber},
  \citenamefont {Lemke}, \citenamefont {Ludlow}, \citenamefont {Sterr},
  \citenamefont {Lisdat},\ and\ \citenamefont {Riehle}}]{Yudin2010}%
  \BibitemOpen
  \bibfield  {author} {\bibinfo {author} {\bibfnamefont {V.~I.}\ \bibnamefont
  {Yudin}}, \bibinfo {author} {\bibfnamefont {A.~V.}\ \bibnamefont
  {Taichenachev}}, \bibinfo {author} {\bibfnamefont {C.~W.}\ \bibnamefont
  {Oates}}, \bibinfo {author} {\bibfnamefont {Z.~W.}\ \bibnamefont {Barber}},
  \bibinfo {author} {\bibfnamefont {N.~D.}\ \bibnamefont {Lemke}}, \bibinfo
  {author} {\bibfnamefont {A.~D.}\ \bibnamefont {Ludlow}}, \bibinfo {author}
  {\bibfnamefont {U.}~\bibnamefont {Sterr}}, \bibinfo {author} {\bibfnamefont
  {C.}~\bibnamefont {Lisdat}},\ and\ \bibinfo {author} {\bibfnamefont
  {F.}~\bibnamefont {Riehle}},\ }\bibfield  {title} {\bibinfo {title}
  {Hyper-{R}amsey spectroscopy of optical clock transitions},\ }\href
  {https://doi.org/10.1103/PhysRevA.82.011804} {\bibfield  {journal} {\bibinfo
  {journal} {Phys. Rev. A}\ }\textbf {\bibinfo {volume} {82}},\ \bibinfo
  {pages} {011804} (\bibinfo {year} {2010})}\BibitemShut {NoStop}%
\bibitem [{\citenamefont {Huntemann}\ \emph {et~al.}(2012)\citenamefont
  {Huntemann}, \citenamefont {Lipphardt}, \citenamefont {Okhapkin},
  \citenamefont {Tamm}, \citenamefont {Peik}, \citenamefont {Taichenachev},\
  and\ \citenamefont {Yudin}}]{Huntemann2012}%
  \BibitemOpen
  \bibfield  {author} {\bibinfo {author} {\bibfnamefont {N.}~\bibnamefont
  {Huntemann}}, \bibinfo {author} {\bibfnamefont {B.}~\bibnamefont
  {Lipphardt}}, \bibinfo {author} {\bibfnamefont {M.}~\bibnamefont {Okhapkin}},
  \bibinfo {author} {\bibfnamefont {C.}~\bibnamefont {Tamm}}, \bibinfo {author}
  {\bibfnamefont {E.}~\bibnamefont {Peik}}, \bibinfo {author} {\bibfnamefont
  {A.~V.}\ \bibnamefont {Taichenachev}},\ and\ \bibinfo {author} {\bibfnamefont
  {V.~I.}\ \bibnamefont {Yudin}},\ }\bibfield  {title} {\bibinfo {title}
  {Generalized {R}amsey excitation scheme with suppressed light shift},\ }\href
  {https://doi.org/10.1103/PhysRevLett.109.213002} {\bibfield  {journal}
  {\bibinfo  {journal} {Phys. Rev. Lett.}\ }\textbf {\bibinfo {volume} {109}},\
  \bibinfo {pages} {213002} (\bibinfo {year} {2012})}\BibitemShut {NoStop}%
\bibitem [{\citenamefont {Zanon-Willette}\ \emph {et~al.}(2022)\citenamefont
  {Zanon-Willette}, \citenamefont {Wilkowski}, \citenamefont {Lefevre},
  \citenamefont {Taichenachev},\ and\ \citenamefont
  {Yudin}}]{Zanon-Willette2022}%
  \BibitemOpen
  \bibfield  {author} {\bibinfo {author} {\bibfnamefont {T.}~\bibnamefont
  {Zanon-Willette}}, \bibinfo {author} {\bibfnamefont {D.}~\bibnamefont
  {Wilkowski}}, \bibinfo {author} {\bibfnamefont {R.}~\bibnamefont {Lefevre}},
  \bibinfo {author} {\bibfnamefont {A.~V.}\ \bibnamefont {Taichenachev}},\ and\
  \bibinfo {author} {\bibfnamefont {V.~I.}\ \bibnamefont {Yudin}},\ }\bibfield
  {title} {\bibinfo {title} {Generalized hyper-{R}amsey-{B}ord\'e matter-wave
  interferometry: {Q}uantum engineering of robust atomic sensors with composite
  pulses},\ }\href {https://doi.org/10.1103/PhysRevResearch.4.023222}
  {\bibfield  {journal} {\bibinfo  {journal} {Phys. Rev. Res.}\ }\textbf
  {\bibinfo {volume} {4}},\ \bibinfo {pages} {023222} (\bibinfo {year}
  {2022})}\BibitemShut {NoStop}%
\bibitem [{\citenamefont {Zanon-Willette}\ \emph {et~al.}(2016)\citenamefont
  {Zanon-Willette}, \citenamefont {de~Clercq},\ and\ \citenamefont
  {Arimondo}}]{Zanon-Willette2016}%
  \BibitemOpen
  \bibfield  {author} {\bibinfo {author} {\bibfnamefont {T.}~\bibnamefont
  {Zanon-Willette}}, \bibinfo {author} {\bibfnamefont {E.}~\bibnamefont
  {de~Clercq}},\ and\ \bibinfo {author} {\bibfnamefont {E.}~\bibnamefont
  {Arimondo}},\ }\bibfield  {title} {\bibinfo {title} {Probe light-shift
  elimination in generalized hyper-{R}amsey quantum clocks},\ }\href
  {https://doi.org/10.1103/PhysRevA.93.042506} {\bibfield  {journal} {\bibinfo
  {journal} {Phys. Rev. A}\ }\textbf {\bibinfo {volume} {93}},\ \bibinfo
  {pages} {042506} (\bibinfo {year} {2016})}\BibitemShut {NoStop}%
\bibitem [{\citenamefont {Hobson}\ \emph {et~al.}(2016)\citenamefont {Hobson},
  \citenamefont {Bowden}, \citenamefont {King}, \citenamefont {Baird},
  \citenamefont {Hill},\ and\ \citenamefont {Gill}}]{Hobson2016}%
  \BibitemOpen
  \bibfield  {author} {\bibinfo {author} {\bibfnamefont {R.}~\bibnamefont
  {Hobson}}, \bibinfo {author} {\bibfnamefont {W.}~\bibnamefont {Bowden}},
  \bibinfo {author} {\bibfnamefont {S.~A.}\ \bibnamefont {King}}, \bibinfo
  {author} {\bibfnamefont {P.~E.~G.}\ \bibnamefont {Baird}}, \bibinfo {author}
  {\bibfnamefont {I.~R.}\ \bibnamefont {Hill}},\ and\ \bibinfo {author}
  {\bibfnamefont {P.}~\bibnamefont {Gill}},\ }\bibfield  {title} {\bibinfo
  {title} {Modified hyper-{R}amsey methods for the elimination of probe shifts
  in optical clocks},\ }\href {https://doi.org/10.1103/PhysRevA.93.010501}
  {\bibfield  {journal} {\bibinfo  {journal} {Phys. Rev. A}\ }\textbf {\bibinfo
  {volume} {93}},\ \bibinfo {pages} {010501} (\bibinfo {year}
  {2016})}\BibitemShut {NoStop}%
\bibitem [{\citenamefont {Cywi{\ifmmode \acute{n}\else \'{n}\fi{}}ski}\ \emph
  {et~al.}(2008)\citenamefont {Cywi{\ifmmode \acute{n}\else \'{n}\fi{}}ski},
  \citenamefont {Lutchyn}, \citenamefont {Nave},\ and\ \citenamefont
  {Das~Sarma}}]{Cywinski2008}%
  \BibitemOpen
  \bibfield  {author} {\bibinfo {author} {\bibfnamefont {{\L{}}.}~\bibnamefont
  {Cywi{\ifmmode \acute{n}\else \'{n}\fi{}}ski}}, \bibinfo {author}
  {\bibfnamefont {R.~M.}\ \bibnamefont {Lutchyn}}, \bibinfo {author}
  {\bibfnamefont {C.~P.}\ \bibnamefont {Nave}},\ and\ \bibinfo {author}
  {\bibfnamefont {S.}~\bibnamefont {Das~Sarma}},\ }\bibfield  {title} {\bibinfo
  {title} {How to enhance dephasing time in superconducting qubits},\ }\href
  {https://doi.org/10.1103/PhysRevB.77.174509} {\bibfield  {journal} {\bibinfo
  {journal} {Phys. Rev. B}\ }\textbf {\bibinfo {volume} {77}},\ \bibinfo
  {pages} {174509} (\bibinfo {year} {2008})}\BibitemShut {NoStop}%
\bibitem [{\citenamefont {Kotler}\ \emph {et~al.}(2011)\citenamefont {Kotler},
  \citenamefont {Akerman}, \citenamefont {Glickman}, \citenamefont {Keselman},\
  and\ \citenamefont {Ozeri}}]{Kotler2011}%
  \BibitemOpen
  \bibfield  {author} {\bibinfo {author} {\bibfnamefont {S.}~\bibnamefont
  {Kotler}}, \bibinfo {author} {\bibfnamefont {N.}~\bibnamefont {Akerman}},
  \bibinfo {author} {\bibfnamefont {Y.}~\bibnamefont {Glickman}}, \bibinfo
  {author} {\bibfnamefont {A.}~\bibnamefont {Keselman}},\ and\ \bibinfo
  {author} {\bibfnamefont {R.}~\bibnamefont {Ozeri}},\ }\bibfield  {title}
  {\bibinfo {title} {Single-ion quantum lock-in amplifier},\ }\href
  {https://doi.org/10.1038/nature10010} {\bibfield  {journal} {\bibinfo
  {journal} {Nature}\ }\textbf {\bibinfo {volume} {473}},\ \bibinfo {pages}
  {61} (\bibinfo {year} {2011})}\BibitemShut {NoStop}%
\bibitem [{\citenamefont {Biercuk}\ \emph {et~al.}(2011)\citenamefont
  {Biercuk}, \citenamefont {Doherty},\ and\ \citenamefont {Uys}}]{Biercuk2011}%
  \BibitemOpen
  \bibfield  {author} {\bibinfo {author} {\bibfnamefont {M.~J.}\ \bibnamefont
  {Biercuk}}, \bibinfo {author} {\bibfnamefont {A.~C.}\ \bibnamefont
  {Doherty}},\ and\ \bibinfo {author} {\bibfnamefont {H.}~\bibnamefont {Uys}},\
  }\bibfield  {title} {\bibinfo {title} {Dynamical decoupling sequence
  construction as a filter-design problem},\ }\href
  {https://doi.org/10.1088/0953-4075/44/15/154002} {\bibfield  {journal}
  {\bibinfo  {journal} {J. Phys. B: At. Mol. Opt. Phys.}\ }\textbf {\bibinfo
  {volume} {44}},\ \bibinfo {pages} {154002} (\bibinfo {year}
  {2011})}\BibitemShut {NoStop}%
\bibitem [{\citenamefont {Shaniv}\ and\ \citenamefont
  {Ozeri}(2017)}]{Shaniv2017}%
  \BibitemOpen
  \bibfield  {author} {\bibinfo {author} {\bibfnamefont {R.}~\bibnamefont
  {Shaniv}}\ and\ \bibinfo {author} {\bibfnamefont {R.}~\bibnamefont {Ozeri}},\
  }\bibfield  {title} {\bibinfo {title} {Quantum lock-in force sensing using
  optical clock {D}oppler velocimetry},\ }\href
  {https://doi.org/10.1038/ncomms14157} {\bibfield  {journal} {\bibinfo
  {journal} {Nat. Commun.}\ }\textbf {\bibinfo {volume} {8}},\ \bibinfo {pages}
  {14157} (\bibinfo {year} {2017})}\BibitemShut {NoStop}%
\bibitem [{\citenamefont {Aharony}\ \emph {et~al.}(2021)\citenamefont
  {Aharony}, \citenamefont {Akerman}, \citenamefont {Ozeri}, \citenamefont
  {Perez}, \citenamefont {Savoray},\ and\ \citenamefont
  {Shaniv}}]{Aharony2021}%
  \BibitemOpen
  \bibfield  {author} {\bibinfo {author} {\bibfnamefont {S.}~\bibnamefont
  {Aharony}}, \bibinfo {author} {\bibfnamefont {N.}~\bibnamefont {Akerman}},
  \bibinfo {author} {\bibfnamefont {R.}~\bibnamefont {Ozeri}}, \bibinfo
  {author} {\bibfnamefont {G.}~\bibnamefont {Perez}}, \bibinfo {author}
  {\bibfnamefont {I.}~\bibnamefont {Savoray}},\ and\ \bibinfo {author}
  {\bibfnamefont {R.}~\bibnamefont {Shaniv}},\ }\bibfield  {title} {\bibinfo
  {title} {Constraining rapidly oscillating scalar dark matter using dynamic
  decoupling},\ }\href {https://doi.org/10.1103/PhysRevD.103.075017} {\bibfield
   {journal} {\bibinfo  {journal} {Phys. Rev. D}\ }\textbf {\bibinfo {volume}
  {103}},\ \bibinfo {pages} {075017} (\bibinfo {year} {2021})}\BibitemShut
  {NoStop}%
\bibitem [{\citenamefont {Zaheer}\ \emph {et~al.}(2025)\citenamefont {Zaheer},
  \citenamefont {Matjelo}, \citenamefont {Hume}, \citenamefont {Safronova},\
  and\ \citenamefont {Leibrandt}}]{Zaheer2025}%
  \BibitemOpen
  \bibfield  {author} {\bibinfo {author} {\bibfnamefont {M.~H.}\ \bibnamefont
  {Zaheer}}, \bibinfo {author} {\bibfnamefont {N.~J.}\ \bibnamefont {Matjelo}},
  \bibinfo {author} {\bibfnamefont {D.~B.}\ \bibnamefont {Hume}}, \bibinfo
  {author} {\bibfnamefont {M.~S.}\ \bibnamefont {Safronova}},\ and\ \bibinfo
  {author} {\bibfnamefont {D.~R.}\ \bibnamefont {Leibrandt}},\ }\bibfield
  {title} {\bibinfo {title} {Quantum metrology algorithms for dark matter
  searches with clocks},\ }\href {https://doi.org/10.1103/PhysRevA.111.012601}
  {\bibfield  {journal} {\bibinfo  {journal} {Phys. Rev. A}\ }\textbf {\bibinfo
  {volume} {111}},\ \bibinfo {pages} {012601} (\bibinfo {year}
  {2025})}\BibitemShut {NoStop}%
\bibitem [{\citenamefont {Dick}\ \emph {et~al.}(1990)\citenamefont {Dick},
  \citenamefont {Prestage}, \citenamefont {Greenhall},\ and\ \citenamefont
  {Maleki}}]{Dick1990}%
  \BibitemOpen
  \bibfield  {author} {\bibinfo {author} {\bibfnamefont {G.~J.}\ \bibnamefont
  {Dick}}, \bibinfo {author} {\bibfnamefont {J.~D.}\ \bibnamefont {Prestage}},
  \bibinfo {author} {\bibfnamefont {C.~A.}\ \bibnamefont {Greenhall}},\ and\
  \bibinfo {author} {\bibfnamefont {L.}~\bibnamefont {Maleki}},\ }\bibfield
  {title} {\bibinfo {title} {Local oscillator induced degradation of
  medium-term stability in passive atomic frequency standards},\ }in\
  \href@noop {} {\emph {\bibinfo {booktitle} {Proceedings of the 22nd Annual
  Precise Time and Time Interval (PTTI) Applications and Planning Meeting}}}\
  (\bibinfo {year} {1990})\ pp.\ \bibinfo {pages} {487--508}\BibitemShut
  {NoStop}%
\bibitem [{\citenamefont {Santarelli}\ \emph {et~al.}(1998)\citenamefont
  {Santarelli}, \citenamefont {Audoin}, \citenamefont {Makdissi}, \citenamefont
  {Laurent}, \citenamefont {Dick},\ and\ \citenamefont
  {Clairon}}]{Santarelli1998}%
  \BibitemOpen
  \bibfield  {author} {\bibinfo {author} {\bibfnamefont {G.}~\bibnamefont
  {Santarelli}}, \bibinfo {author} {\bibfnamefont {C.}~\bibnamefont {Audoin}},
  \bibinfo {author} {\bibfnamefont {A.}~\bibnamefont {Makdissi}}, \bibinfo
  {author} {\bibfnamefont {P.}~\bibnamefont {Laurent}}, \bibinfo {author}
  {\bibfnamefont {G.}~\bibnamefont {Dick}},\ and\ \bibinfo {author}
  {\bibfnamefont {A.}~\bibnamefont {Clairon}},\ }\bibfield  {title} {\bibinfo
  {title} {Frequency stability degradation of an oscillator slaved to a
  periodically interrogated atomic resonator},\ }\href
  {https://doi.org/10.1109/58.710548} {\bibfield  {journal} {\bibinfo
  {journal} {IEEE Trans. Ultrason. Ferroelectr. Freq. Control}\ }\textbf
  {\bibinfo {volume} {45}},\ \bibinfo {pages} {887} (\bibinfo {year}
  {1998})}\BibitemShut {NoStop}%
\bibitem [{\citenamefont {Westergaard}\ \emph {et~al.}(2010)\citenamefont
  {Westergaard}, \citenamefont {Lodewyck},\ and\ \citenamefont
  {Lemonde}}]{Westergaard2010}%
  \BibitemOpen
  \bibfield  {author} {\bibinfo {author} {\bibfnamefont {P.~G.}\ \bibnamefont
  {Westergaard}}, \bibinfo {author} {\bibfnamefont {J.}~\bibnamefont
  {Lodewyck}},\ and\ \bibinfo {author} {\bibfnamefont {P.}~\bibnamefont
  {Lemonde}},\ }\bibfield  {title} {\bibinfo {title} {Minimizing the {D}ick
  effect in an optical lattice clock},\ }\href
  {https://doi.org/10.1109/TUFFC.2010.1457} {\bibfield  {journal} {\bibinfo
  {journal} {IEEE Trans. Ultrason. Ferroelectr. Freq. Control}\ }\textbf
  {\bibinfo {volume} {57}},\ \bibinfo {pages} {623} (\bibinfo {year}
  {2010})}\BibitemShut {NoStop}%
\bibitem [{\citenamefont {Hume}\ and\ \citenamefont
  {Leibrandt}(2016)}]{Hume2016}%
  \BibitemOpen
  \bibfield  {author} {\bibinfo {author} {\bibfnamefont {D.~B.}\ \bibnamefont
  {Hume}}\ and\ \bibinfo {author} {\bibfnamefont {D.~R.}\ \bibnamefont
  {Leibrandt}},\ }\bibfield  {title} {\bibinfo {title} {Probing beyond the
  laser coherence time in optical clock comparisons},\ }\href
  {https://doi.org/10.1103/PhysRevA.93.032138} {\bibfield  {journal} {\bibinfo
  {journal} {Phys. Rev. A}\ }\textbf {\bibinfo {volume} {93}},\ \bibinfo
  {pages} {032138} (\bibinfo {year} {2016})}\BibitemShut {NoStop}%
\bibitem [{\citenamefont {Schioppo}\ \emph {et~al.}(2017)\citenamefont
  {Schioppo}, \citenamefont {Brown}, \citenamefont {McGrew}, \citenamefont
  {Hinkley}, \citenamefont {Fasano}, \citenamefont {Beloy}, \citenamefont
  {Yoon}, \citenamefont {Milani}, \citenamefont {Nicolodi}, \citenamefont
  {Sherman}, \citenamefont {Phillips}, \citenamefont {Oates},\ and\
  \citenamefont {Ludlow}}]{Schioppo2017}%
  \BibitemOpen
  \bibfield  {author} {\bibinfo {author} {\bibfnamefont {M.}~\bibnamefont
  {Schioppo}}, \bibinfo {author} {\bibfnamefont {R.~C.}\ \bibnamefont {Brown}},
  \bibinfo {author} {\bibfnamefont {W.~F.}\ \bibnamefont {McGrew}}, \bibinfo
  {author} {\bibfnamefont {N.}~\bibnamefont {Hinkley}}, \bibinfo {author}
  {\bibfnamefont {R.~J.}\ \bibnamefont {Fasano}}, \bibinfo {author}
  {\bibfnamefont {K.}~\bibnamefont {Beloy}}, \bibinfo {author} {\bibfnamefont
  {T.~H.}\ \bibnamefont {Yoon}}, \bibinfo {author} {\bibfnamefont
  {G.}~\bibnamefont {Milani}}, \bibinfo {author} {\bibfnamefont
  {D.}~\bibnamefont {Nicolodi}}, \bibinfo {author} {\bibfnamefont {J.~A.}\
  \bibnamefont {Sherman}}, \bibinfo {author} {\bibfnamefont {N.~B.}\
  \bibnamefont {Phillips}}, \bibinfo {author} {\bibfnamefont {C.~W.}\
  \bibnamefont {Oates}},\ and\ \bibinfo {author} {\bibfnamefont {A.~D.}\
  \bibnamefont {Ludlow}},\ }\bibfield  {title} {\bibinfo {title} {Ultrastable
  optical clock with two cold-atom ensembles},\ }\href
  {https://doi.org/10.1038/nphoton.2016.231} {\bibfield  {journal} {\bibinfo
  {journal} {Nat. Photonics}\ }\textbf {\bibinfo {volume} {11}},\ \bibinfo
  {pages} {48} (\bibinfo {year} {2017})}\BibitemShut {NoStop}%
\bibitem [{\citenamefont {Schulte}\ \emph {et~al.}(2020)\citenamefont
  {Schulte}, \citenamefont {Lisdat}, \citenamefont {Schmidt}, \citenamefont
  {Sterr},\ and\ \citenamefont {Hammerer}}]{Schulte2020}%
  \BibitemOpen
  \bibfield  {author} {\bibinfo {author} {\bibfnamefont {M.}~\bibnamefont
  {Schulte}}, \bibinfo {author} {\bibfnamefont {C.}~\bibnamefont {Lisdat}},
  \bibinfo {author} {\bibfnamefont {P.~O.}\ \bibnamefont {Schmidt}}, \bibinfo
  {author} {\bibfnamefont {U.}~\bibnamefont {Sterr}},\ and\ \bibinfo {author}
  {\bibfnamefont {K.}~\bibnamefont {Hammerer}},\ }\bibfield  {title} {\bibinfo
  {title} {Prospects and challenges for squeezing-enhanced optical atomic
  clocks},\ }\href {https://doi.org/10.1038/s41467-020-19403-7} {\bibfield
  {journal} {\bibinfo  {journal} {Nat. Commun.}\ }\textbf {\bibinfo {volume}
  {11}},\ \bibinfo {pages} {5955} (\bibinfo {year} {2020})}\BibitemShut
  {NoStop}%
\bibitem [{\citenamefont {Berkeland}\ \emph {et~al.}(1998)\citenamefont
  {Berkeland}, \citenamefont {Miller}, \citenamefont {Bergquist}, \citenamefont
  {Itano},\ and\ \citenamefont {Wineland}}]{Berkeland1998}%
  \BibitemOpen
  \bibfield  {author} {\bibinfo {author} {\bibfnamefont {D.~J.}\ \bibnamefont
  {Berkeland}}, \bibinfo {author} {\bibfnamefont {J.~D.}\ \bibnamefont
  {Miller}}, \bibinfo {author} {\bibfnamefont {J.~C.}\ \bibnamefont
  {Bergquist}}, \bibinfo {author} {\bibfnamefont {W.~M.}\ \bibnamefont
  {Itano}},\ and\ \bibinfo {author} {\bibfnamefont {D.~J.}\ \bibnamefont
  {Wineland}},\ }\bibfield  {title} {\bibinfo {title} {Minimization of ion
  micromotion in a {P}aul trap},\ }\href {https://doi.org/10.1063/1.367318}
  {\bibfield  {journal} {\bibinfo  {journal} {J. Appl. Phys.}\ }\textbf
  {\bibinfo {volume} {83}},\ \bibinfo {pages} {5025} (\bibinfo {year}
  {1998})}\BibitemShut {NoStop}%
\bibitem [{\citenamefont {Keller}\ \emph {et~al.}(2015)\citenamefont {Keller},
  \citenamefont {Partner}, \citenamefont {Burgermeister},\ and\ \citenamefont
  {Mehlst\"aubler}}]{Keller2015}%
  \BibitemOpen
  \bibfield  {author} {\bibinfo {author} {\bibfnamefont {J.}~\bibnamefont
  {Keller}}, \bibinfo {author} {\bibfnamefont {H.~L.}\ \bibnamefont {Partner}},
  \bibinfo {author} {\bibfnamefont {T.}~\bibnamefont {Burgermeister}},\ and\
  \bibinfo {author} {\bibfnamefont {T.~E.}\ \bibnamefont {Mehlst\"aubler}},\
  }\bibfield  {title} {\bibinfo {title} {Precise determination of micromotion
  for trapped-ion optical clocks},\ }\href {https://doi.org/10.1063/1.4930037}
  {\bibfield  {journal} {\bibinfo  {journal} {J. Appl. Phys.}\ }\textbf
  {\bibinfo {volume} {118}},\ \bibinfo {pages} {104501} (\bibinfo {year}
  {2015})}\BibitemShut {NoStop}%
\bibitem [{\citenamefont {Higgins}\ \emph {et~al.}(2021)\citenamefont
  {Higgins}, \citenamefont {Salim}, \citenamefont {Zhang}, \citenamefont
  {Parke}, \citenamefont {Pokorny},\ and\ \citenamefont
  {Hennrich}}]{Higgins2021}%
  \BibitemOpen
  \bibfield  {author} {\bibinfo {author} {\bibfnamefont {G.}~\bibnamefont
  {Higgins}}, \bibinfo {author} {\bibfnamefont {S.}~\bibnamefont {Salim}},
  \bibinfo {author} {\bibfnamefont {C.}~\bibnamefont {Zhang}}, \bibinfo
  {author} {\bibfnamefont {H.}~\bibnamefont {Parke}}, \bibinfo {author}
  {\bibfnamefont {F.}~\bibnamefont {Pokorny}},\ and\ \bibinfo {author}
  {\bibfnamefont {M.}~\bibnamefont {Hennrich}},\ }\bibfield  {title} {\bibinfo
  {title} {Micromotion minimization using {R}amsey interferometry},\ }\href
  {https://doi.org/10.1088/1367-2630/ac3db6} {\bibfield  {journal} {\bibinfo
  {journal} {New J. Phys.}\ }\textbf {\bibinfo {volume} {23}},\ \bibinfo
  {pages} {123028} (\bibinfo {year} {2021})}\BibitemShut {NoStop}%
\bibitem [{\citenamefont {Lindvall}\ \emph {et~al.}(2025)\citenamefont
  {Lindvall}, \citenamefont {Fordell}, \citenamefont {Hanhij\"arvi},
  \citenamefont {Dole\ifmmode~\check{z}\else \v{z}\fi{}al}, \citenamefont
  {Rahm}, \citenamefont {Weyers},\ and\ \citenamefont
  {Wallin}}]{Lindvall2025b}%
  \BibitemOpen
  \bibfield  {author} {\bibinfo {author} {\bibfnamefont {T.}~\bibnamefont
  {Lindvall}}, \bibinfo {author} {\bibfnamefont {T.}~\bibnamefont {Fordell}},
  \bibinfo {author} {\bibfnamefont {K.}~\bibnamefont {Hanhij\"arvi}}, \bibinfo
  {author} {\bibfnamefont {M.}~\bibnamefont {Dole\ifmmode~\check{z}\else
  \v{z}\fi{}al}}, \bibinfo {author} {\bibfnamefont {J.}~\bibnamefont {Rahm}},
  \bibinfo {author} {\bibfnamefont {S.}~\bibnamefont {Weyers}},\ and\ \bibinfo
  {author} {\bibfnamefont {A.}~\bibnamefont {Wallin}},\ }\bibfield  {title}
  {\bibinfo {title} {${}^{88}${Sr}$^{+}$ optical clock with
  $7.9\ifmmode\times\else\texttimes\fi{}{10}^{\ensuremath{-}19}$ systematic
  uncertainty and measurement of its absolute frequency with
  $9.8\ifmmode\times\else\texttimes\fi{}{10}^{\ensuremath{-}17}$ uncertainty},\
  }\href {https://doi.org/10.1103/czlf-bfvp} {\bibfield  {journal} {\bibinfo
  {journal} {Phys. Rev. Appl.}\ }\textbf {\bibinfo {volume} {24}},\ \bibinfo
  {pages} {044082} (\bibinfo {year} {2025})}\BibitemShut {NoStop}%
\bibitem [{\citenamefont {Arnold}\ \emph {et~al.}(2026)\citenamefont {Arnold},
  \citenamefont {Lee}, \citenamefont {Zhao}, \citenamefont {Qin}, \citenamefont
  {Zhang}, \citenamefont {Jayjong},\ and\ \citenamefont
  {Barrett}}]{Arnold2026}%
  \BibitemOpen
  \bibfield  {author} {\bibinfo {author} {\bibfnamefont {K.~J.}\ \bibnamefont
  {Arnold}}, \bibinfo {author} {\bibfnamefont {M.~D.~K.}\ \bibnamefont {Lee}},
  \bibinfo {author} {\bibfnamefont {Q.}~\bibnamefont {Zhao}}, \bibinfo {author}
  {\bibfnamefont {Q.}~\bibnamefont {Qin}}, \bibinfo {author} {\bibfnamefont
  {Z.}~\bibnamefont {Zhang}}, \bibinfo {author} {\bibfnamefont
  {N.}~\bibnamefont {Jayjong}},\ and\ \bibinfo {author} {\bibfnamefont {M.~D.}\
  \bibnamefont {Barrett}},\ }\bibfield  {title} {\bibinfo {title} {Lu$^{+}$
  optical frequency references with accuracy verified at the 19th digit},\
  }\href {https://doi.org/10.1038/s41586-026-11072-8} {\bibfield  {journal}
  {\bibinfo  {journal} {Nature}\ }\textbf {\bibinfo {volume} {658}},\ \bibinfo
  {pages} {60} (\bibinfo {year} {2026})}\BibitemShut {NoStop}%
\bibitem [{\citenamefont {Lizuain}\ \emph {et~al.}(2007)\citenamefont
  {Lizuain}, \citenamefont {Muga},\ and\ \citenamefont
  {Eschner}}]{Lizuain2007}%
  \BibitemOpen
  \bibfield  {author} {\bibinfo {author} {\bibfnamefont {I.}~\bibnamefont
  {Lizuain}}, \bibinfo {author} {\bibfnamefont {J.~G.}\ \bibnamefont {Muga}},\
  and\ \bibinfo {author} {\bibfnamefont {J.}~\bibnamefont {Eschner}},\
  }\bibfield  {title} {\bibinfo {title} {Motional frequency shifts of trapped
  ions in the {L}amb-{D}icke regime},\ }\href
  {https://doi.org/10.1103/PhysRevA.76.033808} {\bibfield  {journal} {\bibinfo
  {journal} {Phys. Rev. A}\ }\textbf {\bibinfo {volume} {76}},\ \bibinfo
  {pages} {033808} (\bibinfo {year} {2007})}\BibitemShut {NoStop}%
\bibitem [{\citenamefont {Itano}(2000)}]{Itano2000}%
  \BibitemOpen
  \bibfield  {author} {\bibinfo {author} {\bibfnamefont {W.~M.}\ \bibnamefont
  {Itano}},\ }\bibfield  {title} {\bibinfo {title} {External-field shifts of
  the $^{199}${H}g$^+$ optical frequency standard},\ }\href
  {https://doi.org/10.6028/jres.105.065} {\bibfield  {journal} {\bibinfo
  {journal} {J. Res. Natl. Inst. Stand. Technol.}\ }\textbf {\bibinfo {volume}
  {105}},\ \bibinfo {pages} {829} (\bibinfo {year} {2000})}\BibitemShut
  {NoStop}%
\bibitem [{\citenamefont {Dub{\'e}}\ \emph {et~al.}(2013)\citenamefont
  {Dub{\'e}}, \citenamefont {Madej}, \citenamefont {Zhou},\ and\ \citenamefont
  {Bernard}}]{Dube2013}%
  \BibitemOpen
  \bibfield  {author} {\bibinfo {author} {\bibfnamefont {P.}~\bibnamefont
  {Dub{\'e}}}, \bibinfo {author} {\bibfnamefont {A.~A.}\ \bibnamefont {Madej}},
  \bibinfo {author} {\bibfnamefont {Z.}~\bibnamefont {Zhou}},\ and\ \bibinfo
  {author} {\bibfnamefont {J.~E.}\ \bibnamefont {Bernard}},\ }\bibfield
  {title} {\bibinfo {title} {Evaluation of systematic shifts of the
  ${}^{88}${S}r${}^{+}$ single-ion optical frequency standard at the
  ${10}^{\ensuremath{-}17}$ level},\ }\href
  {https://doi.org/10.1103/PhysRevA.87.023806} {\bibfield  {journal} {\bibinfo
  {journal} {Phys. Rev. A}\ }\textbf {\bibinfo {volume} {87}},\ \bibinfo
  {pages} {023806} (\bibinfo {year} {2013})}\BibitemShut {NoStop}%
\bibitem [{\citenamefont {Tofful}\ \emph {et~al.}(2024)\citenamefont {Tofful},
  \citenamefont {Baynham}, \citenamefont {Curtis}, \citenamefont {Parsons},
  \citenamefont {Robertson}, \citenamefont {Schioppo}, \citenamefont {Tunesi},
  \citenamefont {Margolis}, \citenamefont {Hendricks}, \citenamefont {Whale},
  \citenamefont {Thompson},\ and\ \citenamefont {Godun}}]{Tofful2024}%
  \BibitemOpen
  \bibfield  {author} {\bibinfo {author} {\bibfnamefont {A.}~\bibnamefont
  {Tofful}}, \bibinfo {author} {\bibfnamefont {C.~F.~A.}\ \bibnamefont
  {Baynham}}, \bibinfo {author} {\bibfnamefont {E.~A.}\ \bibnamefont {Curtis}},
  \bibinfo {author} {\bibfnamefont {A.~O.}\ \bibnamefont {Parsons}}, \bibinfo
  {author} {\bibfnamefont {B.~I.}\ \bibnamefont {Robertson}}, \bibinfo {author}
  {\bibfnamefont {M.}~\bibnamefont {Schioppo}}, \bibinfo {author}
  {\bibfnamefont {J.}~\bibnamefont {Tunesi}}, \bibinfo {author} {\bibfnamefont
  {H.~S.}\ \bibnamefont {Margolis}}, \bibinfo {author} {\bibfnamefont {R.~J.}\
  \bibnamefont {Hendricks}}, \bibinfo {author} {\bibfnamefont {J.}~\bibnamefont
  {Whale}}, \bibinfo {author} {\bibfnamefont {R.~C.}\ \bibnamefont
  {Thompson}},\ and\ \bibinfo {author} {\bibfnamefont {R.~M.}\ \bibnamefont
  {Godun}},\ }\bibfield  {title} {\bibinfo {title} {$^{171}${Y}b$^+$ optical
  clock with 2.2$\times$ $10^{-18}$ systematic uncertainty and absolute
  frequency measurements},\ }\href {https://doi.org/10.1088/1681-7575/ad53cd}
  {\bibfield  {journal} {\bibinfo  {journal} {Metrologia}\ }\textbf {\bibinfo
  {volume} {61}},\ \bibinfo {pages} {045001} (\bibinfo {year}
  {2024})}\BibitemShut {NoStop}%
\bibitem [{\citenamefont {Yudin}\ and\ \citenamefont
  {Taichenachev}(2018)}]{Yudin2018}%
  \BibitemOpen
  \bibfield  {author} {\bibinfo {author} {\bibfnamefont {V.}~\bibnamefont
  {Yudin}}\ and\ \bibinfo {author} {\bibfnamefont {A.}~\bibnamefont
  {Taichenachev}},\ }\bibfield  {title} {\bibinfo {title} {Mass defect effects
  in atomic clocks},\ }\href {https://doi.org/10.1088/1612-202X/aa9aa5}
  {\bibfield  {journal} {\bibinfo  {journal} {Laser Phys. Lett.}\ }\textbf
  {\bibinfo {volume} {15}},\ \bibinfo {pages} {035703} (\bibinfo {year}
  {2018})}\BibitemShut {NoStop}%
\bibitem [{\citenamefont {Sonnleitner}\ and\ \citenamefont
  {Barnett}(2018)}]{Sonnleitner2018}%
  \BibitemOpen
  \bibfield  {author} {\bibinfo {author} {\bibfnamefont {M.}~\bibnamefont
  {Sonnleitner}}\ and\ \bibinfo {author} {\bibfnamefont {S.~M.}\ \bibnamefont
  {Barnett}},\ }\bibfield  {title} {\bibinfo {title} {Mass-energy and anomalous
  friction in quantum optics},\ }\href
  {https://doi.org/10.1103/PhysRevA.98.042106} {\bibfield  {journal} {\bibinfo
  {journal} {Phys. Rev. A}\ }\textbf {\bibinfo {volume} {98}},\ \bibinfo
  {pages} {042106} (\bibinfo {year} {2018})}\BibitemShut {NoStop}%
\bibitem [{\citenamefont {Asano}\ \emph {et~al.}(2024)\citenamefont {Asano},
  \citenamefont {Giese},\ and\ \citenamefont {Di~Pumpo}}]{Asano2024}%
  \BibitemOpen
  \bibfield  {author} {\bibinfo {author} {\bibfnamefont {T.}~\bibnamefont
  {Asano}}, \bibinfo {author} {\bibfnamefont {E.}~\bibnamefont {Giese}},\ and\
  \bibinfo {author} {\bibfnamefont {F.}~\bibnamefont {Di~Pumpo}},\ }\bibfield
  {title} {\bibinfo {title} {Quantum field theory for multipolar composite
  bosons with mass defect and relativistic corrections},\ }\href
  {https://doi.org/10.1103/PRXQuantum.5.020322} {\bibfield  {journal} {\bibinfo
   {journal} {PRX Quantum}\ }\textbf {\bibinfo {volume} {5}},\ \bibinfo {pages}
  {020322} (\bibinfo {year} {2024})}\BibitemShut {NoStop}%
\bibitem [{\citenamefont {Miao}\ \emph {et~al.}(2022)\citenamefont {Miao},
  \citenamefont {Zhang}, \citenamefont {Zheng}, \citenamefont {Qin},
  \citenamefont {Xin}, \citenamefont {Chen}, \citenamefont {Han},\ and\
  \citenamefont {Wang}}]{Miao2022}%
  \BibitemOpen
  \bibfield  {author} {\bibinfo {author} {\bibfnamefont {S.~N.}\ \bibnamefont
  {Miao}}, \bibinfo {author} {\bibfnamefont {J.~W.}\ \bibnamefont {Zhang}},
  \bibinfo {author} {\bibfnamefont {Y.}~\bibnamefont {Zheng}}, \bibinfo
  {author} {\bibfnamefont {H.~R.}\ \bibnamefont {Qin}}, \bibinfo {author}
  {\bibfnamefont {N.~C.}\ \bibnamefont {Xin}}, \bibinfo {author} {\bibfnamefont
  {Y.~T.}\ \bibnamefont {Chen}}, \bibinfo {author} {\bibfnamefont {J.~Z.}\
  \bibnamefont {Han}},\ and\ \bibinfo {author} {\bibfnamefont {L.~J.}\
  \bibnamefont {Wang}},\ }\bibfield  {title} {\bibinfo {title} {Second-order
  {D}oppler frequency shifts of trapped ions in a linear {P}aul trap},\ }\href
  {https://doi.org/10.1103/PhysRevA.106.033121} {\bibfield  {journal} {\bibinfo
   {journal} {Phys. Rev. A}\ }\textbf {\bibinfo {volume} {106}},\ \bibinfo
  {pages} {033121} (\bibinfo {year} {2022})}\BibitemShut {NoStop}%
\bibitem [{\citenamefont {Ufrecht}\ and\ \citenamefont
  {Giese}(2020)}]{Ufrecht20202}%
  \BibitemOpen
  \bibfield  {author} {\bibinfo {author} {\bibfnamefont {C.}~\bibnamefont
  {Ufrecht}}\ and\ \bibinfo {author} {\bibfnamefont {E.}~\bibnamefont
  {Giese}},\ }\bibfield  {title} {\bibinfo {title} {Perturbative operator
  approach to high-precision light-pulse atom interferometry},\ }\href
  {https://doi.org/10.1103/PhysRevA.101.053615} {\bibfield  {journal} {\bibinfo
   {journal} {Phys. Rev. A}\ }\textbf {\bibinfo {volume} {101}},\ \bibinfo
  {pages} {053615} (\bibinfo {year} {2020})}\BibitemShut {NoStop}%
\bibitem [{\citenamefont {Ufrecht}(2019)}]{Ufrecht2019}%
  \BibitemOpen
  \bibfield  {author} {\bibinfo {author} {\bibfnamefont {C.}~\bibnamefont
  {Ufrecht}},\ }\bibfield  {title} {\bibinfo {title} {Theoretical approach to
  high-precision atom interferometry},\ }\href
  {https://doi.org/10.18725/OPARU-17923} {\bibfield  {journal} {\bibinfo
  {journal} {Ph.D. thesis, Universit{\"a}t Ulm}\ } (\bibinfo {year}
  {2019})}\BibitemShut {NoStop}%
\bibitem [{\citenamefont {Viola}\ and\ \citenamefont
  {Lloyd}(1998)}]{Viola1998}%
  \BibitemOpen
  \bibfield  {author} {\bibinfo {author} {\bibfnamefont {L.}~\bibnamefont
  {Viola}}\ and\ \bibinfo {author} {\bibfnamefont {S.}~\bibnamefont {Lloyd}},\
  }\bibfield  {title} {\bibinfo {title} {Dynamical suppression of decoherence
  in two-state quantum systems},\ }\href
  {https://doi.org/10.1103/PhysRevA.58.2733} {\bibfield  {journal} {\bibinfo
  {journal} {Phys. Rev. A}\ }\textbf {\bibinfo {volume} {58}},\ \bibinfo
  {pages} {2733} (\bibinfo {year} {1998})}\BibitemShut {NoStop}%
\bibitem [{\citenamefont {Hahn}(1950)}]{Hahn1950}%
  \BibitemOpen
  \bibfield  {author} {\bibinfo {author} {\bibfnamefont {E.~L.}\ \bibnamefont
  {Hahn}},\ }\bibfield  {title} {\bibinfo {title} {Spin echoes},\ }\href
  {https://doi.org/10.1103/PhysRev.80.580} {\bibfield  {journal} {\bibinfo
  {journal} {Phys. Rev.}\ }\textbf {\bibinfo {volume} {80}},\ \bibinfo {pages}
  {580} (\bibinfo {year} {1950})}\BibitemShut {NoStop}%
\bibitem [{\citenamefont {Di~Pumpo}\ \emph {et~al.}(2024)\citenamefont
  {Di~Pumpo}, \citenamefont {Friedrich},\ and\ \citenamefont
  {Giese}}]{DiPumpo2023b}%
  \BibitemOpen
  \bibfield  {author} {\bibinfo {author} {\bibfnamefont {F.}~\bibnamefont
  {Di~Pumpo}}, \bibinfo {author} {\bibfnamefont {A.}~\bibnamefont
  {Friedrich}},\ and\ \bibinfo {author} {\bibfnamefont {E.}~\bibnamefont
  {Giese}},\ }\bibfield  {title} {\bibinfo {title} {Optimal baseline
  exploitation in vertical dark-matter detectors based on atom
  interferometry},\ }\href {https://doi.org/10.1116/5.0175683} {\bibfield
  {journal} {\bibinfo  {journal} {AVS Quantum Sci.}\ }\textbf {\bibinfo
  {volume} {6}},\ \bibinfo {pages} {014404} (\bibinfo {year}
  {2024})}\BibitemShut {NoStop}%
\bibitem [{\citenamefont {Guevara-Bertsch}(2022)}]{Guevara-Bertsch2022}%
  \BibitemOpen
  \bibfield  {author} {\bibinfo {author} {\bibfnamefont {M.}~\bibnamefont
  {Guevara-Bertsch}},\ }\emph {\bibinfo {title} {Optical clocks with trapped
  ions}},\ \href@noop {} {Ph.D. thesis},\ \bibinfo  {school} {Leopold-Franzens
  University of Innsbruck} (\bibinfo {year} {2022})\BibitemShut {NoStop}%
\bibitem [{\citenamefont {Haslinger}\ \emph {et~al.}(2018)\citenamefont
  {Haslinger}, \citenamefont {Jaffe}, \citenamefont {Xu}, \citenamefont
  {Schwartz}, \citenamefont {Sonnleitner}, \citenamefont {Ritsch-Marte},
  \citenamefont {Ritsch},\ and\ \citenamefont {M{\"u}ller}}]{Haslinger2018}%
  \BibitemOpen
  \bibfield  {author} {\bibinfo {author} {\bibfnamefont {P.}~\bibnamefont
  {Haslinger}}, \bibinfo {author} {\bibfnamefont {M.}~\bibnamefont {Jaffe}},
  \bibinfo {author} {\bibfnamefont {V.}~\bibnamefont {Xu}}, \bibinfo {author}
  {\bibfnamefont {O.}~\bibnamefont {Schwartz}}, \bibinfo {author}
  {\bibfnamefont {M.}~\bibnamefont {Sonnleitner}}, \bibinfo {author}
  {\bibfnamefont {M.}~\bibnamefont {Ritsch-Marte}}, \bibinfo {author}
  {\bibfnamefont {H.}~\bibnamefont {Ritsch}},\ and\ \bibinfo {author}
  {\bibfnamefont {H.}~\bibnamefont {M{\"u}ller}},\ }\bibfield  {title}
  {\bibinfo {title} {Attractive force on atoms due to blackbody radiation},\
  }\href {https://doi.org/10.1038/s41567-017-0004-9} {\bibfield  {journal}
  {\bibinfo  {journal} {Nat. Phys.}\ }\textbf {\bibinfo {volume} {14}},\
  \bibinfo {pages} {257} (\bibinfo {year} {2018})}\BibitemShut {NoStop}%
\bibitem [{\citenamefont {Mintert}\ and\ \citenamefont
  {Wunderlich}(2001)}]{Mintert2001}%
  \BibitemOpen
  \bibfield  {author} {\bibinfo {author} {\bibfnamefont {F.}~\bibnamefont
  {Mintert}}\ and\ \bibinfo {author} {\bibfnamefont {C.}~\bibnamefont
  {Wunderlich}},\ }\bibfield  {title} {\bibinfo {title} {Ion-trap quantum logic
  using long-wavelength radiation},\ }\href
  {https://doi.org/10.1103/PhysRevLett.87.257904} {\bibfield  {journal}
  {\bibinfo  {journal} {Phys. Rev. Lett.}\ }\textbf {\bibinfo {volume} {87}},\
  \bibinfo {pages} {257904} (\bibinfo {year} {2001})}\BibitemShut {NoStop}%
\bibitem [{\citenamefont {Ivanov}(2017)}]{Ivanov2017}%
  \BibitemOpen
  \bibfield  {author} {\bibinfo {author} {\bibfnamefont {P.~A.}\ \bibnamefont
  {Ivanov}},\ }\bibfield  {title} {\bibinfo {title} {Efficient approach for
  quantum sensing field gradients with trapped ions},\ }\href
  {https://doi.org/10.1016/j.optcom.2017.08.055} {\bibfield  {journal}
  {\bibinfo  {journal} {Opt. Commun.}\ }\textbf {\bibinfo {volume} {405}},\
  \bibinfo {pages} {355} (\bibinfo {year} {2017})}\BibitemShut {NoStop}%
\bibitem [{\citenamefont {Damour}\ and\ \citenamefont
  {Donoghue}(2010)}]{Damour2010}%
  \BibitemOpen
  \bibfield  {author} {\bibinfo {author} {\bibfnamefont {T.}~\bibnamefont
  {Damour}}\ and\ \bibinfo {author} {\bibfnamefont {J.~F.}\ \bibnamefont
  {Donoghue}},\ }\bibfield  {title} {\bibinfo {title} {Equivalence principle
  violations and couplings of a light dilaton},\ }\href
  {https://doi.org/10.1103/PhysRevD.82.084033} {\bibfield  {journal} {\bibinfo
  {journal} {Phys. Rev. D}\ }\textbf {\bibinfo {volume} {82}},\ \bibinfo
  {pages} {084033} (\bibinfo {year} {2010})}\BibitemShut {NoStop}%
\bibitem [{\citenamefont {Di~Pumpo}\ \emph {et~al.}(2021)\citenamefont
  {Di~Pumpo}, \citenamefont {Ufrecht}, \citenamefont {Friedrich}, \citenamefont
  {Giese}, \citenamefont {Schleich},\ and\ \citenamefont
  {Unruh}}]{DiPumpo2021}%
  \BibitemOpen
  \bibfield  {author} {\bibinfo {author} {\bibfnamefont {F.}~\bibnamefont
  {Di~Pumpo}}, \bibinfo {author} {\bibfnamefont {C.}~\bibnamefont {Ufrecht}},
  \bibinfo {author} {\bibfnamefont {A.}~\bibnamefont {Friedrich}}, \bibinfo
  {author} {\bibfnamefont {E.}~\bibnamefont {Giese}}, \bibinfo {author}
  {\bibfnamefont {W.~P.}\ \bibnamefont {Schleich}},\ and\ \bibinfo {author}
  {\bibfnamefont {W.~G.}\ \bibnamefont {Unruh}},\ }\bibfield  {title} {\bibinfo
  {title} {Gravitational redshift tests with atomic clocks and atom
  interferometers},\ }\href {https://doi.org/10.1103/PRXQuantum.2.040333}
  {\bibfield  {journal} {\bibinfo  {journal} {PRX Quantum}\ }\textbf {\bibinfo
  {volume} {2}},\ \bibinfo {pages} {040333} (\bibinfo {year}
  {2021})}\BibitemShut {NoStop}%
\bibitem [{\citenamefont {Di~Pumpo}\ \emph {et~al.}(2022)\citenamefont
  {Di~Pumpo}, \citenamefont {Friedrich}, \citenamefont {Geyer}, \citenamefont
  {Ufrecht},\ and\ \citenamefont {Giese}}]{DiPumpo2022}%
  \BibitemOpen
  \bibfield  {author} {\bibinfo {author} {\bibfnamefont {F.}~\bibnamefont
  {Di~Pumpo}}, \bibinfo {author} {\bibfnamefont {A.}~\bibnamefont {Friedrich}},
  \bibinfo {author} {\bibfnamefont {A.}~\bibnamefont {Geyer}}, \bibinfo
  {author} {\bibfnamefont {C.}~\bibnamefont {Ufrecht}},\ and\ \bibinfo {author}
  {\bibfnamefont {E.}~\bibnamefont {Giese}},\ }\bibfield  {title} {\bibinfo
  {title} {Light propagation and atom interferometry in gravity and dilaton
  fields},\ }\href {https://doi.org/10.1103/PhysRevD.105.084065} {\bibfield
  {journal} {\bibinfo  {journal} {Phys. Rev. D}\ }\textbf {\bibinfo {volume}
  {105}},\ \bibinfo {pages} {084065} (\bibinfo {year} {2022})}\BibitemShut
  {NoStop}%
\bibitem [{\citenamefont {Higgins}\ \emph {et~al.}(2017)\citenamefont
  {Higgins}, \citenamefont {Li}, \citenamefont {Pokorny}, \citenamefont
  {Zhang}, \citenamefont {Kress}, \citenamefont {Maier}, \citenamefont {Haag},
  \citenamefont {Bodart}, \citenamefont {Lesanovsky},\ and\ \citenamefont
  {Hennrich}}]{Higgins2017}%
  \BibitemOpen
  \bibfield  {author} {\bibinfo {author} {\bibfnamefont {G.}~\bibnamefont
  {Higgins}}, \bibinfo {author} {\bibfnamefont {W.}~\bibnamefont {Li}},
  \bibinfo {author} {\bibfnamefont {F.}~\bibnamefont {Pokorny}}, \bibinfo
  {author} {\bibfnamefont {C.}~\bibnamefont {Zhang}}, \bibinfo {author}
  {\bibfnamefont {F.}~\bibnamefont {Kress}}, \bibinfo {author} {\bibfnamefont
  {C.}~\bibnamefont {Maier}}, \bibinfo {author} {\bibfnamefont
  {J.}~\bibnamefont {Haag}}, \bibinfo {author} {\bibfnamefont {Q.}~\bibnamefont
  {Bodart}}, \bibinfo {author} {\bibfnamefont {I.}~\bibnamefont {Lesanovsky}},\
  and\ \bibinfo {author} {\bibfnamefont {M.}~\bibnamefont {Hennrich}},\
  }\bibfield  {title} {\bibinfo {title} {Single strontium {R}ydberg ion
  confined in a {P}aul trap},\ }\href
  {https://doi.org/10.1103/PhysRevX.7.021038} {\bibfield  {journal} {\bibinfo
  {journal} {Phys. Rev. X}\ }\textbf {\bibinfo {volume} {7}},\ \bibinfo {pages}
  {021038} (\bibinfo {year} {2017})}\BibitemShut {NoStop}%
\bibitem [{\citenamefont {Higgins}\ \emph {et~al.}(2019)\citenamefont
  {Higgins}, \citenamefont {Pokorny}, \citenamefont {Zhang},\ and\
  \citenamefont {Hennrich}}]{Higgins2019}%
  \BibitemOpen
  \bibfield  {author} {\bibinfo {author} {\bibfnamefont {G.}~\bibnamefont
  {Higgins}}, \bibinfo {author} {\bibfnamefont {F.}~\bibnamefont {Pokorny}},
  \bibinfo {author} {\bibfnamefont {C.}~\bibnamefont {Zhang}},\ and\ \bibinfo
  {author} {\bibfnamefont {M.}~\bibnamefont {Hennrich}},\ }\bibfield  {title}
  {\bibinfo {title} {Highly polarizable {R}ydberg ion in a {P}aul trap},\
  }\href {https://doi.org/10.1103/PhysRevLett.123.153602} {\bibfield  {journal}
  {\bibinfo  {journal} {Phys. Rev. Lett.}\ }\textbf {\bibinfo {volume} {123}},\
  \bibinfo {pages} {153602} (\bibinfo {year} {2019})}\BibitemShut {NoStop}%
\bibitem [{\citenamefont {Lemonde}(2009)}]{Lemonde2009}%
  \BibitemOpen
  \bibfield  {author} {\bibinfo {author} {\bibfnamefont {P.}~\bibnamefont
  {Lemonde}},\ }\bibfield  {title} {\bibinfo {title} {Optical lattice clocks},\
  }\href {https://doi.org/10.1140/epjst/e2009-01043-5} {\bibfield  {journal}
  {\bibinfo  {journal} {Eur. Phys. J.: Spec. Top.}\ }\textbf {\bibinfo {volume}
  {172}},\ \bibinfo {pages} {81} (\bibinfo {year} {2009})}\BibitemShut
  {NoStop}%
\bibitem [{\citenamefont {Derevianko}\ and\ \citenamefont
  {Katori}(2011)}]{Derevianko2011}%
  \BibitemOpen
  \bibfield  {author} {\bibinfo {author} {\bibfnamefont {A.}~\bibnamefont
  {Derevianko}}\ and\ \bibinfo {author} {\bibfnamefont {H.}~\bibnamefont
  {Katori}},\ }\bibfield  {title} {\bibinfo {title} {Colloquium: {P}hysics of
  optical lattice clocks},\ }\href {https://doi.org/10.1103/RevModPhys.83.331}
  {\bibfield  {journal} {\bibinfo  {journal} {Rev. Mod. Phys.}\ }\textbf
  {\bibinfo {volume} {83}},\ \bibinfo {pages} {331} (\bibinfo {year}
  {2011})}\BibitemShut {NoStop}%
\bibitem [{\citenamefont {Ahlheit}\ \emph {et~al.}(2025)\citenamefont
  {Ahlheit}, \citenamefont {Nill}, \citenamefont {Svirskiy}, \citenamefont
  {de~Haan}, \citenamefont {Schroers}, \citenamefont {Alt}, \citenamefont
  {Stiesdal}, \citenamefont {Lesanovsky},\ and\ \citenamefont
  {Hofferberth}}]{Ahlheit2025}%
  \BibitemOpen
  \bibfield  {author} {\bibinfo {author} {\bibfnamefont {L.}~\bibnamefont
  {Ahlheit}}, \bibinfo {author} {\bibfnamefont {C.}~\bibnamefont {Nill}},
  \bibinfo {author} {\bibfnamefont {D.}~\bibnamefont {Svirskiy}}, \bibinfo
  {author} {\bibfnamefont {J.}~\bibnamefont {de~Haan}}, \bibinfo {author}
  {\bibfnamefont {S.}~\bibnamefont {Schroers}}, \bibinfo {author}
  {\bibfnamefont {W.}~\bibnamefont {Alt}}, \bibinfo {author} {\bibfnamefont
  {N.}~\bibnamefont {Stiesdal}}, \bibinfo {author} {\bibfnamefont
  {I.}~\bibnamefont {Lesanovsky}},\ and\ \bibinfo {author} {\bibfnamefont
  {S.}~\bibnamefont {Hofferberth}},\ }\bibfield  {title} {\bibinfo {title}
  {Magic running- and standing-wave optical traps for {R}ydberg atoms},\ }\href
  {https://doi.org/10.1103/PhysRevA.111.013115} {\bibfield  {journal} {\bibinfo
   {journal} {Phys. Rev. A}\ }\textbf {\bibinfo {volume} {111}},\ \bibinfo
  {pages} {013115} (\bibinfo {year} {2025})}\BibitemShut {NoStop}%
\bibitem [{\citenamefont {Damour}\ and\ \citenamefont
  {Polyakov}(1994)}]{Damour1994}%
  \BibitemOpen
  \bibfield  {author} {\bibinfo {author} {\bibfnamefont {T.}~\bibnamefont
  {Damour}}\ and\ \bibinfo {author} {\bibfnamefont {A.}~\bibnamefont
  {Polyakov}},\ }\bibfield  {title} {\bibinfo {title} {The string dilation and
  a least coupling principle},\ }\href
  {https://doi.org/10.1016/0550-3213(94)90143-0} {\bibfield  {journal}
  {\bibinfo  {journal} {Nucl. Phys. B}\ }\textbf {\bibinfo {volume} {423}},\
  \bibinfo {pages} {532 } (\bibinfo {year} {1994})}\BibitemShut {NoStop}%
\bibitem [{\citenamefont {Alves}\ and\ \citenamefont
  {Bezerra}(2000)}]{Alves2000}%
  \BibitemOpen
  \bibfield  {author} {\bibinfo {author} {\bibfnamefont {M.}~\bibnamefont
  {Alves}}\ and\ \bibinfo {author} {\bibfnamefont {V.}~\bibnamefont
  {Bezerra}},\ }\bibfield  {title} {\bibinfo {title} {Dilaton gravity with a
  nonminimally coupled scalar field},\ }\href
  {https://doi.org/10.1142/S0218271800000815} {\bibfield  {journal} {\bibinfo
  {journal} {Int. J. Mod. Phys. D}\ }\textbf {\bibinfo {volume} {09}},\
  \bibinfo {pages} {697} (\bibinfo {year} {2000})}\BibitemShut {NoStop}%
\bibitem [{\citenamefont {Damour}(2012)}]{Damour2012}%
  \BibitemOpen
  \bibfield  {author} {\bibinfo {author} {\bibfnamefont {T.}~\bibnamefont
  {Damour}},\ }\bibfield  {title} {\bibinfo {title} {Theoretical aspects of the
  equivalence principle},\ }\href
  {https://doi.org/10.1088/0264-9381/29/18/184001} {\bibfield  {journal}
  {\bibinfo  {journal} {Class. Quantum Grav.}\ }\textbf {\bibinfo {volume}
  {29}},\ \bibinfo {pages} {184001} (\bibinfo {year} {2012})}\BibitemShut
  {NoStop}%
\bibitem [{\citenamefont {Hees}\ \emph {et~al.}(2018)\citenamefont {Hees},
  \citenamefont {Minazzoli}, \citenamefont {Savalle}, \citenamefont {Stadnik},\
  and\ \citenamefont {Wolf}}]{Hees2018}%
  \BibitemOpen
  \bibfield  {author} {\bibinfo {author} {\bibfnamefont {A.}~\bibnamefont
  {Hees}}, \bibinfo {author} {\bibfnamefont {O.}~\bibnamefont {Minazzoli}},
  \bibinfo {author} {\bibfnamefont {E.}~\bibnamefont {Savalle}}, \bibinfo
  {author} {\bibfnamefont {Y.~V.}\ \bibnamefont {Stadnik}},\ and\ \bibinfo
  {author} {\bibfnamefont {P.}~\bibnamefont {Wolf}},\ }\bibfield  {title}
  {\bibinfo {title} {Violation of the equivalence principle from light scalar
  dark matter},\ }\href {https://doi.org/10.1103/PhysRevD.98.064051} {\bibfield
   {journal} {\bibinfo  {journal} {Phys. Rev. D}\ }\textbf {\bibinfo {volume}
  {98}},\ \bibinfo {pages} {064051} (\bibinfo {year} {2018})}\BibitemShut
  {NoStop}%
\bibitem [{\citenamefont {Derr}\ and\ \citenamefont {Giese}(2023)}]{Derr2023}%
  \BibitemOpen
  \bibfield  {author} {\bibinfo {author} {\bibfnamefont {D.}~\bibnamefont
  {Derr}}\ and\ \bibinfo {author} {\bibfnamefont {E.}~\bibnamefont {Giese}},\
  }\bibfield  {title} {\bibinfo {title} {Clock transitions versus {B}ragg
  diffraction in atom-interferometric dark-matter detection},\ }\href
  {https://doi.org/10.1116/5.0176666} {\bibfield  {journal} {\bibinfo
  {journal} {AVS Quantum Sci.}\ }\textbf {\bibinfo {volume} {5}},\ \bibinfo
  {pages} {044404} (\bibinfo {year} {2023})}\BibitemShut {NoStop}%
\bibitem [{\citenamefont {Zhao}\ \emph {et~al.}(2024)\citenamefont {Zhao},
  \citenamefont {Liu},\ and\ \citenamefont {Mei}}]{Zhao2024}%
  \BibitemOpen
  \bibfield  {author} {\bibinfo {author} {\bibfnamefont {W.}~\bibnamefont
  {Zhao}}, \bibinfo {author} {\bibfnamefont {H.}~\bibnamefont {Liu}},\ and\
  \bibinfo {author} {\bibfnamefont {X.}~\bibnamefont {Mei}},\ }\bibfield
  {title} {\bibinfo {title} {Ultralight scalar and axion dark matter detection
  with long-baseline atom interferometers},\ }\href
  {https://doi.org/10.1088/0256-307X/41/12/129501} {\bibfield  {journal}
  {\bibinfo  {journal} {Chin. Phys. Lett.}\ }\textbf {\bibinfo {volume} {41}},\
  \bibinfo {pages} {129501} (\bibinfo {year} {2024})}\BibitemShut {NoStop}%
\bibitem [{\citenamefont {Kim}\ and\ \citenamefont {Perez}(2024)}]{Kim2024}%
  \BibitemOpen
  \bibfield  {author} {\bibinfo {author} {\bibfnamefont {H.}~\bibnamefont
  {Kim}}\ and\ \bibinfo {author} {\bibfnamefont {G.}~\bibnamefont {Perez}},\
  }\bibfield  {title} {\bibinfo {title} {Oscillations of atomic energy levels
  induced by {QCD} axion dark matter},\ }\href
  {https://doi.org/10.1103/PhysRevD.109.015005} {\bibfield  {journal} {\bibinfo
   {journal} {Phys. Rev. D}\ }\textbf {\bibinfo {volume} {109}},\ \bibinfo
  {pages} {015005} (\bibinfo {year} {2024})}\BibitemShut {NoStop}%
\bibitem [{\citenamefont {Flambaum}\ and\ \citenamefont
  {Samsonov}(2023)}]{Flambaum2023}%
  \BibitemOpen
  \bibfield  {author} {\bibinfo {author} {\bibfnamefont {V.~V.}\ \bibnamefont
  {Flambaum}}\ and\ \bibinfo {author} {\bibfnamefont {I.~B.}\ \bibnamefont
  {Samsonov}},\ }\bibfield  {title} {\bibinfo {title} {Fluctuations of atomic
  energy levels due to axion dark matter},\ }\href
  {https://doi.org/10.1103/PhysRevD.108.075022} {\bibfield  {journal} {\bibinfo
   {journal} {Phys. Rev. D}\ }\textbf {\bibinfo {volume} {108}},\ \bibinfo
  {pages} {075022} (\bibinfo {year} {2023})}\BibitemShut {NoStop}%
\bibitem [{\citenamefont {Bauer}\ \emph {et~al.}(2025)\citenamefont {Bauer},
  \citenamefont {Chakraborti},\ and\ \citenamefont {Rostagni}}]{Bauer2025}%
  \BibitemOpen
  \bibfield  {author} {\bibinfo {author} {\bibfnamefont {M.}~\bibnamefont
  {Bauer}}, \bibinfo {author} {\bibfnamefont {S.}~\bibnamefont {Chakraborti}},\
  and\ \bibinfo {author} {\bibfnamefont {G.}~\bibnamefont {Rostagni}},\
  }\bibfield  {title} {\bibinfo {title} {Axion bounds from quantum
  technology},\ }\href {https://doi.org/10.1007/JHEP05(2025)023} {\bibfield
  {journal} {\bibinfo  {journal} {{J. High Energy Phys.}}\ }\textbf {\bibinfo
  {volume} {2025}},\ \bibinfo {pages} {23} (\bibinfo {year}
  {2025})}\BibitemShut {NoStop}%
\bibitem [{\citenamefont {Letchumanan}\ \emph {et~al.}(2004)\citenamefont
  {Letchumanan}, \citenamefont {Gill}, \citenamefont {Riis},\ and\
  \citenamefont {Sinclair}}]{Letchumanan2004}%
  \BibitemOpen
  \bibfield  {author} {\bibinfo {author} {\bibfnamefont {V.}~\bibnamefont
  {Letchumanan}}, \bibinfo {author} {\bibfnamefont {P.}~\bibnamefont {Gill}},
  \bibinfo {author} {\bibfnamefont {E.}~\bibnamefont {Riis}},\ and\ \bibinfo
  {author} {\bibfnamefont {A.~G.}\ \bibnamefont {Sinclair}},\ }\bibfield
  {title} {\bibinfo {title} {Optical {R}amsey spectroscopy of a single trapped
  $^{88}${Sr}$^{+}$ ion},\ }\href {https://doi.org/10.1103/PhysRevA.70.033419}
  {\bibfield  {journal} {\bibinfo  {journal} {Phys. Rev. A}\ }\textbf {\bibinfo
  {volume} {70}},\ \bibinfo {pages} {033419} (\bibinfo {year}
  {2004})}\BibitemShut {NoStop}%
\bibitem [{\citenamefont {Abramowitz}\ and\ \citenamefont
  {Stegun}(1964)}]{Abramowitz1964}%
  \BibitemOpen
  \bibinfo {editor} {\bibfnamefont {M.}~\bibnamefont {Abramowitz}}\ and\
  \bibinfo {editor} {\bibfnamefont {I.~A.}\ \bibnamefont {Stegun}},\ eds.,\
  \href@noop {} {\emph {\bibinfo {title} {Handbook of Mathematical Functions
  with Formulas, Graphs, and Mathematical Tables}}},\ \bibinfo {series}
  {Applied Mathematics Series}, Vol.~\bibinfo {volume} {55}\ (\bibinfo
  {publisher} {U.S. Government Printing Office},\ \bibinfo {address}
  {Washington, D.C.},\ \bibinfo {year} {1964})\BibitemShut {NoStop}%
\bibitem [{\citenamefont {Meixner}\ and\ \citenamefont
  {Sch\"{a}fke}(1954)}]{Meixner1954}%
  \BibitemOpen
  \bibfield  {author} {\bibinfo {author} {\bibfnamefont {J.}~\bibnamefont
  {Meixner}}\ and\ \bibinfo {author} {\bibfnamefont {F.~W.}\ \bibnamefont
  {Sch\"{a}fke}},\ }\href {https://doi.org/10.1007/978-3-662-00941-3} {\emph
  {\bibinfo {title} {Mathieusche Funktionen und Sph\"{a}roidfunktionen}}},\
  Grundlehren der mathematischen Wissenschaften\ (\bibinfo  {publisher}
  {Springer Berlin},\ \bibinfo {address} {Heidelberg},\ \bibinfo {year}
  {1954})\BibitemShut {NoStop}%
\bibitem [{\citenamefont {Steinel}\ \emph {et~al.}(2023)\citenamefont
  {Steinel}, \citenamefont {Shao}, \citenamefont {Filzinger}, \citenamefont
  {Lipphardt}, \citenamefont {Brinkmann}, \citenamefont {Didier}, \citenamefont
  {Mehlst{\"a}ubler}, \citenamefont {Lindvall}, \citenamefont {Peik},\ and\
  \citenamefont {Huntemann}}]{Steinel2023}%
  \BibitemOpen
  \bibfield  {author} {\bibinfo {author} {\bibfnamefont {M.}~\bibnamefont
  {Steinel}}, \bibinfo {author} {\bibfnamefont {H.}~\bibnamefont {Shao}},
  \bibinfo {author} {\bibfnamefont {M.}~\bibnamefont {Filzinger}}, \bibinfo
  {author} {\bibfnamefont {B.}~\bibnamefont {Lipphardt}}, \bibinfo {author}
  {\bibfnamefont {M.}~\bibnamefont {Brinkmann}}, \bibinfo {author}
  {\bibfnamefont {A.}~\bibnamefont {Didier}}, \bibinfo {author} {\bibfnamefont
  {T.~E.}\ \bibnamefont {Mehlst{\"a}ubler}}, \bibinfo {author} {\bibfnamefont
  {T.}~\bibnamefont {Lindvall}}, \bibinfo {author} {\bibfnamefont
  {E.}~\bibnamefont {Peik}},\ and\ \bibinfo {author} {\bibfnamefont
  {N.}~\bibnamefont {Huntemann}},\ }\bibfield  {title} {\bibinfo {title}
  {Evaluation of a $^{88}${Sr}$^{+}$ optical clock with a direct measurement of
  the blackbody radiation shift and determination of the clock frequency},\
  }\href {https://doi.org/10.1103/PhysRevLett.131.083002} {\bibfield  {journal}
  {\bibinfo  {journal} {Phys. Rev. Lett.}\ }\textbf {\bibinfo {volume} {131}},\
  \bibinfo {pages} {083002} (\bibinfo {year} {2023})}\BibitemShut {NoStop}%
\bibitem [{\citenamefont {Wineland}\ \emph {et~al.}(1998)\citenamefont
  {Wineland}, \citenamefont {Monroe}, \citenamefont {Itano}, \citenamefont
  {Leibfried}, \citenamefont {King},\ and\ \citenamefont
  {Meekhof}}]{Wineland1998}%
  \BibitemOpen
  \bibfield  {author} {\bibinfo {author} {\bibfnamefont {D.~J.}\ \bibnamefont
  {Wineland}}, \bibinfo {author} {\bibfnamefont {C.}~\bibnamefont {Monroe}},
  \bibinfo {author} {\bibfnamefont {W.~M.}\ \bibnamefont {Itano}}, \bibinfo
  {author} {\bibfnamefont {D.}~\bibnamefont {Leibfried}}, \bibinfo {author}
  {\bibfnamefont {B.~E.}\ \bibnamefont {King}},\ and\ \bibinfo {author}
  {\bibfnamefont {D.~M.}\ \bibnamefont {Meekhof}},\ }\bibfield  {title}
  {\bibinfo {title} {Experimental issues in coherent quantum-state manipulation
  of trapped atomic ions},\ }\href {https://doi.org/10.6028/jres.103.019}
  {\bibfield  {journal} {\bibinfo  {journal} {J. Res. Natl. Inst. Stand.
  Technol.}\ }\textbf {\bibinfo {volume} {103}},\ \bibinfo {pages} {259}
  (\bibinfo {year} {1998})}\BibitemShut {NoStop}%
\bibitem [{\citenamefont {Zhang}\ \emph {et~al.}(2020)\citenamefont {Zhang},
  \citenamefont {Zhou}, \citenamefont {Cheng}, \citenamefont {Chen},
  \citenamefont {Luo}, \citenamefont {Xu}, \citenamefont {Cao}, \citenamefont
  {Duan},\ and\ \citenamefont {Hu}}]{Zhang2020}%
  \BibitemOpen
  \bibfield  {author} {\bibinfo {author} {\bibfnamefont {K.}~\bibnamefont
  {Zhang}}, \bibinfo {author} {\bibfnamefont {M.-K.}\ \bibnamefont {Zhou}},
  \bibinfo {author} {\bibfnamefont {Y.}~\bibnamefont {Cheng}}, \bibinfo
  {author} {\bibfnamefont {L.-L.}\ \bibnamefont {Chen}}, \bibinfo {author}
  {\bibfnamefont {Q.}~\bibnamefont {Luo}}, \bibinfo {author} {\bibfnamefont
  {W.-J.}\ \bibnamefont {Xu}}, \bibinfo {author} {\bibfnamefont {L.-S.}\
  \bibnamefont {Cao}}, \bibinfo {author} {\bibfnamefont {X.-C.}\ \bibnamefont
  {Duan}},\ and\ \bibinfo {author} {\bibfnamefont {Z.-K.}\ \bibnamefont {Hu}},\
  }\bibfield  {title} {\bibinfo {title} {Testing the universality of free fall
  by comparing the atoms in different hyperfine states with {B}ragg
  diffraction},\ }\href {https://doi.org/10.1088/0256-307X/37/4/043701}
  {\bibfield  {journal} {\bibinfo  {journal} {Chin. Phys. Lett.}\ }\textbf
  {\bibinfo {volume} {37}},\ \bibinfo {pages} {043701} (\bibinfo {year}
  {2020})}\BibitemShut {NoStop}%
\bibitem [{\citenamefont {Di~Pumpo}\ \emph {et~al.}(2023)\citenamefont
  {Di~Pumpo}, \citenamefont {Friedrich}, \citenamefont {Ufrecht},\ and\
  \citenamefont {Giese}}]{DiPumpo2023}%
  \BibitemOpen
  \bibfield  {author} {\bibinfo {author} {\bibfnamefont {F.}~\bibnamefont
  {Di~Pumpo}}, \bibinfo {author} {\bibfnamefont {A.}~\bibnamefont {Friedrich}},
  \bibinfo {author} {\bibfnamefont {C.}~\bibnamefont {Ufrecht}},\ and\ \bibinfo
  {author} {\bibfnamefont {E.}~\bibnamefont {Giese}},\ }\bibfield  {title}
  {\bibinfo {title} {Universality-of-clock-rates test using atom interferometry
  with ${T}^{3}$ scaling},\ }\href
  {https://doi.org/10.1103/PhysRevD.107.064007} {\bibfield  {journal} {\bibinfo
   {journal} {Phys. Rev. D}\ }\textbf {\bibinfo {volume} {107}},\ \bibinfo
  {pages} {064007} (\bibinfo {year} {2023})}\BibitemShut {NoStop}%
\bibitem [{\citenamefont {Uhlenbeck}\ and\ \citenamefont
  {Ornstein}(1930)}]{Uhlenbeck1930}%
  \BibitemOpen
  \bibfield  {author} {\bibinfo {author} {\bibfnamefont {G.~E.}\ \bibnamefont
  {Uhlenbeck}}\ and\ \bibinfo {author} {\bibfnamefont {L.~S.}\ \bibnamefont
  {Ornstein}},\ }\bibfield  {title} {\bibinfo {title} {On the theory of the
  {B}rownian motion},\ }\href {https://doi.org/10.1103/PhysRev.36.823}
  {\bibfield  {journal} {\bibinfo  {journal} {Phys. Rev.}\ }\textbf {\bibinfo
  {volume} {36}},\ \bibinfo {pages} {823} (\bibinfo {year} {1930})}\BibitemShut
  {NoStop}%
\bibitem [{\citenamefont {Gardiner}(2004)}]{Gardiner2004}%
  \BibitemOpen
  \bibfield  {author} {\bibinfo {author} {\bibfnamefont {C.~W.}\ \bibnamefont
  {Gardiner}},\ }\href@noop {} {\emph {\bibinfo {title} {Handbook of Stochastic
  Methods for Physics, Chemistry and the Natural Sciences}}},\ \bibinfo
  {edition} {3rd}\ ed.,\ \bibinfo {series} {Springer Series in Synergetics},
  Vol.~\bibinfo {volume} {13}\ (\bibinfo  {publisher} {Springer-Verlag},\
  \bibinfo {address} {Berlin},\ \bibinfo {year} {2004})\BibitemShut {NoStop}%
\bibitem [{\citenamefont {S\'anchez~Vel\'azquez}\ \emph
  {et~al.}(2025)\citenamefont {S\'anchez~Vel\'azquez}, \citenamefont {Steiner},
  \citenamefont {Freund}, \citenamefont {Guevara-Bertsch}, \citenamefont
  {Marciniak}, \citenamefont {Monz},\ and\ \citenamefont
  {Bermudez}}]{Sanchez2025}%
  \BibitemOpen
  \bibfield  {author} {\bibinfo {author} {\bibfnamefont {J.~M.}\ \bibnamefont
  {S\'anchez~Vel\'azquez}}, \bibinfo {author} {\bibfnamefont {A.}~\bibnamefont
  {Steiner}}, \bibinfo {author} {\bibfnamefont {R.}~\bibnamefont {Freund}},
  \bibinfo {author} {\bibfnamefont {M.}~\bibnamefont {Guevara-Bertsch}},
  \bibinfo {author} {\bibfnamefont {C.~D.}\ \bibnamefont {Marciniak}}, \bibinfo
  {author} {\bibfnamefont {T.}~\bibnamefont {Monz}},\ and\ \bibinfo {author}
  {\bibfnamefont {A.}~\bibnamefont {Bermudez}},\ }\bibfield  {title} {\bibinfo
  {title} {Dynamical quantum maps for single-qubit gates under universal
  non-{M}arkovian noise},\ }\href
  {https://doi.org/10.1103/PhysRevResearch.7.013008} {\bibfield  {journal}
  {\bibinfo  {journal} {Phys. Rev. Res.}\ }\textbf {\bibinfo {volume} {7}},\
  \bibinfo {pages} {013008} (\bibinfo {year} {2025})}\BibitemShut {NoStop}%
\bibitem [{\citenamefont {Filzinger}\ \emph {et~al.}(2023)\citenamefont
  {Filzinger}, \citenamefont {D{\"o}rscher}, \citenamefont {Lange},
  \citenamefont {Klose}, \citenamefont {Steinel}, \citenamefont {Benkler},
  \citenamefont {Peik}, \citenamefont {Lisdat},\ and\ \citenamefont
  {Huntemann}}]{Filzinger2023}%
  \BibitemOpen
  \bibfield  {author} {\bibinfo {author} {\bibfnamefont {M.}~\bibnamefont
  {Filzinger}}, \bibinfo {author} {\bibfnamefont {S.}~\bibnamefont
  {D{\"o}rscher}}, \bibinfo {author} {\bibfnamefont {R.}~\bibnamefont {Lange}},
  \bibinfo {author} {\bibfnamefont {J.}~\bibnamefont {Klose}}, \bibinfo
  {author} {\bibfnamefont {M.}~\bibnamefont {Steinel}}, \bibinfo {author}
  {\bibfnamefont {E.}~\bibnamefont {Benkler}}, \bibinfo {author} {\bibfnamefont
  {E.}~\bibnamefont {Peik}}, \bibinfo {author} {\bibfnamefont {C.}~\bibnamefont
  {Lisdat}},\ and\ \bibinfo {author} {\bibfnamefont {N.}~\bibnamefont
  {Huntemann}},\ }\bibfield  {title} {\bibinfo {title} {Improved limits on the
  coupling of ultralight bosonic dark matter to photons from optical atomic
  clock comparisons},\ }\href {https://doi.org/10.1103/PhysRevLett.130.253001}
  {\bibfield  {journal} {\bibinfo  {journal} {Phys. Rev. Lett.}\ }\textbf
  {\bibinfo {volume} {130}},\ \bibinfo {pages} {253001} (\bibinfo {year}
  {2023})}\BibitemShut {NoStop}%
\bibitem [{\citenamefont {Arvanitaki}\ \emph {et~al.}(2018)\citenamefont
  {Arvanitaki}, \citenamefont {Graham}, \citenamefont {Hogan}, \citenamefont
  {Rajendran},\ and\ \citenamefont {Van~Tilburg}}]{Arvanitaki2018}%
  \BibitemOpen
  \bibfield  {author} {\bibinfo {author} {\bibfnamefont {A.}~\bibnamefont
  {Arvanitaki}}, \bibinfo {author} {\bibfnamefont {P.~W.}\ \bibnamefont
  {Graham}}, \bibinfo {author} {\bibfnamefont {J.~M.}\ \bibnamefont {Hogan}},
  \bibinfo {author} {\bibfnamefont {S.}~\bibnamefont {Rajendran}},\ and\
  \bibinfo {author} {\bibfnamefont {K.}~\bibnamefont {Van~Tilburg}},\
  }\bibfield  {title} {\bibinfo {title} {Search for light scalar dark matter
  with atomic gravitational wave detectors},\ }\href
  {https://doi.org/10.1103/PhysRevD.97.075020} {\bibfield  {journal} {\bibinfo
  {journal} {Phys. Rev. D}\ }\textbf {\bibinfo {volume} {97}},\ \bibinfo
  {pages} {075020} (\bibinfo {year} {2018})}\BibitemShut {NoStop}%
\bibitem [{\citenamefont {Derevianko}\ and\ \citenamefont
  {Pospelov}(2014)}]{Derevianko2014}%
  \BibitemOpen
  \bibfield  {author} {\bibinfo {author} {\bibfnamefont {A.}~\bibnamefont
  {Derevianko}}\ and\ \bibinfo {author} {\bibfnamefont {M.}~\bibnamefont
  {Pospelov}},\ }\bibfield  {title} {\bibinfo {title} {Hunting for topological
  dark matter with atomic clocks},\ }\href {https://doi.org/10.1038/nphys3137}
  {\bibfield  {journal} {\bibinfo  {journal} {Nat. Phys.}\ }\textbf {\bibinfo
  {volume} {10}},\ \bibinfo {pages} {933} (\bibinfo {year} {2014})}\BibitemShut
  {NoStop}%
\bibitem [{\citenamefont {Filzinger}\ \emph {et~al.}(2025)\citenamefont
  {Filzinger}, \citenamefont {Caddell}, \citenamefont {Jani}, \citenamefont
  {Steinel}, \citenamefont {Giani}, \citenamefont {Huntemann},\ and\
  \citenamefont {Roberts}}]{Filzinger2025}%
  \BibitemOpen
  \bibfield  {author} {\bibinfo {author} {\bibfnamefont {M.}~\bibnamefont
  {Filzinger}}, \bibinfo {author} {\bibfnamefont {A.~R.}\ \bibnamefont
  {Caddell}}, \bibinfo {author} {\bibfnamefont {D.}~\bibnamefont {Jani}},
  \bibinfo {author} {\bibfnamefont {M.}~\bibnamefont {Steinel}}, \bibinfo
  {author} {\bibfnamefont {L.}~\bibnamefont {Giani}}, \bibinfo {author}
  {\bibfnamefont {N.}~\bibnamefont {Huntemann}},\ and\ \bibinfo {author}
  {\bibfnamefont {B.~M.}\ \bibnamefont {Roberts}},\ }\bibfield  {title}
  {\bibinfo {title} {Ultralight dark matter search with space-time separated
  atomic clocks and cavities},\ }\href
  {https://doi.org/10.1103/PhysRevLett.134.031001} {\bibfield  {journal}
  {\bibinfo  {journal} {Phys. Rev. Lett.}\ }\textbf {\bibinfo {volume} {134}},\
  \bibinfo {pages} {031001} (\bibinfo {year} {2025})}\BibitemShut {NoStop}%
\bibitem [{\citenamefont {Schlippert}\ \emph {et~al.}(2020)\citenamefont
  {Schlippert}, \citenamefont {Meiners}, \citenamefont {Rengelink},
  \citenamefont {Schubert}, \citenamefont {Tell}, \citenamefont {Wodey},
  \citenamefont {Zipfel}, \citenamefont {Ertmer},\ and\ \citenamefont
  {Rasel}}]{Schlippert2020}%
  \BibitemOpen
  \bibfield  {author} {\bibinfo {author} {\bibfnamefont {D.}~\bibnamefont
  {Schlippert}}, \bibinfo {author} {\bibfnamefont {C.}~\bibnamefont {Meiners}},
  \bibinfo {author} {\bibfnamefont {R.~J.}\ \bibnamefont {Rengelink}}, \bibinfo
  {author} {\bibfnamefont {C.}~\bibnamefont {Schubert}}, \bibinfo {author}
  {\bibfnamefont {D.}~\bibnamefont {Tell}}, \bibinfo {author} {\bibfnamefont
  {{\'{E}}.}~\bibnamefont {Wodey}}, \bibinfo {author} {\bibfnamefont {K.~H.}\
  \bibnamefont {Zipfel}}, \bibinfo {author} {\bibfnamefont {W.}~\bibnamefont
  {Ertmer}},\ and\ \bibinfo {author} {\bibfnamefont {E.~M.}\ \bibnamefont
  {Rasel}},\ }\bibinfo {title} {Matter-wave interferometry for inertial sensing
  and tests of fundamental physics},\ in\ \href
  {https://doi.org/10.1142/9789811213984_0010} {\emph {\bibinfo {booktitle}
  {{CPT} and Lorentz Symmetry}}},\ \bibinfo {editor} {edited by\ \bibinfo
  {editor} {\bibfnamefont {R.}~\bibnamefont {Lehnert}}}\ (\bibinfo  {publisher}
  {World Scientific},\ \bibinfo {address} {Singapore},\ \bibinfo {year}
  {2020})\ pp.\ \bibinfo {pages} {37--40}\BibitemShut {NoStop}%
\bibitem [{\citenamefont {Canuel}\ \emph {et~al.}(2018)\citenamefont {Canuel}
  \emph {et~al.}}]{Canuel2018}%
  \BibitemOpen
  \bibfield  {author} {\bibinfo {author} {\bibfnamefont {B.}~\bibnamefont
  {Canuel}} \emph {et~al.},\ }\bibfield  {title} {\bibinfo {title} {Exploring
  gravity with the {MIGA} large scale atom interferometer},\ }\href
  {https://doi.org/10.1038/s41598-018-32165-z} {\bibfield  {journal} {\bibinfo
  {journal} {Sci. Rep.}\ }\textbf {\bibinfo {volume} {8}},\ \bibinfo {pages}
  {14064} (\bibinfo {year} {2018})}\BibitemShut {NoStop}%
\bibitem [{\citenamefont {Zhan}\ \emph {et~al.}(2020)\citenamefont {Zhan} \emph
  {et~al.}}]{Zhan2020}%
  \BibitemOpen
  \bibfield  {author} {\bibinfo {author} {\bibfnamefont {M.-S.}\ \bibnamefont
  {Zhan}} \emph {et~al.},\ }\bibfield  {title} {\bibinfo {title} {{ZAIGA}:
  {Z}haoshan long-baseline atom interferometer gravitation antenna},\ }\href
  {https://doi.org/10.1142/S0218271819400054} {\bibfield  {journal} {\bibinfo
  {journal} {Int. J. Mod. Phys. D}\ }\textbf {\bibinfo {volume} {29}},\
  \bibinfo {pages} {1940005} (\bibinfo {year} {2020})}\BibitemShut {NoStop}%
\bibitem [{\citenamefont {Abe}\ \emph {et~al.}(2021)\citenamefont {Abe} \emph
  {et~al.}}]{Abe2021}%
  \BibitemOpen
  \bibfield  {author} {\bibinfo {author} {\bibfnamefont {M.}~\bibnamefont
  {Abe}} \emph {et~al.},\ }\bibfield  {title} {\bibinfo {title} {Matter-wave
  atomic gradiometer interferometric sensor {(MAGIS-100)}},\ }\href
  {https://doi.org/10.1088/2058-9565/abf719} {\bibfield  {journal} {\bibinfo
  {journal} {Quantum Sci. Technol.}\ }\textbf {\bibinfo {volume} {6}},\
  \bibinfo {pages} {044003} (\bibinfo {year} {2021})}\BibitemShut {NoStop}%
\bibitem [{\citenamefont {Graham}\ \emph {et~al.}(2013)\citenamefont {Graham},
  \citenamefont {Hogan}, \citenamefont {Kasevich},\ and\ \citenamefont
  {Rajendran}}]{Graham2013}%
  \BibitemOpen
  \bibfield  {author} {\bibinfo {author} {\bibfnamefont {P.~W.}\ \bibnamefont
  {Graham}}, \bibinfo {author} {\bibfnamefont {J.~M.}\ \bibnamefont {Hogan}},
  \bibinfo {author} {\bibfnamefont {M.~A.}\ \bibnamefont {Kasevich}},\ and\
  \bibinfo {author} {\bibfnamefont {S.}~\bibnamefont {Rajendran}},\ }\bibfield
  {title} {\bibinfo {title} {New method for gravitational wave detection with
  atomic sensors},\ }\href {https://doi.org/10.1103/PhysRevLett.110.171102}
  {\bibfield  {journal} {\bibinfo  {journal} {Phys. Rev. Lett.}\ }\textbf
  {\bibinfo {volume} {110}},\ \bibinfo {pages} {171102} (\bibinfo {year}
  {2013})}\BibitemShut {NoStop}%
\bibitem [{\citenamefont {Yu}\ and\ \citenamefont {Tinto}(2011)}]{Yu2011}%
  \BibitemOpen
  \bibfield  {author} {\bibinfo {author} {\bibfnamefont {N.}~\bibnamefont
  {Yu}}\ and\ \bibinfo {author} {\bibfnamefont {M.}~\bibnamefont {Tinto}},\
  }\bibfield  {title} {\bibinfo {title} {Gravitational wave detection with
  single-laser atom interferometers},\ }\href
  {https://doi.org/10.1007/s10714-010-1055-8} {\bibfield  {journal} {\bibinfo
  {journal} {Gen. Relativ. Gravit.}\ }\textbf {\bibinfo {volume} {43}},\
  \bibinfo {pages} {1943} (\bibinfo {year} {2011})}\BibitemShut {NoStop}%
\bibitem [{\citenamefont {Kolkowitz}\ \emph {et~al.}(2016)\citenamefont
  {Kolkowitz}, \citenamefont {Pikovski}, \citenamefont {Langellier},
  \citenamefont {Lukin}, \citenamefont {Walsworth},\ and\ \citenamefont
  {Ye}}]{Kolkowitz2016}%
  \BibitemOpen
  \bibfield  {author} {\bibinfo {author} {\bibfnamefont {S.}~\bibnamefont
  {Kolkowitz}}, \bibinfo {author} {\bibfnamefont {I.}~\bibnamefont {Pikovski}},
  \bibinfo {author} {\bibfnamefont {N.}~\bibnamefont {Langellier}}, \bibinfo
  {author} {\bibfnamefont {M.~D.}\ \bibnamefont {Lukin}}, \bibinfo {author}
  {\bibfnamefont {R.~L.}\ \bibnamefont {Walsworth}},\ and\ \bibinfo {author}
  {\bibfnamefont {J.}~\bibnamefont {Ye}},\ }\bibfield  {title} {\bibinfo
  {title} {Gravitational wave detection with optical lattice atomic clocks},\
  }\href {https://doi.org/10.1103/PhysRevD.94.124043} {\bibfield  {journal}
  {\bibinfo  {journal} {Phys. Rev. D}\ }\textbf {\bibinfo {volume} {94}},\
  \bibinfo {pages} {124043} (\bibinfo {year} {2016})}\BibitemShut {NoStop}%
\bibitem [{\citenamefont {Graham}\ \emph {et~al.}(2016)\citenamefont {Graham},
  \citenamefont {Kaplan}, \citenamefont {Mardon}, \citenamefont {Rajendran},\
  and\ \citenamefont {Terrano}}]{Graham2016}%
  \BibitemOpen
  \bibfield  {author} {\bibinfo {author} {\bibfnamefont {P.~W.}\ \bibnamefont
  {Graham}}, \bibinfo {author} {\bibfnamefont {D.~E.}\ \bibnamefont {Kaplan}},
  \bibinfo {author} {\bibfnamefont {J.}~\bibnamefont {Mardon}}, \bibinfo
  {author} {\bibfnamefont {S.}~\bibnamefont {Rajendran}},\ and\ \bibinfo
  {author} {\bibfnamefont {W.~A.}\ \bibnamefont {Terrano}},\ }\bibfield
  {title} {\bibinfo {title} {Dark matter direct detection with
  accelerometers},\ }\href {https://doi.org/10.1103/PhysRevD.93.075029}
  {\bibfield  {journal} {\bibinfo  {journal} {Phys. Rev. D}\ }\textbf {\bibinfo
  {volume} {93}},\ \bibinfo {pages} {075029} (\bibinfo {year}
  {2016})}\BibitemShut {NoStop}%
\bibitem [{\citenamefont {Biedermann}\ \emph {et~al.}(2013)\citenamefont
  {Biedermann}, \citenamefont {Takase}, \citenamefont {Wu}, \citenamefont
  {Deslauriers}, \citenamefont {Roy},\ and\ \citenamefont
  {Kasevich}}]{Biedermann2013}%
  \BibitemOpen
  \bibfield  {author} {\bibinfo {author} {\bibfnamefont {G.~W.}\ \bibnamefont
  {Biedermann}}, \bibinfo {author} {\bibfnamefont {K.}~\bibnamefont {Takase}},
  \bibinfo {author} {\bibfnamefont {X.}~\bibnamefont {Wu}}, \bibinfo {author}
  {\bibfnamefont {L.}~\bibnamefont {Deslauriers}}, \bibinfo {author}
  {\bibfnamefont {S.}~\bibnamefont {Roy}},\ and\ \bibinfo {author}
  {\bibfnamefont {M.~A.}\ \bibnamefont {Kasevich}},\ }\bibfield  {title}
  {\bibinfo {title} {Zero-dead-time operation of interleaved atomic clocks},\
  }\href {https://doi.org/10.1103/PhysRevLett.111.170802} {\bibfield  {journal}
  {\bibinfo  {journal} {Phys. Rev. Lett.}\ }\textbf {\bibinfo {volume} {111}},\
  \bibinfo {pages} {170802} (\bibinfo {year} {2013})}\BibitemShut {NoStop}%
\bibitem [{\citenamefont {Bertoldi}\ \emph {et~al.}(2019)\citenamefont
  {Bertoldi}, \citenamefont {Minardi},\ and\ \citenamefont
  {Prevedelli}}]{Bertoldi2019}%
  \BibitemOpen
  \bibfield  {author} {\bibinfo {author} {\bibfnamefont {A.}~\bibnamefont
  {Bertoldi}}, \bibinfo {author} {\bibfnamefont {F.}~\bibnamefont {Minardi}},\
  and\ \bibinfo {author} {\bibfnamefont {M.}~\bibnamefont {Prevedelli}},\
  }\bibfield  {title} {\bibinfo {title} {Phase shift in atom interferometers:
  {C}orrections for nonquadratic potentials and finite-duration laser pulses},\
  }\href {https://doi.org/10.1103/PhysRevA.99.033619} {\bibfield  {journal}
  {\bibinfo  {journal} {Phys. Rev. A}\ }\textbf {\bibinfo {volume} {99}},\
  \bibinfo {pages} {033619} (\bibinfo {year} {2019})}\BibitemShut {NoStop}%
\bibitem [{\citenamefont {Sanner}\ \emph {et~al.}(2018)\citenamefont {Sanner},
  \citenamefont {Huntemann}, \citenamefont {Lange}, \citenamefont {Tamm},\ and\
  \citenamefont {Peik}}]{Sanner2018}%
  \BibitemOpen
  \bibfield  {author} {\bibinfo {author} {\bibfnamefont {C.}~\bibnamefont
  {Sanner}}, \bibinfo {author} {\bibfnamefont {N.}~\bibnamefont {Huntemann}},
  \bibinfo {author} {\bibfnamefont {R.}~\bibnamefont {Lange}}, \bibinfo
  {author} {\bibfnamefont {C.}~\bibnamefont {Tamm}},\ and\ \bibinfo {author}
  {\bibfnamefont {E.}~\bibnamefont {Peik}},\ }\bibfield  {title} {\bibinfo
  {title} {Autobalanced {R}amsey spectroscopy},\ }\href
  {https://doi.org/10.1103/PhysRevLett.120.053602} {\bibfield  {journal}
  {\bibinfo  {journal} {Phys. Rev. Lett.}\ }\textbf {\bibinfo {volume} {120}},\
  \bibinfo {pages} {053602} (\bibinfo {year} {2018})}\BibitemShut {NoStop}%
\bibitem [{\citenamefont {Brownnutt}\ \emph {et~al.}(2015)\citenamefont
  {Brownnutt}, \citenamefont {Kumph}, \citenamefont {Rabl},\ and\ \citenamefont
  {Blatt}}]{Brownnutt2015}%
  \BibitemOpen
  \bibfield  {author} {\bibinfo {author} {\bibfnamefont {M.}~\bibnamefont
  {Brownnutt}}, \bibinfo {author} {\bibfnamefont {M.}~\bibnamefont {Kumph}},
  \bibinfo {author} {\bibfnamefont {P.}~\bibnamefont {Rabl}},\ and\ \bibinfo
  {author} {\bibfnamefont {R.}~\bibnamefont {Blatt}},\ }\bibfield  {title}
  {\bibinfo {title} {Ion-trap measurements of electric-field noise near
  surfaces},\ }\href {https://doi.org/10.1103/RevModPhys.87.1419} {\bibfield
  {journal} {\bibinfo  {journal} {Rev. Mod. Phys.}\ }\textbf {\bibinfo {volume}
  {87}},\ \bibinfo {pages} {1419} (\bibinfo {year} {2015})}\BibitemShut
  {NoStop}%
\bibitem [{\citenamefont {Clements}\ \emph {et~al.}(2020)\citenamefont
  {Clements}, \citenamefont {Kim}, \citenamefont {Cui}, \citenamefont {Hankin},
  \citenamefont {Brewer}, \citenamefont {Valencia}, \citenamefont {Chen},
  \citenamefont {Chou}, \citenamefont {Leibrandt},\ and\ \citenamefont
  {Hume}}]{Clements2020}%
  \BibitemOpen
  \bibfield  {author} {\bibinfo {author} {\bibfnamefont {E.~R.}\ \bibnamefont
  {Clements}}, \bibinfo {author} {\bibfnamefont {M.~E.}\ \bibnamefont {Kim}},
  \bibinfo {author} {\bibfnamefont {K.}~\bibnamefont {Cui}}, \bibinfo {author}
  {\bibfnamefont {A.~M.}\ \bibnamefont {Hankin}}, \bibinfo {author}
  {\bibfnamefont {S.~M.}\ \bibnamefont {Brewer}}, \bibinfo {author}
  {\bibfnamefont {J.}~\bibnamefont {Valencia}}, \bibinfo {author}
  {\bibfnamefont {J.-S.}\ \bibnamefont {Chen}}, \bibinfo {author}
  {\bibfnamefont {C.-W.}\ \bibnamefont {Chou}}, \bibinfo {author}
  {\bibfnamefont {D.~R.}\ \bibnamefont {Leibrandt}},\ and\ \bibinfo {author}
  {\bibfnamefont {D.~B.}\ \bibnamefont {Hume}},\ }\bibfield  {title} {\bibinfo
  {title} {Lifetime-limited interrogation of two independent
  {$^{27}\mathrm{Al}^{+}$} clocks using correlation spectroscopy},\ }\href
  {https://doi.org/10.1103/PhysRevLett.125.243602} {\bibfield  {journal}
  {\bibinfo  {journal} {Phys. Rev. Lett.}\ }\textbf {\bibinfo {volume} {125}},\
  \bibinfo {pages} {243602} (\bibinfo {year} {2020})}\BibitemShut {NoStop}%
\bibitem [{\citenamefont {Schlosshauer}(2019)}]{Schlosshauer2019}%
  \BibitemOpen
  \bibfield  {author} {\bibinfo {author} {\bibfnamefont {M.}~\bibnamefont
  {Schlosshauer}},\ }\bibfield  {title} {\bibinfo {title} {Quantum
  decoherence},\ }\href {https://doi.org/10.1016/j.physrep.2019.10.001}
  {\bibfield  {journal} {\bibinfo  {journal} {Phys. Rep.}\ }\textbf {\bibinfo
  {volume} {831}},\ \bibinfo {pages} {1} (\bibinfo {year} {2019})}\BibitemShut
  {NoStop}%
\bibitem [{\citenamefont {H{\"a}nsel}\ \emph {et~al.}(2001)\citenamefont
  {H{\"a}nsel}, \citenamefont {Reichel}, \citenamefont {Hommelhoff},\ and\
  \citenamefont {H{\"a}nsch}}]{Haensel2001}%
  \BibitemOpen
  \bibfield  {author} {\bibinfo {author} {\bibfnamefont {W.}~\bibnamefont
  {H{\"a}nsel}}, \bibinfo {author} {\bibfnamefont {J.}~\bibnamefont {Reichel}},
  \bibinfo {author} {\bibfnamefont {P.}~\bibnamefont {Hommelhoff}},\ and\
  \bibinfo {author} {\bibfnamefont {T.~W.}\ \bibnamefont {H{\"a}nsch}},\
  }\bibfield  {title} {\bibinfo {title} {Trapped-atom interferometer in a
  magnetic microtrap},\ }\href {https://doi.org/10.1103/PhysRevA.64.063607}
  {\bibfield  {journal} {\bibinfo  {journal} {Phys. Rev. A}\ }\textbf {\bibinfo
  {volume} {64}},\ \bibinfo {pages} {063607} (\bibinfo {year}
  {2001})}\BibitemShut {NoStop}%
\bibitem [{\citenamefont {Roati}\ \emph {et~al.}(2004)\citenamefont {Roati},
  \citenamefont {de~Mirandes}, \citenamefont {Ferlaino}, \citenamefont {Ott},
  \citenamefont {Modugno},\ and\ \citenamefont {Inguscio}}]{Roati2004}%
  \BibitemOpen
  \bibfield  {author} {\bibinfo {author} {\bibfnamefont {G.}~\bibnamefont
  {Roati}}, \bibinfo {author} {\bibfnamefont {E.}~\bibnamefont {de~Mirandes}},
  \bibinfo {author} {\bibfnamefont {F.}~\bibnamefont {Ferlaino}}, \bibinfo
  {author} {\bibfnamefont {H.}~\bibnamefont {Ott}}, \bibinfo {author}
  {\bibfnamefont {G.}~\bibnamefont {Modugno}},\ and\ \bibinfo {author}
  {\bibfnamefont {M.}~\bibnamefont {Inguscio}},\ }\bibfield  {title} {\bibinfo
  {title} {Atom interferometry with trapped {F}ermi gases},\ }\href
  {https://doi.org/10.1103/PhysRevLett.92.230402} {\bibfield  {journal}
  {\bibinfo  {journal} {Phys. Rev. Lett.}\ }\textbf {\bibinfo {volume} {92}},\
  \bibinfo {pages} {230402} (\bibinfo {year} {2004})}\BibitemShut {NoStop}%
\bibitem [{\citenamefont {Petrucciani}\ \emph {et~al.}(2026)\citenamefont
  {Petrucciani}, \citenamefont {Santoni}, \citenamefont {Mazzinghi},
  \citenamefont {Trypogeorgos}, \citenamefont {Cataliotti}, \citenamefont
  {Inguscio}, \citenamefont {Modugno}, \citenamefont {Smerzi}, \citenamefont
  {Pezz{\`e}},\ and\ \citenamefont {Fattori}}]{Petrucciani2026}%
  \BibitemOpen
  \bibfield  {author} {\bibinfo {author} {\bibfnamefont {T.}~\bibnamefont
  {Petrucciani}}, \bibinfo {author} {\bibfnamefont {A.}~\bibnamefont
  {Santoni}}, \bibinfo {author} {\bibfnamefont {C.}~\bibnamefont {Mazzinghi}},
  \bibinfo {author} {\bibfnamefont {D.}~\bibnamefont {Trypogeorgos}}, \bibinfo
  {author} {\bibfnamefont {F.~S.}\ \bibnamefont {Cataliotti}}, \bibinfo
  {author} {\bibfnamefont {M.}~\bibnamefont {Inguscio}}, \bibinfo {author}
  {\bibfnamefont {G.}~\bibnamefont {Modugno}}, \bibinfo {author} {\bibfnamefont
  {A.}~\bibnamefont {Smerzi}}, \bibinfo {author} {\bibfnamefont
  {L.}~\bibnamefont {Pezz{\`e}}},\ and\ \bibinfo {author} {\bibfnamefont
  {M.}~\bibnamefont {Fattori}},\ }\bibfield  {title} {\bibinfo {title}
  {{M}ach-{Z}ehnder atom interferometry with non-interacting trapped
  {B}ose-{E}instein condensates},\ }\href
  {https://doi.org/10.1038/s41467-026-69692-7} {\bibfield  {journal} {\bibinfo
  {journal} {Nat. Commun.}\ }\textbf {\bibinfo {volume} {17}},\ \bibinfo
  {pages} {3948} (\bibinfo {year} {2026})}\BibitemShut {NoStop}%
\bibitem [{\citenamefont {Arlt}\ \emph {et~al.}(2005)\citenamefont {Arlt},
  \citenamefont {Birkl}, \citenamefont {Rasel},\ and\ \citenamefont
  {Ertmer}}]{Arlt2005}%
  \BibitemOpen
  \bibfield  {author} {\bibinfo {author} {\bibfnamefont {J.}~\bibnamefont
  {Arlt}}, \bibinfo {author} {\bibfnamefont {G.}~\bibnamefont {Birkl}},
  \bibinfo {author} {\bibfnamefont {E.}~\bibnamefont {Rasel}},\ and\ \bibinfo
  {author} {\bibfnamefont {W.}~\bibnamefont {Ertmer}},\ }\bibfield  {title}
  {\bibinfo {title} {Atom optics, guided atoms, and atom interferometry}\
  }(\bibinfo  {publisher} {Academic Press},\ \bibinfo {year} {2005})\ pp.\
  \bibinfo {pages} {55--89}\BibitemShut {NoStop}%
\bibitem [{\citenamefont {Wu}\ \emph {et~al.}(2007)\citenamefont {Wu},
  \citenamefont {Su},\ and\ \citenamefont {Prentiss}}]{Wu2007}%
  \BibitemOpen
  \bibfield  {author} {\bibinfo {author} {\bibfnamefont {S.}~\bibnamefont
  {Wu}}, \bibinfo {author} {\bibfnamefont {E.}~\bibnamefont {Su}},\ and\
  \bibinfo {author} {\bibfnamefont {M.}~\bibnamefont {Prentiss}},\ }\bibfield
  {title} {\bibinfo {title} {Demonstration of an area-enclosing guided-atom
  interferometer for rotation sensing},\ }\href
  {https://doi.org/10.1103/PhysRevLett.99.173201} {\bibfield  {journal}
  {\bibinfo  {journal} {Phys. Rev. Lett.}\ }\textbf {\bibinfo {volume} {99}},\
  \bibinfo {pages} {173201} (\bibinfo {year} {2007})}\BibitemShut {NoStop}%
\bibitem [{\citenamefont {Duspayev}\ and\ \citenamefont
  {Raithel}(2021)}]{Duspayev2021}%
  \BibitemOpen
  \bibfield  {author} {\bibinfo {author} {\bibfnamefont {A.}~\bibnamefont
  {Duspayev}}\ and\ \bibinfo {author} {\bibfnamefont {G.}~\bibnamefont
  {Raithel}},\ }\bibfield  {title} {\bibinfo {title} {Tractor atom
  interferometry},\ }\href {https://doi.org/10.1103/PhysRevA.104.013307}
  {\bibfield  {journal} {\bibinfo  {journal} {Phys. Rev. A}\ }\textbf {\bibinfo
  {volume} {104}},\ \bibinfo {pages} {013307} (\bibinfo {year}
  {2021})}\BibitemShut {NoStop}%
\bibitem [{\citenamefont {Meltzer}\ and\ \citenamefont
  {Sagi}(2024)}]{Meltzer2024}%
  \BibitemOpen
  \bibfield  {author} {\bibinfo {author} {\bibfnamefont {I.}~\bibnamefont
  {Meltzer}}\ and\ \bibinfo {author} {\bibfnamefont {Y.}~\bibnamefont {Sagi}},\
  }\bibfield  {title} {\bibinfo {title} {Atomic clock interferometry using
  optical tweezers},\ }\href {https://doi.org/10.1103/PhysRevA.110.032602}
  {\bibfield  {journal} {\bibinfo  {journal} {Phys. Rev. A}\ }\textbf {\bibinfo
  {volume} {110}},\ \bibinfo {pages} {032602} (\bibinfo {year}
  {2024})}\BibitemShut {NoStop}%
\bibitem [{\citenamefont {Cronin}\ \emph {et~al.}(2009)\citenamefont {Cronin},
  \citenamefont {Schmiedmayer},\ and\ \citenamefont {Pritchard}}]{Cronin2009}%
  \BibitemOpen
  \bibfield  {author} {\bibinfo {author} {\bibfnamefont {A.~D.}\ \bibnamefont
  {Cronin}}, \bibinfo {author} {\bibfnamefont {J.}~\bibnamefont
  {Schmiedmayer}},\ and\ \bibinfo {author} {\bibfnamefont {D.~E.}\ \bibnamefont
  {Pritchard}},\ }\bibfield  {title} {\bibinfo {title} {Optics and
  interferometry with atoms and molecules},\ }\href
  {https://doi.org/10.1103/RevModPhys.81.1051} {\bibfield  {journal} {\bibinfo
  {journal} {Rev. Mod. Phys.}\ }\textbf {\bibinfo {volume} {81}},\ \bibinfo
  {pages} {1051} (\bibinfo {year} {2009})}\BibitemShut {NoStop}%
\bibitem [{\citenamefont {Bott}\ \emph {et~al.}(2026)\citenamefont {Bott},
  \citenamefont {Di~Pumpo},\ and\ \citenamefont
  {Giese}}]{BottScriptMathematica2026}%
  \BibitemOpen
  \bibfield  {author} {\bibinfo {author} {\bibfnamefont {A.}~\bibnamefont
  {Bott}}, \bibinfo {author} {\bibfnamefont {F.}~\bibnamefont {Di~Pumpo}},\
  and\ \bibinfo {author} {\bibfnamefont {E.}~\bibnamefont {Giese}},\ }\href
  {https://github.com/ABott79/Phase-shifts-and-fluctuations-from-motional-relativistic-and-dark-matter-effects-in-ion-trap-clocks}
  {\bibinfo {title} {{Phase shifts and fluctuations from motional,
  relativistic, and dark-matter effects in ion-trap clocks}}} (\bibinfo {year}
  {2026}),\ \bibinfo {note} {version v1.0.0-submitted; to be archived on
  {Zenodo} upon publication}\BibitemShut {NoStop}%
\bibitem [{\citenamefont {McLachlan}(1947)}]{McLachlan1947}%
  \BibitemOpen
  \bibfield  {author} {\bibinfo {author} {\bibfnamefont {N.~W.}\ \bibnamefont
  {McLachlan}},\ }\href@noop {} {\emph {\bibinfo {title} {Theory and
  Application of Mathieu Functions}}}\ (\bibinfo  {publisher} {Clarendon},\
  \bibinfo {address} {Oxford},\ \bibinfo {year} {1947})\BibitemShut {NoStop}%
\bibitem [{\citenamefont {Sutherland}\ \emph {et~al.}(2021)\citenamefont
  {Sutherland}, \citenamefont {Burd}, \citenamefont {Slichter}, \citenamefont
  {Libby},\ and\ \citenamefont {Leibfried}}]{Sutherland2021}%
  \BibitemOpen
  \bibfield  {author} {\bibinfo {author} {\bibfnamefont {R.~T.}\ \bibnamefont
  {Sutherland}}, \bibinfo {author} {\bibfnamefont {S.~C.}\ \bibnamefont
  {Burd}}, \bibinfo {author} {\bibfnamefont {D.~H.}\ \bibnamefont {Slichter}},
  \bibinfo {author} {\bibfnamefont {S.~B.}\ \bibnamefont {Libby}},\ and\
  \bibinfo {author} {\bibfnamefont {D.}~\bibnamefont {Leibfried}},\ }\bibfield
  {title} {\bibinfo {title} {Motional squeezing for trapped ion transport and
  separation},\ }\href {https://doi.org/10.1103/PhysRevLett.127.083201}
  {\bibfield  {journal} {\bibinfo  {journal} {Phys. Rev. Lett.}\ }\textbf
  {\bibinfo {volume} {127}},\ \bibinfo {pages} {083201} (\bibinfo {year}
  {2021})}\BibitemShut {NoStop}%
\bibitem [{\citenamefont {Blanes}\ \emph {et~al.}(2010)\citenamefont {Blanes},
  \citenamefont {Casas}, \citenamefont {Oteo},\ and\ \citenamefont
  {Ros}}]{Blanes2010}%
  \BibitemOpen
  \bibfield  {author} {\bibinfo {author} {\bibfnamefont {S.}~\bibnamefont
  {Blanes}}, \bibinfo {author} {\bibfnamefont {F.}~\bibnamefont {Casas}},
  \bibinfo {author} {\bibfnamefont {J.~A.}\ \bibnamefont {Oteo}},\ and\
  \bibinfo {author} {\bibfnamefont {J.}~\bibnamefont {Ros}},\ }\bibfield
  {title} {\bibinfo {title} {A pedagogical approach to the {M}agnus
  expansion},\ }\href {https://doi.org/10.1088/0143-0807/31/4/020} {\bibfield
  {journal} {\bibinfo  {journal} {Eur. J. Phys.}\ }\textbf {\bibinfo {volume}
  {31}},\ \bibinfo {pages} {907} (\bibinfo {year} {2010})}\BibitemShut
  {NoStop}%
\bibitem [{\citenamefont {Magnus}(1954)}]{Magnus1954}%
  \BibitemOpen
  \bibfield  {author} {\bibinfo {author} {\bibfnamefont {W.}~\bibnamefont
  {Magnus}},\ }\bibfield  {title} {\bibinfo {title} {On the exponential
  solution of differential equations for a linear operator},\ }\href
  {https://doi.org/10.1002/cpa.3160070404} {\bibfield  {journal} {\bibinfo
  {journal} {Commun. Pure Appl. Math.}\ }\textbf {\bibinfo {volume} {7}},\
  \bibinfo {pages} {649} (\bibinfo {year} {1954})}\BibitemShut {NoStop}%
\bibitem [{\citenamefont {Cram{\'e}r}(1999)}]{Cramer1999}%
  \BibitemOpen
  \bibfield  {author} {\bibinfo {author} {\bibfnamefont {H.}~\bibnamefont
  {Cram{\'e}r}},\ }\href@noop {} {\emph {\bibinfo {title} {Mathematical Methods
  of Statistics}}}\ (\bibinfo  {publisher} {Princeton University Press},\
  \bibinfo {address} {Princeton, NJ},\ \bibinfo {year} {1999})\BibitemShut
  {NoStop}%
\end{thebibliography}%
\end{document}